\documentclass[%
aps,pre,amsmath,amssymb,showpacs,preprint,longbibliography,tightenlines
]{revtex4-2}

\usepackage{lineno}
\newcommand*\patchAmsMathEnvironmentForLineno[1]{
  \expandafter\let\csname old#1\expandafter\endcsname\csname #1\endcsname
  \expandafter\let\csname oldend#1\expandafter\endcsname\csname end#1\endcsname
  \renewenvironment{#1}
     {\linenomath\csname old#1\endcsname}
     {\csname oldend#1\endcsname\endlinenomath}}
\newcommand*\patchBothAmsMathEnvironmentsForLineno[1]{
  \patchAmsMathEnvironmentForLineno{#1}
  \patchAmsMathEnvironmentForLineno{#1*}}
\AtBeginDocument{
\patchBothAmsMathEnvironmentsForLineno{equation}
\patchBothAmsMathEnvironmentsForLineno{align}
\patchBothAmsMathEnvironmentsForLineno{flalign}
\patchBothAmsMathEnvironmentsForLineno{alignat}
\patchBothAmsMathEnvironmentsForLineno{gather}
\patchBothAmsMathEnvironmentsForLineno{multline}
}

\usepackage{mathtools}
\usepackage{color}
\usepackage{bigints}
\usepackage{multirow}
\usepackage{graphicx}
\usepackage{dcolumn}
\usepackage{bm}
\usepackage{subfigure}
\usepackage{hyperref}
\usepackage{multirow}
\usepackage{mathtools}

\newif\iffigure
\figurefalse
\figuretrue

\allowdisplaybreaks[3]
\usepackage{cprotect}

\usepackage {cancel}

\usepackage{changes}
\def\red#1{\textcolor{black}{#1}}

\def\spi{\sqrt{\pi}}

\def\dd{\mathrm{d}}

\def\w{\mathrm{w}}
\def\e{\mathrm{e}}

\def\Kn{\mathit{Kn}}
\def\all{\mathrm{all\;}}
\def\RR{\mathbb{R}}
\def\rnd#1#2{\frac{\partial #1}{\partial #2}}
\def\drnd#1#2{\dfrac{\partial #1}{\partial #2}}
\def\trnd#1#2{\partial #1/\partial #2}

\def\dif{\mathrm{dif}}

\def\hf{\frac{1}{2}}

\def\z{\zeta}
\def\zn{\z_n}
\def\zrho{\z_{\rho}}
\def\bpsid{\overline{\psi}_\delta}
\def\I{\mathcal{I}}
\def\dif#1#2{\frac{\dd #1}{\dd #2}}
\def\ddif#1#2{\dfrac{\dd #1}{\dd #2}}
\def\tdif#1#2{\dd #1/\dd #2}

\def\G{\mathrm{G}}
\def\K{\mathrm{K}}

\def\p#1#2{\varphi_{#1}^{(#2)}}
\def\pj#1{\varphi_{j}^{(#1)}}
\def\c#1#2{c_{#1}^{(#2)}}
\def\b#1#2{b_{#1}^{(#2)}}

\def\zero#1{\left(#1\right)_0}

\def\ITJ{\mathcal{I}_{\mathrm{TJ}}\,}
\def\ITS{\mathcal{I}_{\mathrm{TS}}\,}
\def\ISS{\mathcal{I}_{\mathrm{SS}}\,}

\def\tpsi{\widetilde{\psi}_\chi}
\def\psid{\psi_\delta}
\def\talpha{\widetilde{\alpha}}
\def\bz{\bar{\z}}

\def\gKL{g_{\mathrm{KL}}}
\def\IKL{\I_{\mathrm{KL}}}
\def\uKL{u_{\mathrm{KL}}}
\def\alphaKL{\alpha_{\mathrm{KL}}}
\def\gTO{g_{\mathrm{TO}}}
\def\ITO{\I_{\mathrm{TO}}}
\def\uTO{u_{\mathrm{TO}}}
\def\alphaTO{\alpha_{\mathrm{TO}}}

\def\zmax{\z_{\mathrm{max}}}
\def\etamax{\eta_{\mathrm{max}}}

\def\gsf{\text{generalized slip-flow}}
\def\slip{K_{\mathrm{TS}}}

\def\bnum{b_{2,\mathrm{num}}^{(1)}}

\def\q{\phantom{1}}
\def\ss{\phantom{.25}}
\def\s{\phantom{5}}
\def\m{\phantom{-}}

\def\SM{Supplemental Material}

\begin{document}

\preprint{\today, ver.~\number\time}

\newcommand{\titleB}{
Slip-flow theory for thermo-osmosis\\ based on a kinetic model with near-wall potential
}

\title{
\titleB
}

\author{Tetsuro Tsuji}
\email{tsuji.tetsuro.7x@kyoto-u.ac.jp; corresponding author}
\author{Koichiro Takita}%
\author{Satoshi Taguchi}%
\affiliation{%
Graduate School of Informatics, Kyoto University, Kyoto 606-8501, Japan
}%

\date{\today\\\vspace{2em}}

\begin{abstract}
\noindent 
In this paper, thermal-slip coefficients in slip boundary conditions of the Stokes equation are derived using the generalized slip-flow theory, with special interest in the role of near-wall potential in micro- and nanoscale flows. 
As the model of fluids and fluid-solid interaction, we employ the model Boltzmann equation for dilute gases and the diffuse-reflection boundaries with near-wall potential, respectively. It is found that, when the mean free path of gas molecules and the effective range of potential are of the same order of magnitude, the thermal-slip boundary condition can be derived in the near-continuum limit. In the derived slip-flow theory, the thermal-slip coefficient and the boundary-layer corrections (i.e., Knudsen-layer corrections) are determined by solving the kinetic boundary-layer problems (i.e., Knudsen-layer problems) that include external-force terms and inhomogeneous terms both driven by the potential. As \red{an} application of the slip-flow theory, thermo-osmosis between two parallel plates with uniform temperature gradients is analyzed. The results of the slip-flow theory are validated by comparing them with those of the direct numerical analysis of the same problem. Furthermore, it is found that thermo-osmosis and thermal slip on the plates are significantly affected by the features of the near-wall potential; even the gas-flow direction can be reversed when the near-wall potential is repulsive. Such a flow reversal is qualitatively similar to thermo-osmosis in liquids reported in existing molecular dynamics simulation.  
\end{abstract}


\maketitle
\titlepage
\thispagestyle{empty}


\section{\label{sec:intro}Introduction}
In conventional fluid mechanics based on the incompressible Navier-Stokes equations, the temperature field of a fluid is decoupled from the flow field. 
Therefore, the temperature variation of the fluid cannot induce fluid flows without the help of external forces such as temperature-dependent buoyancy forces. 
However, in micro- and nanoscale flows, the temperature variation of boundaries (i.e., channel walls) can induce fluid flows, which we call thermo-osmosis \cite{Derjaguin1987} or thermo-osmotic flows. 
Thermo-osmosis is induced mainly by surface effects, and thus is practically significant in micro- and nanoscale confinements \cite{Bregulla2016} where external body forces are negligible. 
[However, it should be remarked that, when being coupled with electric effects such as ionic motion, there may arise a thermally-induced body force due to the variation of electric permittivity (see, e.g., Ref.~\cite{Dietzel2017}).] 
Thermo-osmosis is a conversion from heat to fluid motion and can be applied to various technologies. For instance, in the case of gases, where thermo-osmosis is often called thermal transpiration, thermally-driven pumps without mechanically-moving parts (the so-called Knudsen pump \cite{Wang2020}) have been developed and prototypes have been put into practice using microfabrication techniques \cite{An2014}. 
In the case of liquids, laser-heating-induced thermo-osmosis near a liquid-solid interface has been applied to the optothermal manipulation of nanomaterials \cite{Fraenzl2022}.

From a macroscopic viewpoint, both for gases and liquids, thermo-osmosis can be regarded as a thermal-slip boundary condition of fluid-dynamic equations (e.g., the Stokes or Navier--Stokes equations), that is, 
\begin{align}
\bm{v}\cdot \bm{t} = - \slip\, \bm{t}\cdot \nabla T \quad (\text{on boundaries}),\label{eq:thermo-osmotic-slip}
\end{align}
where $\bm{v}$ represents the flow velocity, $\bm{t}$ is a unit tangent on boundaries, $\slip$ is a thermal-slip coefficient, and $\nabla T$ is the temperature gradient of the fluid on the boundaries. 
The condition Eq.~\eqref{eq:thermo-osmotic-slip} is sometimes called thermo-osmotic slip  \cite{Ganti2017,Anzini2019,Tsuji2023}. 
The thermal-slip coefficient $\slip$ is usually negative in experiments, i.e., the flow is directed from cold to hot (see, e.g., Ref.~\cite{Sone1991} for gases and Refs.~\cite{Bregulla2016,Tsuji2023} for liquids). 
The concern in this paper is the derivation of Eq.~\eqref{eq:thermo-osmotic-slip} together with the determination of the thermal-slip coefficient $\slip$, based on the molecular-scale description of fluids that includes interaction potential between fluid molecules and solid molecules.

Let us first take an overview of the current situation on the thermal-slip boundary condition [Eq.~\eqref{eq:thermo-osmotic-slip}] for gases. 
In the case of a gas, the ratio between the mean free path $\ell_0$ of gas molecules and a reference length scale $L_0$ (e.g., the channel dimension) is defined as the Knudsen number $(\Kn = \ell_0/L_0)$, which represents the degree of gas rarefaction. When a gas is rarefied, i.e., $\Kn$ is not negligibly small, the gas is out of thermal equilibrium and a conventional fluid-mechanics framework becomes inapplicable; the Boltzmann equation based on the kinetic theory of gases must be employed \cite{Cercignani2000,Sone2007}. However, in the regime of near-local thermal equilibrium where $\Kn$ is small, it is possible to use an asymptotic analysis of the Boltzmann equation (or its simplified models) to systematically derive fluid-dynamic equations and associated slip and jump boundary conditions that describe the gas behavior \cite{Sone2007,Takata2012}. We refer to this framework as the $\gsf$ theory in this paper. The thermal-slip boundary condition [Eq.~\eqref{eq:thermo-osmotic-slip}] can be obtained as one of the outcomes of the $\gsf$ theory. A key significance of the $\gsf$ theory is that, for a given set of inter-molecular collision models and molecular scattering laws at boundaries, the thermal-slip coefficient  $\slip$ can be determined by a systematic procedure. However, since the Boltzmann equation is formulated for dilute gases, the $\gsf$ theory is not directly applicable to dense gases or liquids. 
Also note that thermally-induced gas flows still occur for moderate or large Knudsen numbers. However, such flows lies outside the scope of fluid-dynamic equations and require an analysis based on the Boltzmann equation \cite{Cercignani2000,Sone2007,Ohwada1989a} (see also experimental results for straight channels \cite{RojasCardenas2011,Yamaguchi2016}).  

Regarding liquids, the derivation of the thermal-slip boundary condition [Eq.~\eqref{eq:thermo-osmotic-slip}] is more complicated due to the dense nature of the medium. A theoretical approach was proposed by Derjaguin et al. \cite{Derjaguin1987} using the linear-response theory, and, more recently, a microscopic theory of thermo-osmosis was presented \cite{Anzini2019,Anzini2022,Anzini2025}. 
However, to compute the thermal-slip coefficient $\slip$, these theories require microscopic information such as excess enthalpy profiles near boundaries, which are difficult to measure in experiments. 
In this respect, the prediction of thermo-osmosis for liquids is more difficult than for gases, since the $\gsf$ theory for gases uses the information that is experimentally accessible; for instance, the molecular scattering laws at solid surfaces can be measured using molecular-beam experiments \cite{Kinefuchi2017}.
Therefore, up to now, molecular dynamics (MD) simulation has been the primary practical tool for investigating thermo-osmosis in liquids, and MD simulations have been carried out for various systems such as Lennard-Jones (LJ) fluids \cite{Ganti2017,Fu2017,Wang2020a,Wang2021,Fan2024,Qi2024}, water \cite{Fu2017,Chen2021a}, and aqueous electrolytes \cite{Herrero2022,Chen2023b,Ouadfel2024}. A remarkable trend observed in MD studies is the direction of thermo-osmosis is reversible depending on hydrophilic properties at liquid-solid interfaces. 
To be more precise, the thermal-slip coefficient tends to be $\slip<0$ (or $\slip>0$) for hydrophilic (or hydrophobic) surfaces \cite{Fu2017,Wang2020a,Wang2021,Fan2024,Qi2024}.  
Since the hydrophilic properties are strongly influenced by the interaction potential between fluid and solid molecules, the near-wall potential is considered to play a crucial role in thermo-osmosis. This paper therefore focuses on elucidating the impact of the near-wall interaction potential on the thermal-slip boundary condition [Eq.~\eqref{eq:thermo-osmotic-slip}]. (We also note that the sign reversal of thermo-osmosis was also discussed recently for a near-critical binary fluid mixture \cite{Yabunaka2024}.) 
Now, the central question addressed here is how to obtain the thermal-slip coefficient $\slip$ under the effect of near-wall fluid-solid interactions.

The influence of the near-wall interaction potential on slip coefficients has been extensively investigated in the literature, with primary focus on the so-called shear slip. 
Here, the shear slip is described by the boundary condition $\bm{v}\cdot \bm{t}\approx L_S \dot{\gamma}$, where $L_S$ is the slip length and $\dot{\gamma}$ is the shear rate. 
For instance, using MD simulations (or in experiments), the effects on the slip length $L_S$ caused by the shear-rate dependence \cite{Thompson1997} (or \cite{Craig2001}), the wetting properties at liquid-solid interfaces \cite{Barrat1999a} (or \cite{Zhu2001,Baudry2001,Leger2002,CottinBizonne2002}), and the electrical charge distribution \cite{Xie2020a} (or \cite{Chen2022d}) have been reported. 
Various theoretical frameworks and models to predict $L_S$ have also been elaborated: the results of equilibrium and/or nonequilibrium MD simulations were utilized to obtain $L_s$ \cite{Barrat1999,Hansen2011,Hadjiconstantinou2022,CorralCasas2024} and the modeling of molecular motion in a thin layer adjacent to solid surfaces was proposed \cite{Lichter2004,Wang2019,Hadjiconstantinou2021,Shan2022} (see also the recent review article \cite{Hadjiconstantinou2024}). 
However, the effect of the near-wall potential on thermal slip has not yet been investigated using the existing theoretical frameworks mentioned above. In this paper, given that the $\gsf$ theory has succeeded in obtaining slip coefficients for gases, we try to apply the $\gsf$ theory to account for the effect of the near-wall potential on the thermal-slip coefficient. This serves as a primary step toward developing the model of thermal-slip boundary condition for both gases and liquids based on molecular-scale descriptions. 
(Note that the $\gsf$ theory also provides shear-slip and temperature-jump coefficients as we will discuss later.) As a proof-of-concept study, we adopt a simplified physical setting using the Bhatnager-Gross-Krook (BGK) model \cite{Bhatnagar1954,Welander1954} of the Boltzmann equation to describe the inter-molecular interaction, along with diffuse reflection as the molecular scattering law at boundaries. The equation is supplemented by a force field representing a near-wall interaction potential. In particular, we focus on the case in which the bulk flow is not directly affected by the near-wall potential; only the thermal-slip boundary condition and the boundary-layer corrections are modified by the potential. 

The paper is organized as follows. 
In Sec.~\ref{sec:formula}, we describe the physical setting and formulate the problem in detail. 
In Sec.~\ref{sec:slip-flow-theory}, we develop a slip-flow theory under near-wall interactions based on the $\gsf$ theory. 
In Sec.~\ref{sec:thermo-osmosis}, we apply the slip-flow theory to the problem of thermo-osmosis, which is a special case of the problem defined in Sec.~\ref{sec:formula}, and compare the results with the direct numerical analysis of the same problem. This allows us to validate the slip-flow theory and examine the effects of the near-wall potential on thermo-osmosis. 
In Sec.~\ref{sec:discussion}, we provide a brief comparison with previous studies and highlight some key features of the present slip-flow theory. 
Finally, Sec.~\ref{sec:conclusion} summarizes the main findings and provides future perspectives.

\section{\label{sec:formula}Formulation}
\subsection{Description of the problem\label{sec:problem}}
\begin{figure}[bt]
    \centering
    \includegraphics[width=\textwidth]{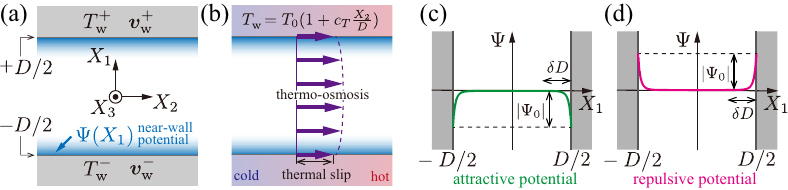}
    \caption{Schematic of the problem. (a) General setting of the slip-flow theory (Sec.~\ref{sec:slip-flow-theory}). (b) Application of the slip-flow theory to the thermo-osmosis problem (Sec.~\ref{sec:thermo-osmosis}). (c,d) Schematic examples of $\Psi$ for (c) purely-attractive and (d) purely-repulsive potential.}
    \label{fig:problem}
\end{figure}

Let us consider an ideal monatomic gas contained between two parallel plates. Cartesian coordinates $X_i$ $(i=1,\,2,\,3)$ are introduced such that the $X_1$ direction is perpendicular to the plates, as shown in Fig.~\ref{fig:problem}(a). The plates are placed at $X_1=\pm D/2$, where $D$ is the distance between the plates. 
The volume average of the gas density is denoted by $\rho_0$. Then, the mean free path of gas molecules at a reference equilibrium state with density $\rho_0$ and temperature $T_0$ is defined as $\ell_0$, where $T_0$ is the reference temperature of the plates. 
The Knudsen number $\Kn$, which characterizes the degree of gas rarefaction, is \red{defined by} $\Kn=\ell_0/D$. 

The velocity and the temperature of the plates at $X_1=\pm D/2$ are kept at $v_{\w i}^\pm$ ($i=1$, $2$, $3$) and $T_\w^\pm$, respectively. 
The velocities of the plates are tangential to the surfaces, meaning that $v_{\w 1}^\pm =0$, while $v_{\w 2}^\pm$ and $v_{\w 3}^\pm$ are constants since the plates are rigid walls. The magnitude of the plate velocity $|v_{\w i}^\pm|$ is small compared with the thermal speed $v_0=\sqrt{2RT_0}$, where $R$ is the specific gas constant, i.e., 
\begin{align}
c_v^\pm=\frac{|v_{\w i}^\pm|}{v_0} \ll 1. \label{eq:cv}
\end{align}
The temperatures of the plates may vary with $X_2$ and $X_3$, but the magnitude of the temperature gradient is assumed to be small, i.e., 
\begin{align}
c_T^\pm=\frac{D}{T_0}\left|t_i\rnd{T_\w^\pm}{X_i}\right|\ll 1, \label{eq:cT}
\end{align}
where $t_i$ is a unit tangent on the plates. 
When the plates are stationary and have infinitesimal temperature gradients in the $X_2$ direction, that is, 
\begin{align}
|v_{\w i}^\pm|=0, \quad T_{\w}^\pm = T_\w(X_2)\equiv T_0\left(1+c_T\frac{X_2}{D}\right) \quad \left(c_T=\frac{D}{T_0}\frac{\dd T_\w}{\dd X_2}\right), \label{eq:assumption-linear}
\end{align}
($c_T \ll 1$), 
the problem is reduced to the fundamental setting of thermo-osmosis, as shown in Fig.~\ref{fig:problem}(b). 
We will mainly analyze this thermo-osmosis setting later in Sec.~\ref{sec:thermo-osmosis}\red{, but Secs.~\ref{sec:formula} and \ref{sec:slip-flow-theory} consider a more general case with $|v_{\w i}^\pm|\neq 0$. }

Gas molecules are subject to external forces per unit mass exerted by \red{a} potential field $\Psi$ near the surfaces. 
The potential $\Psi$ is assumed to depend only on $X_1$, i.e., $\Psi=\Psi(X_1)$, and is symmetric with respect to $X_1=0$. 
The characteristic value of the potential (e.g., at the plates) is given by $\Psi_0\equiv 2RT_0 U$, where $U$ is a dimensionless parameter of the order of unity.
The potential decays to zero as the distance from the plate increases; the characteristic length of its effective range is denoted by $\delta D$ with $\delta(\geq 0)$ a dimensionless parameter. 
For instance, two specific cases, the purely-attractive and purely-repulsive potentials, are shown in Figs.~\ref{fig:problem}(c) and \ref{fig:problem}(d), respectively. These cases will be further discussed in Sec.~\ref{sec:thermo-osmosis}. 

We investigate the steady gas flow under the following assumptions. (i)
The gas behavior is described by the BGK model of the Boltzmann equation \cite{Bhatnagar1954,Welander1954}.
(ii) Gas molecules undergo diffuse reflection \cite{Sone2007} at the boundaries, i.e., at the surfaces of the plates. Specifically, molecules impinging on a boundary are reflected according to the half-range Maxwellian distribution with velocity $v_{\w i}^\pm$ and temperature $T_\w^\pm$ together with the condition that there is no net mass flux across the boundary. 
(iii) The magnitudes of the velocity and temperature gradient of the plates are small [see Eqs.~\eqref{eq:cv} and \eqref{eq:cT}], allowing for linearization of the system around a local equilibrium state defined by setting $c_v^\pm=c_T^\pm=0$. In the absence of interaction potential (that is, when $\delta =0$ or $U=0$), the problem is formally reduced to a classical two-surface problem of a rarefied gas. \red{Therefore, without the potential, the slip-flow theory developed here becomes identical with the one described in the literature (e.g.~Ref.~\cite{Sone2007}).} In this study, we are interested in the effect of a finite interaction potential characterized by $\delta(>0)$, representing long-range fluid-solid interaction, in the fluid-dynamic limit ($\Kn\ll 1$).


\subsection{Basic equations}

Let $f(X_i,\xi_i)$ denote the molecular velocity distribution function, where $\xi_i$ is the molecular velocity vector; let $\rho(X_i)$, $v_i(X_i)$, and $T(X_i)$ denote the density, flow velocity, and temperature of the gas. Then, the BGK model incorporating the effect of \red{a} potential is written as 
\begin{subequations}\label{eq:bgk}
\begin{align}
&
\xi_i\rnd{f}{X_i} - \ddif{\Psi}{X_1}\rnd {f}{\xi_1} = A_c \rho (f_\e -f ), \\
&
f_\e = \frac{\rho}{(2\pi R T)^{3/2}}\exp\left(-\frac{(\xi_i-v_i)^2}{2RT}\right), \label{e:BGK_localM}\\
&
\begin{bmatrix}
\rho\\\rho v_i\\
3\rho R T
\end{bmatrix}
=
\int_{\all \xi_i} 
\begin{bmatrix}
1\\ \xi_i\\
(\xi_i-v_i)^2
\end{bmatrix} f\; \dd \bm{\xi}, \label{e:bgk_macro}
\end{align}
\end{subequations}
where $A_c$ is a constant such that $A_c \rho$ represents the collision frequency, and $\dd \bm{\xi}$ is a short-hand notation of $\dd \xi_1 \dd \xi_2 \dd \xi_3$. \red{Note that in this model, $f$ is coupled with $\rho$, $v_i$, and $T$ through Eqs.~\eqref{e:BGK_localM} and \eqref{e:bgk_macro}, while $\Psi$ and $A_c$ are prescribed. The relations between $f$ and $v_i$ and $T$ are nonlinear.} The reference density $\rho_0$ and the reference equilibrium distribution $f_0$ are expressed as $\rho_0=\int_{-D/2}^{D/2} \rho(X_1)\dd X_1$ and $f_0= \rho_0(2\pi R T_0)^{-3/2}\exp(-\xi_i^2/(2RT_0))$, respectively. 
The potential $\Psi(X_1)$ has a characteristic value of $\Psi_0(=2RT_0U)$ near the plates and decays to zero as the distance from the plates increases. 
The full detail of $\Psi$ will be specified [cf. Eq.~\eqref{eq:potential-dim}] in the analysis in Sec.~\ref{sec:thermo-osmosis}. 

The diffuse-reflection boundary condition \cite{Sone2007} is given by
\begin{subequations}\label{eq:diffuse}
\begin{align}
&f = \frac{\sigma_{\w\pm}}{(2\pi R T_\w^\pm)^{3/2}} \exp\left(-\frac{(\xi_i-v_{\w i}^\pm)^2}{2RT_\w^\pm}\right)\quad \left(X_1=\pm \frac{D}{2},\;\xi_1\lessgtr 0\right),
\end{align}
\red{with}
\begin{align}
\sigma_{\w\pm} = \pm\left(\frac{2\pi}{RT_\w^\pm}\right)^{1/2}\int_{\xi_1 \gtrless 0 }
\xi_1 f \;\dd\bm{\xi}. \label{eq:sigma_w}
\end{align}
\end{subequations}
\red{Here, $\sigma_{\w+}$ and $\sigma_{\w -}$ are the quantities determined by $f$ via Eq.~\eqref{eq:sigma_w}, which ensures that there are no net mass fluxes across the boundaries (the impermeability condition), i.e., $\int \xi_1 f \dd \bm{\xi}=0$ at $X_1=\pm D/2$ \cite{Sone2007}.} 
\red{It should be remarked that the kinetic boundary condition, Eq.~\eqref{eq:diffuse}, is defined for reflected molecules, i.e., $\xi_1<0$ at $X_1=D/2$ or $\xi_1>0$ at $X_1=-D/2$, and that no condition is imposed for the molecules arriving at the walls.} Other macroscopic quantities, such as the pressure, stress tensor, and heat flux of the gas, are defined as 
\begin{align}
\begin{bmatrix}
p\\
p_{ij}\\
q_{i}
\end{bmatrix}
=
\int_{\all \xi_i} 
\begin{bmatrix}
\frac{1}{3}(\xi_i-v_i)^2\\ 
(\xi_i-v_i)(\xi_j-v_j)\\
\frac{1}{2}(\xi_i-v_i)(\xi_j-v_j)^2
\end{bmatrix} f\; \dd \bm{\xi}. 
\end{align}


\subsection{Linearized problem}
The smallness assumption [Eqs.~\eqref{eq:cv} and \eqref{eq:cT}] allows us to linearize the problem described by Eqs.~\eqref{eq:bgk} and \eqref{eq:diffuse}. 
First, let us consider the case where $c_v^\pm=c_T^\pm=0$ while the potential remains nonzero. In this case, the solution to Eqs.~\eqref{eq:bgk} and \eqref{eq:diffuse} is given by the local equilibrium $f=f_0^\ast$, where 
\begin{subequations}\label{eq:f0ast}
\begin{align}
&f_0^\ast(X_1,\xi_i) = \frac{\rho_0^\ast(X_1)}{(2\pi R T_0)^{3/2}} \exp\left(-\frac{\xi_i^2}{2RT_0}\right)\left(=\frac{\rho_0^\ast}{\rho_0}f_0\right), 
\\
&
\rho_0^\ast(X_1) = \frac{\exp(-\Psi(X_1)/R T_0)}{\overline{\Psi}} \rho_0, \label{eq:ref-density}\\
&
\overline{\Psi} = \int_{-D/2}^{D/2} \exp\left(-\frac{\Psi(X_1)}{R T_0}\right) \dd X_1. 
\end{align}
\end{subequations}
In the following, we consider the perturbation from the local Maxwellian distribution $f_0^\ast$. To this end, we first introduce dimensionless quantities as 
\begin{equation}\label{eq:nondimensional}
\begin{split}
&
x_i=X_i/D, \quad
\z_i = \xi_i/v_0, \quad 
\phi=f/f_0^\ast - 1, \quad 
\psid = \Psi/(2RT_0U), \\
&\tau_\w^\pm = T_\w^\pm/T_0 -1, \quad 
u_{\w i}^\pm = v_{\w i}^\pm/v_0, \quad 
\omega = \rho/\rho_0^\ast - 1, \quad 
u_i=v_i/v_0, \\
& \tau = T/T_0 - 1, \quad 
P=p/p_0 - 1, \quad 
P_{ij}=p_{ij}/p_0 - \delta_{ij}, \quad 
Q_{i}=q_{i}/(p_0v_0), \quad 
\end{split}
\end{equation}
where $p_0=\rho_0R T_0$ denotes the reference pressure, \red{$v_0=\sqrt{2RT_0}$ is taken as the reference speed,} and the subscript $\delta$ of $\psid$ explicitly indicates that the (dimensionless) potential includes the parameter $\delta$. 
Substituting Eq.~\eqref{eq:nondimensional} into Eqs.~\eqref{eq:bgk} and \eqref{eq:diffuse} and neglecting the higher-order terms of perturbation, i.e., $\phi$, $\omega$, $u_i$, $\tau$, $u_{\w i}^\pm$, and $\tau_\w^\pm$, we obtain the following linearized BGK equation: 
\begin{subequations}\label{eq:bgk-nd}
\begin{align}
&
\z_i\rnd{\phi}{x_i} - U \ddif{\psid}{x_1}\rnd{\phi}{\z_1} = \frac{\alpha}{k} (\phi_\e - \phi), \label{eq:bgk-nd-a}\\
&
\phi_\e = \omega + 2 \z_i u_i + \left(\z^2-\frac{3}{2}\right)\tau, \label{eq:bgk-nd-b}\\
&
\begin{bmatrix}
\omega\\ u_i\\
\tau
\end{bmatrix}
=
\int_{\RR^3} 
\begin{bmatrix}
1\\ \z_i\\
\frac{2}{3}\left(\z^2-\frac{3}{2}\right)
\end{bmatrix} \phi E\; \dd \bm{\z}, \label{eq:bgk-nd-c}\\
&
k = \frac{\sqrt{\pi}}{2}\Kn,\quad \Kn = \frac{\ell_0}{D}, \quad 
\ell_0 = \frac{2}{\spi}\frac{v_0}{A_c \rho_0}, \label{eq:bgk-nd-d}
\end{align}
\end{subequations}
where $E=\pi^{-1/2}\exp(-\z^2)$, $\z=|\z_i|=(\zeta_j^2)^{1/2}$, and $\dd \bm{\z}$ is a short-hand notation of $\dd \z_1 \dd \z_2 \dd \z_3$. The auxiliary functions associated with the potential are given by 
\begin{align}
\alpha = \frac{\exp(-2U\psid)}{\bpsid}  \quad 
\left(\bpsid = \int_{-1/2}^{1/2} \exp(-2U\psid(x_1)) \;\dd x_1\right). 
\label{eq:alpha}
\end{align}
Here, $\alpha$ represents the dimensionless reference density corresponding to Eq.~\eqref{eq:ref-density}. 
\red{Once the functional form of the potential $\psid$ is specified, Eq.~\eqref{eq:bgk-nd} forms a system of equations for unknown functions $\phi$, $\omega$, $u_i$, and $\tau$, while $U$, $\alpha$, and $k$ are external parameters.}
In the following, we use $k$ in place of $\Kn$ [\red{see} Eq.~\eqref{eq:bgk-nd-d}] for notational simplicity. The usual BGK model is formally recovered by setting $U=0$ and $\alpha = 1$. 
The boundary condition is linearized in the same manner as 
\begin{subequations}\label{eq:bc-nd}
\begin{align}
&
\phi = \phi_{\w}^\pm \quad \left(x_1=\pm \frac{1}{2},\;\z_1\lessgtr0\right),
\label{eq:bc-nd-a}
\end{align}
\red{with}
\begin{align}
\phi_{\w}^\pm = 2\z_iu_{\w i}^\pm + (\z^2-2)\tau_\w^\pm  \pm 2\spi\int_{\z_1\gtrless 0} 
\z_1 \phi E \; \dd \bm{\z}. \label{eq:bc-nd-b}
\end{align}
\end{subequations}
\red{The $\phi_{\w}^\pm$ are functions of $\zeta_i$, in which $u_{\w i}^\pm$ and $\tau_\w^\pm$ are prescribed. Note that $\phi_{\w}^\pm$ depends on $\phi$ through the integral on the right-hand side (i.e., the last term).
} 
The other (dimensionless) macroscopic quantities are defined as 
\begin{align}
P=\omega + \tau, \quad 
P_{ij}
=
2\int_{\RR^3} 
\z_i \z_j\phi E\; \dd \bm{\z}, 
\quad 
Q_i
=
\int_{\RR^3} 
\z_i \z^2
\phi E\; \dd \bm{\z} - \frac{5}{2} u_i. \label{eq:macro}
\end{align}
In the following, we analyze Eqs.~\eqref{eq:bgk-nd}--\eqref{eq:bc-nd} based on the generalized slip-flow theory, presented in Sec.~\ref{sec:slip-flow-theory}, to obtain the slip and jump coefficients in the presence of the near-wall potential $\Psi$ (or $\psid$) of the plates. Then, in Sec.~\ref{sec:thermo-osmosis}, we compare the slip-flow theory with the direct numerical analysis of Eqs.~\eqref{eq:bgk-nd}--\eqref{eq:bc-nd} for the thermo-osmosis problem. 


\section{Slip-flow theory with near-wall potential}\label{sec:slip-flow-theory}
\subsection{Assumptions}
\begin{figure}[bt]
    \centering
    \includegraphics[width=0.7\textwidth]{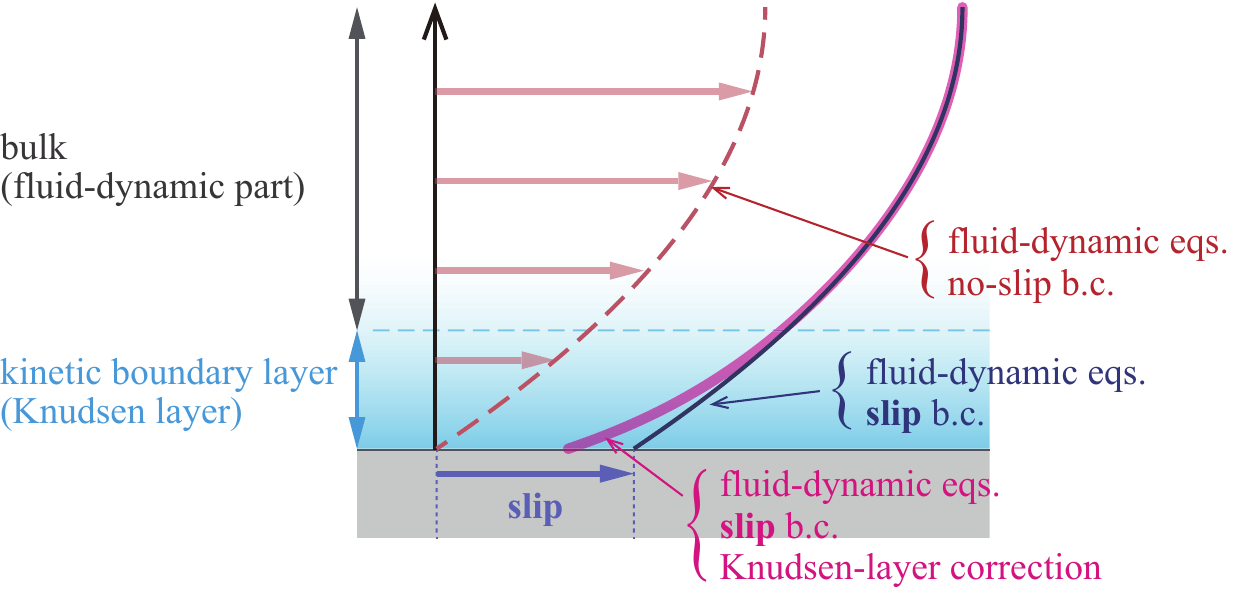}
    \caption{Schematics of the slip-flow theory. As an illustrative example, a parallel flow near a stationary boundary is considered. The entire region is described by the fluid-dynamic equations with no-slip boundary conditions (red). Near the boundary, the Knudsen layer plays two roles: (i) the slip boundary condition modifies the fluid-dynamic part (blue) and (ii) the Knudsen-layer correction modifies the macroscopic quantities near the boundary. The Knudsen-layer correction rapidly decays as the distance from the boundary increases.}
    \label{fig:slipflowtheory}
\end{figure}

Hereafter, we assume $k\ll 1$ and apply the generalized slip-flow theory to investigate the gas behavior. In this framework, the whole gas domain is decomposed into the bulk region and the kinetic boundary layer adjacent to the boundary (i.e., the Knudsen layer). 
The schematics of the bulk and the Knudsen layer are shown in Fig.~\ref{fig:slipflowtheory}. 
In short, using the generalized slip-flow theory, the fluid-dynamic part in the entire region, i.e., the bulk and the Knudsen layer, will be described by the fluid-dynamic equations of the macroscopic quantities (e.g., the Stokes equation); the boundary-layer analysis (i.e., the Knudsen-layer analysis) provides appropriate slip and jump boundary conditions for the fluid-dynamic equations, along with corrections that are significant only in the boundary layer.

We further assume that the range of interaction, $\delta$, is as small as the Knudsen number $k$, meaning we consider the scaling
\begin{align}
\chi \equiv \delta/k = O(1), \quad k \ll 1. \label{eq:same-order}
\end{align}
The validity of this scaling assumption Eq.~\eqref{eq:same-order} will be examined later in Sec.~\ref{sec:adequateness}. 
Under the scaling Eq.~\eqref{eq:same-order}, we perform an asymptotic analysis based on a power series expansion in $k$, with particular attention to the differences between the cases with and without the near-wall potential $\psid$. 
In Eq.~\eqref{eq:bgk-nd-a}, these differences appear in the terms $U(\tdif{\psid}{x_1})(\trnd{\phi}{\z_1})$ and $\alpha(\neq 1)$. 
However, under the assumption Eq.~\eqref{eq:same-order}, these differences are only significant within the boundary layer, which has a length scale of $k$. Consequently, we neglect these terms in the analysis of fluid-dynamic part, while retaining them as inhomogeneous terms in the Knudsen-layer analysis. In this way, the fluid-dynamic part remains unchanged from the generalized slip-flow theory without the near-wall potential. However, the scaling Eq.~\eqref{eq:same-order} suggests that the range of the potential (e.g., the molecular size) is of the same order as the mean free path, implying that the effect of finite volume fraction should become relevant. In this study, we neglect finite-volume effects to focus on the influence of the near-wall potential.

When $\chi$ is small, say $\chi=O(k)$ [or equivalently $\delta=O(k^2)$], a different situation arises: the range of the potential is localized within an inner layer that is thinner than the Knudsen layer. In this case, the kinetic boundary condition can be derived \cite{Aoki2022} by analyzing the inner layer problem, where gas molecules interact with solid molecules through attractive-repulsive potential. The inner layer problem was further mathematically justified \cite{Aoki2024}. 

\subsection{Main results}
We describe only the main results here, leaving the details and the derivation in \SM~\ref{sec:slip-flow-theory-detail}. 

The macroscopic quantities $h$ ($h=\omega$, $u_i$, $\tau$, etc) are decomposed as 
\begin{align}
h=h_\G + h_\K.    
\end{align} 
Here, $h_\G$, referred to as the Hilbert part hereafter, describes the overall gas behavior throughout the gas region, while $h_\K$, referred to as the Knudsen-layer correction, corresponds to the corrections to the Hilbert part required near the plate surfaces. The Hilbert part $h_\G$ is moderately varying, i.e., $\trnd{h_{\G}}{x_i}=O(h_\G)$. On the other hand, $h_\K$ varies in the scale of the Knudsen number (or the mean free path) in the direction normal to the boundary and decays fast as the distance from the plates increases. 

The results of the slip-flow theory are summarized as follows. 
First, neglecting the terms of $O(k^2)$, the Hilbert part is governed by the following fluid-dynamic equations:
\begin{subequations}\label{eq:Stokeseqs}
\begin{align}
&\rnd{u_{i\G}}{x_i}=0\quad (\text{the equation of continuity}), \\
&\rnd{P_\G}{x_i} = k \rnd{^2u_{i\G}}{x_j^2} \quad (\text{the Stokes equation}), \label{eq:Stokes-maintext}\\
&\rnd{^2\tau_\G}{x_j^2} = 0 \quad (\text{the heat equation}), \\
&\omega_\G = P_\G - \tau_\G \quad (\text{the equation of state}),
\end{align}
\end{subequations}
where the dimensionless viscosity in Eq.~\eqref{eq:Stokes-maintext} is reduced to $k$ in the case of the BGK model. 
The system is supplemented by the slip and jump boundary conditions
\begin{subequations}\label{eq:slipbcs}
\begin{align}
&(u_{i\G}-u_{\w i }^\pm)t_i=\mp k\b11 t_i \drnd{u_{i\G}}{x_1}+k\b21t_i\drnd{\tau_{\G}}{x_i}  \quad 
\left(x_1=\pm\frac{1}{2}\right)\quad (\text{slip boundary condition}),
\label{eq:slip-coef}\\
&\tau_{\G}-\tau_{\w}^\pm=\mp k\c10 \drnd{\tau_{\G}}{x_1} \quad 
\left(x_1=\pm\frac{1}{2}\right)  \quad (\text{jump boundary condition}),
\label{eq:jump-coef}
\end{align}
and the impermeability condition
\begin{align}
    u_{1\G} = 0  \quad \left(x_1=\pm\frac{1}{2}\right).
\end{align}
\end{subequations}
Here, $t_i$ denotes the unit tangent vector on the boundaries, and $\b11$, $\b21$, and $\c10$ represent the shear-slip, thermal-slip, and temperature-jump coefficients, respectively. Notations for these constants follow those of Ref.~\cite{Takata2012}. 
The slip and jump coefficients are predefined constants that depend on the inter-molecular collision model, the molecular scattering law at the boundaries, and the near-wall potential, and can be obtained by solving the Knudsen-layer problems summarized in Appendix~\ref{sec:KLother}. These problems are formulated as boundary-value problems for linearized BGK-type equations with external forces, and the derivation of the problems is fully described in \SM~\ref{sec:KL}. Because the potential considered here is symmetric about the channel center ($x_1=0$), the coefficients $\b11$, $\b21$, and $\c10$ take identical values at both boundaries $x_1=\pm1/2$. The expressions for the stress tensor and the heat flux are given in \SM~\ref{sec:fluid-dynamic-part}.

Next, we describe the Knudsen-layer corrections $h_\K$. 
We introduce a stretched coordinate in the Knudsen layer near the plate at $x_1=1/2$ as \begin{align}
\eta = \frac{1/2-x_1}{k}.     
\end{align}
Then, the potential near the plate in the stretched coordinate is denoted by $\tpsi(\eta)$, where the subscript $\chi$ indicates explicitly that the potential includes the parameter $\chi$ [Eq.~\eqref{eq:same-order}]. The overall potential $\psid$ is divided as $\psid=\psid^+ + \psid^-$, where $\psid^\pm(x_1)$ represents the potential by the plate at $x_1=\pm1/2$. We define  $\tpsi(\eta)=\psid^+(x_1)$,
that is, the contribution from $\psid^-$ is discarded in $\tpsi(\eta)$ since the potentials are assumed to be localized near the plates at $x_1=\pm 1/2$.
The description of the region $x_1\geq 0$ is sufficient due to the symmetry with respect to $x_1=0$, and the effect of the potential $\psid^-$ is negligible in this region. 
Then, the Knudsen-layer corrections near the plate $x_1=1/2$ are given as
\begin{subequations}\label{eq:KLcorrections}
\begin{align}
&u_{i\K}t_i =   -kY_1^{(1)}\left(\eta\right)\left.t_i\drnd{u_{i\G}}{x_1}\right|_{x_1=1/2}
 +kY_2^{(1)}\left(\eta\right)\left.t_i\drnd{\tau_{\G}}{x_i} \right|_{x_1=1/2}  
,\\
&u_{1\K} =0, \\
&\tau_{\K} = -k\Theta_1^{(0)}\left(\eta\right)\left.\drnd{\tau_{\G}}{x_1}\right|_{x_1=1/2}  
, \\
&\omega_{\K} = 2U\tpsi(\eta)\tau_\G|_{x_1=1/2} -k\Omega_1^{(0)}\left(\eta\right)\left.\drnd{\tau_{\G}}{x_1}\right|_{x_1=1/2}, \\
&P_\K=\omega_\K + \tau_\K, 
\end{align}
\end{subequations}
where $h_\K=h_\K(\eta,\,x_2,\,x_3)$ with $h=u_{i\K}$, $\tau_\K$, $\omega_\K$, and $P_\K$. 
The functions $Y_1^{(1)}(\eta)$, $Y_2^{(1)}(\eta)$, $\Omega_1^{(0)}(\eta)$, and $\Theta_1^{(0)}(\eta)$ represent the so-called Knudsen-layer functions. 
These Knudsen-layer functions are predefined functions determined alongside the slip and jump coefficients $\b11$, $\b21$, and $\c10$ by solving the Knudsen-layer problems. 
Note that $Y_1^{(1)}(\eta)$, $Y_2^{(1)}(\eta)$, $\Omega_1^{(0)}(\eta)$, and $\Theta_1^{(0)}(\eta)$ all decays as $\eta\to\infty$, as we will see later in Sec.~\ref{sec:effect-of-U} for $Y_2^{(1)}(\eta)$.

The strategy of the flow-behavior analysis is as follows. (i) First, we solve the Knudsen-layer problems [Eqs.~\eqref{eq:KLproblem-decomposed21-maintext}, \eqref{eq:KLproblem-decomposed10-maintext}, and \eqref{eq:KLproblem-decomposed11-maintext}]. This process is done numerically in principle, but the recipe has been presented in the literature \cite{Ohwada1989}. Note that the Knudsen-layer problems does not include $k$ as the physical parameter. (ii) Second, under the specific setting of boundary values $u_{\w i}^\pm$ and $\tau_\w^\pm$, we solve the fluid-dynamic equations Eq.~\eqref{eq:Stokeseqs} with the slip and jump boundary conditions Eq.~\eqref{eq:slipbcs} to obtain the Hilbert part. This process can be done either analytically or numerically, depending on the situations. In any case, the cost for solving the fluid-dynamics equations is expected to be much smaller than that for solving the original system Eqs.~\eqref{eq:bgk-nd}--\eqref{eq:bc-nd}. (iii) Finally, we add the Knudsen-layer corrections Eq.~\eqref{eq:KLcorrections} to the Hilbert part. If the flow behavior in the bulk is only interested, the process (iii) may be omitted. \red{In this paper, we will discuss the flow field near the boundary and include the Knudsen-layer corrections as we will see in Sec.~\ref{sec:thermo-osmosis}.}

We leave a couple of remarks. (i) We consider the expansion up to the first-order in $k$ and neglect the contributions of $O(k^2)$ in this paper. However, it is possible to extend the $\gsf$ theory to incorporate higher-order effects, including the effects of curved surfaces \cite{Sone2007}. 
(ii) The analysis presented so far is also applicable to the Boltzmann equation \cite{Sone2007}. 
In \red{this} study, we have adopted the BGK model for simplicity, as it enables us to understand the framework of the generalized slip-flow theory more easily. A known limitation of the BGK model, however, is that it fixes  
the Prandtl number of the gas to unity. 
This drawback can be amended by employing more sophisticated (but complex) models, such as the ellipsoidal-statistical (ES) model \cite{Holway1966} or the Shakhov model \cite{Shakhov1972}. 
The framework developed in the present paper is expected to be extendable to these models in a straightforward manner.


\section{Analysis of thermo-osmosis problem}\label{sec:thermo-osmosis}

In this section, we investigate the thermo-osmosis problem as defined by Eq.~\eqref{eq:assumption-linear} and illustrated in Fig.~\ref{fig:problem}(b). 
To analyze this problem, we employ two approaches: (i) the slip-flow theory developed in Sec.~\ref{sec:slip-flow-theory} and (ii) direct numerical analysis. We then compare both results for small $\Kn$ (or $k$) to validate the slip-flow theory. Note that the approach (i) provides an explicit analytical solution, although the numerical value of the thermal-slip coefficient (i.e., $\b21$), which is problem-independent, must be determined numerically.


\subsection{Detailed setting}
First, we set the boundary parameters as
\begin{align}
v_{\w i}^\pm = 0, \quad 
\tau_{\w }^\pm = \tau_\w \equiv c_T x_2,
\label{eq:thermo-osmosis-problem}
\end{align}
[see Eq.~\eqref{eq:assumption-linear}].
To obtain the flow field, the potential $\Psi(X_1)$ must be specified. 
We use the following pseudo-Sutherland-type potential [see Figs.~\ref{fig:problem}(b,\,c)]: 
\begin{align}
\begin{cases}
\Psi = \Psi^+ + \Psi^-,\\ 
\Psi^{\pm}(X_1) = - \Psi_0 \left(\dfrac{\delta D}{\pm (1/2+\delta )D - X_1 }\right)^6, 
\end{cases}
\label{eq:potential-dim}
\end{align}
where $\Psi^{+}$ and $\Psi^{-}$ represent the potentials near the plates located at $X_1=D/2$ and $-D/2$, respectively, and $\Psi_0 = 2RT_0 U$. 
When $U$ is positive (or negative), the potential is purely attractive (or repulsive).
\red{Physically, an attractive potential with the power $r_0^{-6}$, where $r_0$ is the distance from the wall, corresponds to the van der Waals interaction potential \cite{Israelachvili2011}.} The interaction parameter $\delta$ determines the effective range of the potential and is assumed to be small. The corresponding dimensionless potential shape, $\psid(x_1)$, and its corresponding form in the Knudsen-layer analysis, $\tpsi(\eta)$, are given by
\begin{subequations}
\begin{align}
& 
\begin{cases}
\psid=\psid^++\psid^-, \quad 
 \psid^{\pm}(x_1) = -\left(\dfrac{\delta}{\pm(1/2+\delta)-x_1}\right)^6, 
\\
\ddif{\psid}{x_1} = 
\dfrac{6}{1/2+\delta-x_1}\psid^+ +
\dfrac{6}{-(1/2+\delta)-x_1}\psid^-, 
\end{cases}\label{eq:psi-nd}
\\
&\tpsi(\eta)= -\left(\frac{\chi}{\chi+\eta}\right)^6, \quad 
\ddif{\tpsi}{\eta} = -\frac{6}{\chi+\eta}\tpsi. \label{eq:psi-nd-KL}
\end{align}
\end{subequations}
Note that $\tpsi$ depends on $\chi$ rather than $\delta$ [see Eq.~\eqref{eq:psi-nd-KL}].

\begin{figure}[tb]
    \centering
    \includegraphics[width=\textwidth]{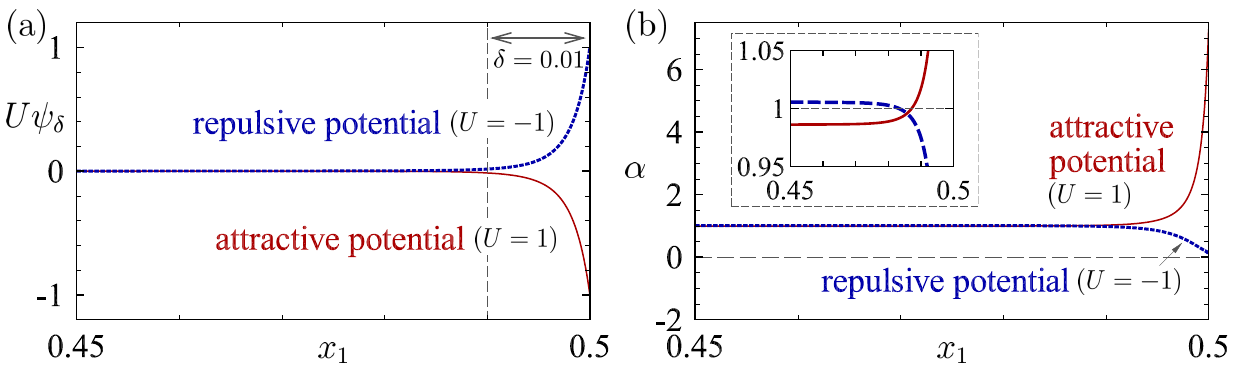}
    \caption{(a) Dimensionless potential $U\psid(x_1)$ [Eq.~\eqref{eq:psi-nd}] for $\delta=0.01$ near the plate at $x_1=1/2$. The cases of $U=1$ and $U=-1$ are shown by red-solid and blue-dash curves, respectively. (b) Corresponding reference density profile $\alpha$ [Eq.~\eqref{eq:alpha}], where the inset shows the magnification near $\alpha=1$.}
    \label{fig:potential}
\end{figure}

The profiles of $U\psid(x_1)$ for $\delta=0.01$ and $U=\pm 1$ are presented in Fig.~\ref{fig:potential}(a). As shown in the figure, the effective range of the potential is restricted in the region $x_1\in[1/2-\delta,1/2]$. The reference density $\alpha$ [Eq.~\eqref{eq:alpha}] for these cases are shown in Fig.~\ref{fig:potential}(b). 
For the attractive (or repulsive) potential, the reference density increases (or decreases) near the plate at $x_1=1/2$. The bulk density is obtained as $\alpha\approx\bpsid^{-1}$, with $\bpsid\approx 1.014$ and $0.995$ for $U=1$ and $U=-1$, respectively.

\subsection{Similarity solution}\label{sec:similarity}

Since the unknown function $\phi$ depends on $(x_1,x_2,\z_1,\z_2,\z_3)$, the analysis of the problem requires substantial computational resources due to the high-dimensionality. 
A \red{useful} approach to simplifying the problem is to introduce a similarity solution \cite{Sone2007}. 
Based on the similarity solution for the thermal transpiration \cite{Ohwada1989}, the following form of the similarity solution \red{can be applied}:
\begin{align}
\phi=c_T\left[\left(\z^2-\frac{5}{2}\right)x_2+2U\psid(x_1)x_2+\z_2\phi_T(x_1,\z_1,\z_\rho)\right], \quad \z_\rho = (\z_2^2+\z_3^2)^{1/2}. 
\label{eq:similarity}
\end{align}
\red{Without any approximation, this} form is compatible with Eqs.~\eqref{eq:bgk-nd}--\eqref{eq:bc-nd} in the presence of the potential, where $\phi_T$ is an alternative unknown function that depends only on $(x_1,\z_1,\z_\rho)$. 
The macroscopic quantities 
are then expressed as
\begin{subequations}\label{eq:macro-similarity-approach-ii}
\begin{align}
&\frac{u_2}{c_T} = u_T(x_1), \quad u_1=u_3=0, 
\label{eq:macro-similarity-approach-ii-a}
\\
&\frac{\omega}{c_T}  = -[1 - 2  U \psid(x_1)] x_2, \quad 
\frac{\tau}{c_T}= x_2, \quad 
\frac{P}{c_T} = 2  U \psid(x_1) x_2. 
\label{eq:similarity-macro-2}
\end{align}
\end{subequations}
The functions $\phi_T(x_1,\,\z_1,\,\zrho)$ and $u_T(x_1)$ remain unknowns. 
Therefore, we are mainly interested in obtaining the flow field $u_T$ in the following. 
\red{Here, it should be remarked that temperature perturbation, $\tau$, is uniform in $x_1$ and linearly depends on $x_2$, being identical to the temperature \red{variation} of the wall $(\tau=\tau_\w)$. Physically, this means that the effect of flow on the temperature profile is negligible and pure thermal conduction is dominant.}

\subsection{Approach (i): the slip-flow theory}\label{sec:slip-flow-theory-TOproblem}

Under the condition~\eqref{eq:thermo-osmosis-problem}, the heat equation with Eq.~\eqref{eq:jump-coef} yields $\tau_\G=c_T x_2$. 
Then, the thermal-slip boundary condition turns out to be 
\begin{align}
u_{2\G} = \b21 \rnd{\tau_{\G}}{x_2} = c_T \b21, \quad 
u_{1\G}=u_{3\G} =0, \quad 
\left(x_1=\pm\hf\right).
\label{eq:slip-coef3}
\end{align}
The Stokes equation with the thermal-slip boundary condition [Eq.~\eqref{eq:slip-coef}] then gives the solution $u_{2\G}=c_T \b21$, $u_{1\G}=u_{3\G}=0$, $P_\G=const$, indicating that thermo-osmosis is a plug flow. The additive constant in $P_\G$ should be zero because of Eq.~\eqref{eq:similarity-macro-2}.
The equation of states leads to $\omega_\G=-c_T x_2$. All of these are for the Hilbert part. Because of the symmetry at $x_1=0$, we consider only the region $0\leq x_1\leq 1/2$ hereafter. Then, using the results of Hilbert part, we can immediately compute the Knudsen-layer corrections Eq.~\eqref{eq:KLcorrections} for $0\leq x_1 \leq 1/2$: 
$\tau_\K=0$, $\omega_{\K} = P_{\K} = 2 U \tpsi(\eta) c_T x_2$, $u_{1\K}=u_{3\K}=0$, $u_{2\K}=k Y_2^{(1)}(\eta)$ with $\eta=(1/2-x_1)/k$. 

Summarizing these results, the nontrivial flow fields are given by
\begin{align}
&\frac{u_2}{c_T} = u_T =   k\left[ \b21 + Y_2^{(1)}\left(\frac{1}{k}\left(\hf-x_1\right)\right) 
\right] + O(k^2), 
\label{eq:slip-flow-sol}
\end{align}
Since $\b21$ and $Y_2^{(1)}$ have been numerically obtained by solving Eq.~\eqref{eq:KLproblem-decomposed21-maintext} 
(see \SM~\ref{sec:numerical-KL} for the detail of numerical analysis), Eq.~\eqref{eq:slip-flow-sol} provides an explicit form of the approximate solution to the thermo-osmosis problem [Eqs.~\eqref{eq:bgk-nd}--\eqref{eq:bc-nd}, \eqref{eq:thermo-osmosis-problem}, \eqref{eq:psi-nd}] for small $k$. \red{Note that Eq.~\eqref{eq:slip-flow-sol} is uniform (i.e., plug-like) if the Knudsen-layer correction $Y_2^{(1)}$, describing spatial variations near the walls, is neglected.}

It should be remarked that the forms of $u_1$, $u_3$, $\omega$, $\tau$, and $P$ described above are consistent with those in Eq.~\eqref{eq:macro-similarity-approach-ii}, up to the first order in $k$. 


\subsection{Approach (ii): the numerical analysis}\label{sec:numerical}
We solve the thermo-osmosis problem [Eqs.~\eqref{eq:bgk-nd}--\eqref{eq:bc-nd}, \eqref{eq:thermo-osmosis-problem}, \eqref{eq:psi-nd}] numerically by applying the similarity solution $\phi_T$ [Eq.~\eqref{eq:similarity}]. 
By substituting Eq.~\eqref{eq:similarity} into the BGK equations [Eqs.~\eqref{eq:bgk-nd}--\eqref{eq:bc-nd}], it turns out that the similarity solution $\phi_T$ solves the following boundary-value problem
\begin{subequations}\label{eq:bgk-numerical-similarity}
\begin{align}
&\z_1\rnd{\phi_T}{x_1} -U\ddif{\psid}{x_1}\rnd{\phi_T}{\z_1}
= 
\frac{\alpha}{k}(
2u_T-
\phi_T)
- \left(\z_1^2+\zrho^2-\frac{5}{2}\right) 
- 2 U \psid, \\ 
&u_T=\frac{1}{2}\int_{\RR^3} \zrho^2 \phi_T E \; \dd \bm{\z}, \\ 
&\phi_T = 0 \quad \left(x_1=\pm\hf,\;\z_1\lessgtr0\right), 
\end{align}
\end{subequations}
where Eq.~\eqref{eq:bgk-numerical-similarity} includes the Knudsen number $k$ as a physical parameter. 
The detail of the numerical analysis is given in \SM~\ref{sec:numerical-TO}.

Let us denote by $\bnum$ the thermal-slip coefficient $\b21$ estimated from the result of numerical analysis. 
Recalling Eq.~\eqref{eq:slip-flow-sol}, we can set $\bnum$ as
\begin{align}
\bnum = \frac{u_2|_{x_1=0}}{k c_T} + O(k) \quad \text{as}\quad k\to 0. \label{eq:bnum}
\end{align}
Note that $u_2$ in Eq.~\eqref{eq:bnum} is evaluated at $x_1=0$ (i.e., the channel center) at which the Knudsen-layer function $Y_2^{(1)}$ vanishes. By comparing $\b21$ obtained from the Knudsen-layer problem [Eq.~\eqref{eq:KLproblem-decomposed21-maintext}] and $\bnum$ obtained from Eq.~\eqref{eq:bnum}, we can assess the validity of the slip-flow theory developed in Sec.~\ref{sec:slip-flow-theory}.


\subsection{Results of thermo-osmosis problem}

Numerical method used in the subsequent results is summarized in \SM~\ref{sec:numerical-appendix}. In particular, numerical parameters, including the number of lattice points, are presented in \SM~\ref{sec:numerical-overview}. Although the velocity distribution functions in the present problem are known to exhibit discontinuities in the $(x_1,\,\z_1)$ space (see, e.g., the similar case for convex boundaries \cite{Sone1992}), the present scheme naively neglects the discontinuity. To compensate, we use a relatively large number of lattice points and assess the accuracy in \SM~\ref{sec:accuracy}. For highly accurate numerical analysis, however, the discontinuities should be treated properly, e.g., using a hybrid scheme \cite{Sugimoto1992} or the method of characteristics \cite{Tsuji2013}. 

\begin{figure}[bt]
    \centering
    \includegraphics[width=0.9\textwidth]{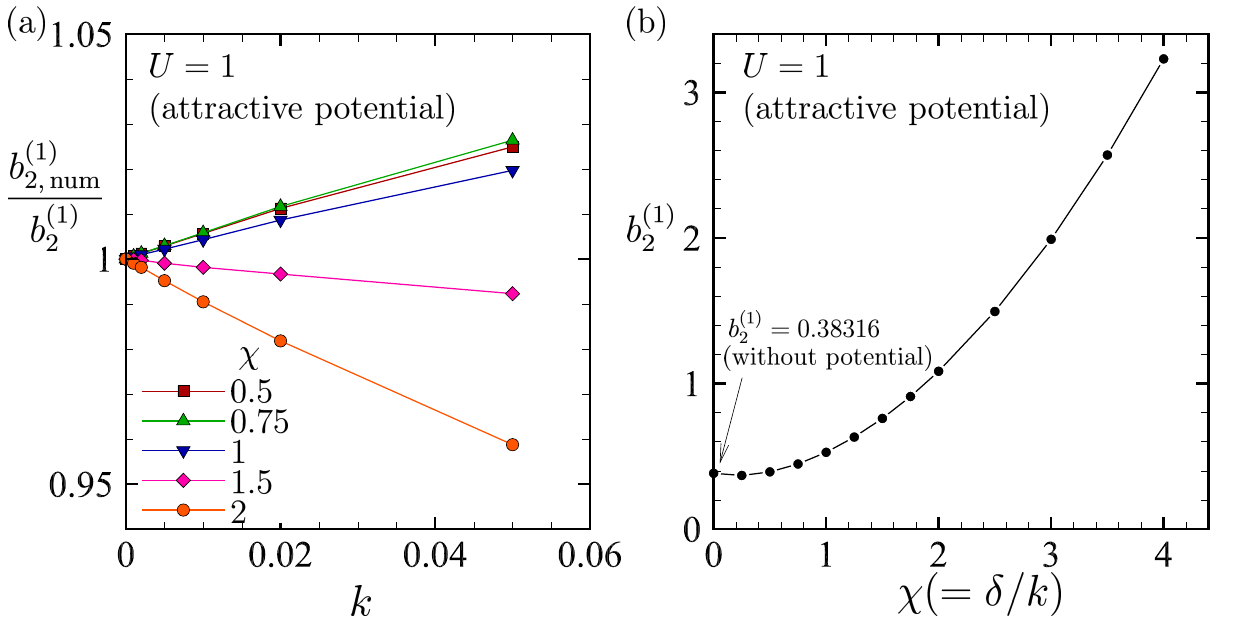}
    \caption{Thermal-slip coefficients for attractive potential $U=1$. (a) The ratio $\bnum/\b21$ between the thermal-slip coefficient obtained by the numerical analysis $\bnum$ and that by the slip-flow theory $\b21$. (b) The thermal-slip coefficient $\b21$ as a function of the range of the potential $\chi$. }
    \label{fig:U=1}
 \end{figure}
\subsubsection{Comparison of the thermal-slip coefficients between the slip-flow theory and the numerical analysis}\label{sec:compare-slip}
In this section, we compare the thermal-slip coefficient $\b21$ and $\bnum$ for representative cases of attractive $(U=1)$ and repulsive $(U=-1)$ potentials with various values of $\chi(=\delta/k)$ that characterizes the effective range of the potential. 
\red{Recall that we are interested in developing a slip-flow theory for small $k$ and thus we choose $k\leq 0.05$ below. Recall also that the value of $\chi$ is assumed to be $O(1)$ quantity by the assumption Eq.~\eqref{eq:same-order}. }

First, let us consider the case of $U=1$ (attractive potential). 
For a fixed $\chi>0$, Fig.~\ref{fig:U=1}(a) shows that the ratio $\bnum/\b21$ varies linearly with $k$, \red{that is, $\bnum/\b21 \approx 1 + \lambda_\chi k$, where $\lambda_\chi$ is a slope depending on $\chi$. These trends are consistent with Eqs.~\eqref{eq:slip-flow-sol} and \eqref{eq:bnum}, although it is seen that the higher-order analysis is necessary to obtain the information of slope $\lambda_\chi$. To further check the trends of Fig.~\ref{fig:U=1}(a), the numerical values of $\bnum$ and $\b21$ for some $k$ and $\chi$ are provided in Table~\ref{tab:U=pm1}}. 
\red{Without the potentials (i.e., $U=0$), the present thermo-osmosis problem is reduced to a classical thermal transpiration problem (see, e.g., Refs.~\cite{Niimi1971,Ohwada1989a}). The thermal-slip coefficient $\b21$ in Table~\ref{tab:U=pm1} for this case, $\b21=0.38316$, is the same as that reported in Ref.~\cite{Sone2007} up to five decimal points, ensuring the accuracy of the present numerical analysis.}
As the range of the potential $\chi$ increases, Fig.~\ref{fig:U=1}(b) shows that $\b21$ also increases except for $\chi=0.25$ at which a slight decrease is observed.

 \begin{figure}[bt]
    \centering
    \includegraphics[width=0.9\textwidth]{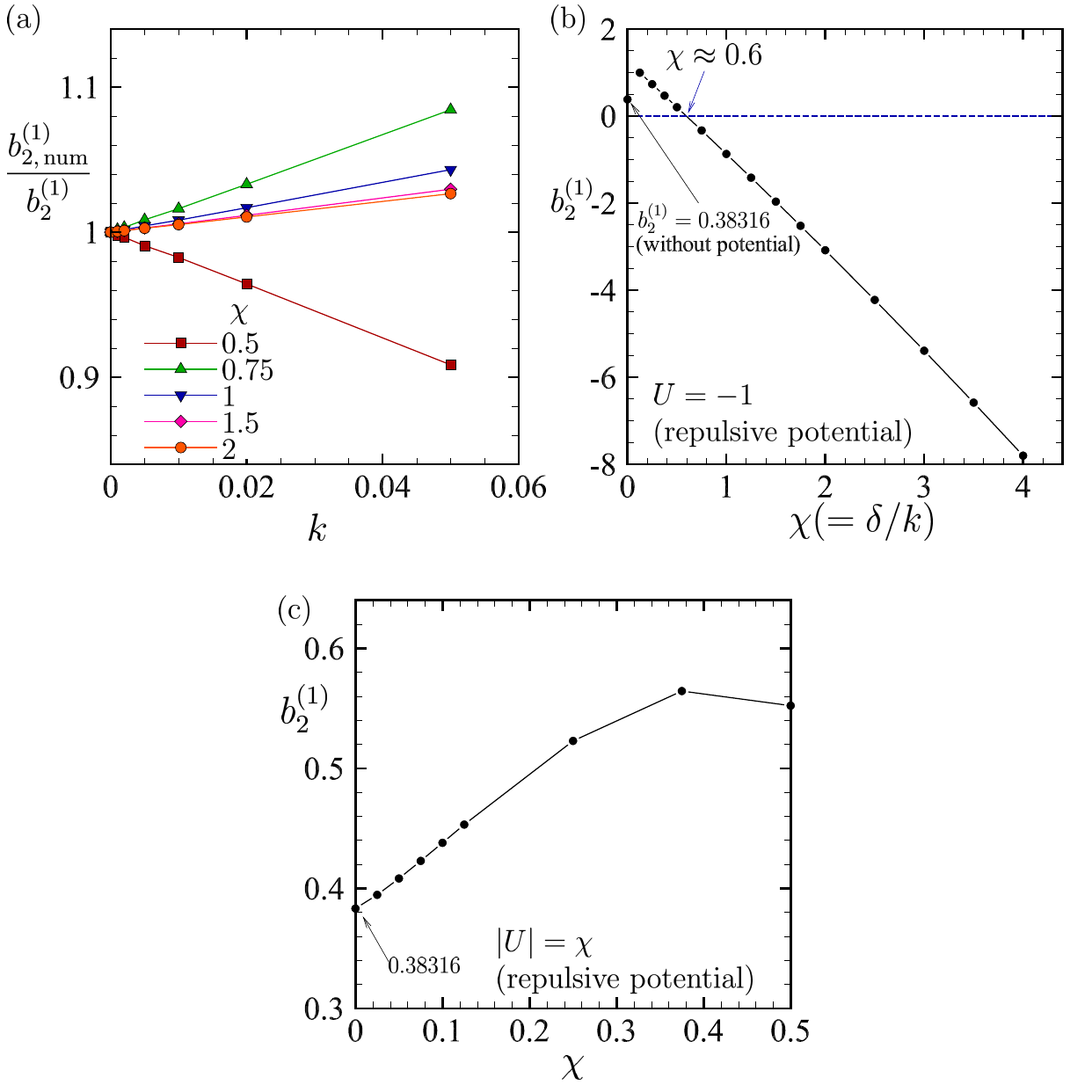}
    \caption{Thermal-slip coefficients for repulsive potential. (a) The ratio $\bnum/\b21$ between the thermal-slip coefficient obtained by the numerical analysis $\bnum$ and that by the slip-flow theory $\b21$ for $U=-1$. (b)(c) The thermal-slip coefficient $\b21$ as a function of the range of the potential $\chi$ for (b) $U=-1$ and (c) $U<0$ with $|U|=\chi$.}
    \label{fig:U=-1}
\end{figure}

Next, we consider the case with $U=-1$ (repulsive potential). 
Figure~\ref{fig:U=-1}(a), corresponding to Fig.~\ref{fig:U=1}(a), shows that $\bnum\to \b21$ as $k\to0$ for various values of $\chi$ \red{(see also Table~\ref{tab:U=pm1})}. 
As the range of the potential $\chi$ increases, Fig.~\ref{fig:U=-1}(b) shows that $\b21$ decreases for $\chi>0$ (the case of $\chi\to0$ will be discussed below), and eventually changes the sign at $\chi\approx 0.6$. Thus, the effect of the potential on the thermal-slip coefficient is more pronounced in the repulsive case than in the attractive one, as increasing $\chi$ ultimately reverses the flow direction. Moreover, the rate of increase in the magnitude $|\b21|$ with respect to $\chi$ is steeper than that observed for $U=1$ (attractive potential). A more quantitative discussion of this behavior will be provided later in Sec.~\ref{sec:effect-of-U}.

\begin{table}[tb]
\begin{center}
\caption{\red{Thermal-slip coefficients $\bnum$ and $\b21$ for $U=0$ (no potential), $U=1$ (attractive potential), and $U=-1$ (repulsive potential) for various sets of $k$ and $\chi$. See also Figs.~\ref{fig:U=1}(a) and \ref{fig:U=-1}(a) for corresponding plots. 
More complete data set, including the values for other parameter values, is found in \SM~\ref{sec:values-of-b21}.}}
{\tabcolsep = 0.65em 
\renewcommand{\arraystretch}{1.3}
\red{
\begin{tabular}{ccccccccc}\hline\hline
& & \multicolumn{4}{c}{$\bnum$} & $\b21$ \\ 
     \cline{3-6}
    $U$&$\chi$ & $k=0.05$ & $k=0.01$ & $k=0.005$ & $k=0.001$ & --- \\ \hline
    $U=0\m$&---& $\m0.38213$& $\m0.38316 $& $\m0.38316 $& $\m0.38317 $& $\m0.38316 $\\
    \hline
    $U=1\m$&$\chi=0.5$\s& $\m0.40254 $& $\m0.39498 $& $\m0.39388 $& $\m0.39301 $& $\m0.39275 $\\
    $U=1\m$&$\chi=1$\ss & $\m0.53745 $& $\m0.52933 $& $\m0.52822 $& $\m0.52733 $& $\m0.52706 $\\
    $U=1\m$&$\chi=2$\ss & $\m1.0389\q$& $\m1.0733\q$& $\m1.0784\q$& $\m1.0826\q$& $\m1.0836\q$\\
    \hline
    $U=-1$ &$\chi=0.5$\s& $\m0.18523 $ & $\m0.20032 $ & $\m0.20197 $ & $\m0.20343 $ & $\m0.20388 $\\
    $U=-1$ &$\chi=1$\ss & $ -0.90738 $ & $ -0.87713 $ & $ -0.87369 $ & $ -0.87074 $ & $ -0.86994 $\\
    $U=-1$ &$\chi=2$\ss & $ -3.1643\q$ & $ -3.0984\q$ & $ -3.0907\q$ & $ -3.0842\q$ & $ -3.0825\q$\\
    \hline \hline
\end{tabular}}
\\
}
\label{tab:U=pm1}
\end{center}
\end{table}

It is worth noting that, for $U=-1$ [Fig.~\ref{fig:U=-1}(b)], the thermal-slip coefficient $\b21$ does not converge to the value corresponding to the case without potential $(\chi=0)$ as $\chi\to0$. This singular behavior is not observed in the attractive case ($U=1$) [Fig.~\ref{fig:U=1}(b)]. To clarify this singular behavior, we show in 
Fig.~\ref{fig:U=-1}(c) the values of $\b21$ versus $\chi$ for $|U|/\chi=1$ ($U<0$) instead of $U=-1$. Then, $\b21$ for $\chi\to0$ smoothly approaches that for the case without potential. 
The singularity comes from the fact that, in the Knudsen-layer problem [Eq.~\eqref{eq:KLproblem-decomposed21-maintext}], we can estimate
$U (\tdif{\tpsi}{\eta})|_{\eta=0} \approx U/\chi$. 
Hence, the limit $\chi\to0$ leads to the blow up of the term $U (\tdif{\tpsi}{\eta})$ when $U$ is finite, although its support vanishes. 
In the case of $U=-1$, it is expected that the blow up of $U (\tdif{\tpsi}{\eta})$ with the vanishing support results in a nonvanishing effect on $\b21$. 
The behavior of $\b21$ observed in Fig.~\ref{fig:U=-1}(c) supports the above discussion. 

\begin{figure}
    \centering
    \includegraphics[width=0.9\textwidth]{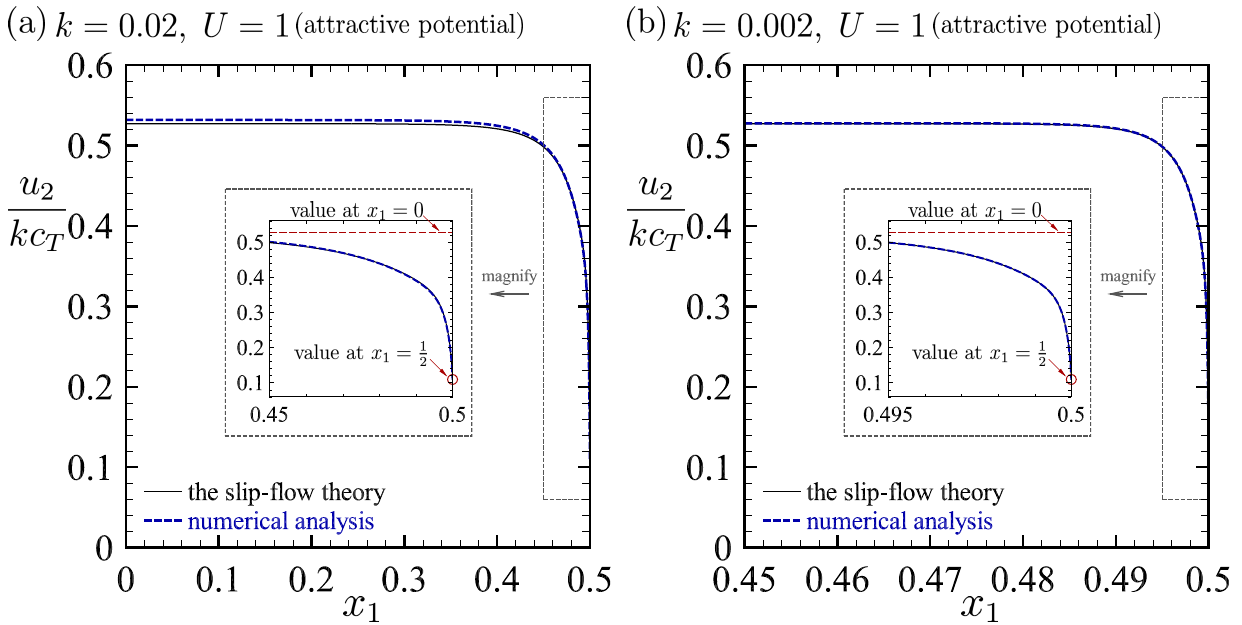}
    \caption{Profiles of thermo-osmosis under the near-wall attractive potential: $u_2=u_2(x_1)$ obtained from the slip-flow theory [black-solid curve; Eq.~\eqref{eq:slip-flow-sol}] and from the numerical analysis [blue-dashed curve; Eq.~\eqref{eq:macro-similarity-approach-ii-a}]. The profiles are normalized by $k c_T$, where $k \approx$ Knudsen number and $c_T$ is the magnitude of (dimensionless) temperature gradient. (a) $k=0.02$ and (b) $k=0.002$ with $\chi=1$ and $U=1$. The insets show the magnification near the plate $x_1=1/2$.}
    \label{fig:u2-U=1}
 \end{figure}
 \begin{figure}
    \includegraphics[width=0.9\textwidth]{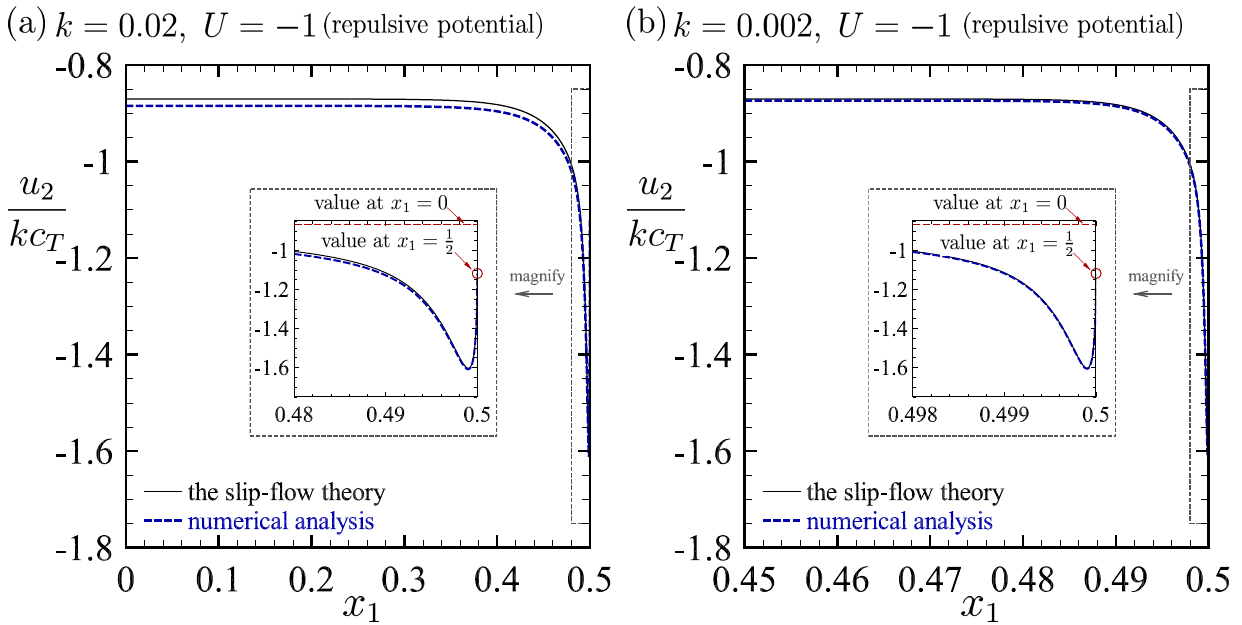}
    \caption{Profiles of thermo-osmosis under the near-wall repulsive potential: $u_2=u_2(x_1)$ obtained from the slip-flow theory [black-solid curve; Eq.~\eqref{eq:slip-flow-sol}] and from the numerical analysis [blue-dashed curve; Eq.~\eqref{eq:macro-similarity-approach-ii-a}]. The profiles are normalized by $k c_T$, where $k \approx$ Knudsen number and $c_T$ is the magnitude of (dimensionless) temperature gradient. (a) $k=0.02$ and (b) $k=0.002$ with $\chi=1$ and $U=-1$. The insets show the magnification near the plate $x_1=1/2$.}
    \label{fig:u2-U=-1}
\end{figure}


\subsubsection{Comparison of the thermo-osmotic flow profiles between the slip-flow theory and the numerical analysis}\label{sec:compare-profile}
In this section, we discuss the spatial profiles of the thermo-osmosis flow, i.e., $u_2$ [see Fig.~\ref{fig:problem}(b)]. 
As in the previous section, we compare the results obtained from the slip-flow theory [Eq.~\eqref{eq:slip-flow-sol}] with those from the numerical analysis [Eqs.~\eqref{eq:macro-similarity-approach-ii-a} and \eqref{eq:bgk-numerical-similarity}]. 

For $U=1$ (attractive potential) and $\chi=1$, we show the profiles of $u_2(x_1)$ in Fig.~\ref{fig:u2-U=1} for (a) $k=0.02$ and (b) $k=0.002$. In panel (b), the $x_1$ range is restricted to $x_1\in[0.45,0.5]$, since $u_2$ remains nearly constant for $x_1\in[0,0.45]$. Recall that $c_T$ denotes the magnitude of the dimensionless temperature gradient of the plates [see Eq.~\eqref{eq:cT}], and the velocity profiles are normalized by $k c_T$. 
Note that $u_2\to0$ as $k\to0$. 
For both values of $k$ presented in the figure, the slip-flow theory accurately reproduces the results of the numerical analysis, including regions near the plate surface (see the insets). The discrepancy between the two approaches is smaller for smaller $k$, which is consistent with the fact that the slip-flow theory is an asymptotic approximation valid in the limit $k \to 0$.

For $U=-1$ (repulsive potential) and $\chi=1$, the corresponding velocity profiles are shown in Fig.~\ref{fig:u2-U=-1} for (a) $k=0.02$ and (b) $k=0.002$. 
The main differences from the attractive case ($U=1$; see Fig.~\ref{fig:u2-U=1}) are found in the sign of $u_2$, which is now negative when $U=-1$ (see Fig.~\ref{fig:u2-U=-1}), and in the behavior of $u_2$ near $x_1=1/2$. 
As discussed in Sec.~\ref{sec:compare-slip}, thermo-osmosis can be directed from hot to cold, i.e., $u_2<0$, in the presence of a repulsive potential. The profiles of $u_2$ near the plate $x_1=1/2$ exhibit singular behavior, where $|\partial u_2/\partial x_1|$ appears to diverge at the boundary $x_1=1/2$. 
This singularity is not observed in the attractive case ($U=1$; see Fig.~\ref{fig:u2-U=1}). 
Such singular (or regular) behavior of macroscopic quantities under a repulsive (or attractive) near-wall potential can be attributed to the fact that the velocity distribution function on the plate is discontinuous (or continuous) in the velocity variables. 
This observation is reminiscent of the singular or regular behavior of the macroscopic quantities near convex or concave boundaries \cite{Sone1992,Takata2017}, although we do not explore this connection in detail here.


\subsubsection{Effect of near-wall potential on thermo-osmosis, thermal slip, and Knudsen-layer correction}\label{sec:effect-of-U}
\begin{figure}[bt]
    \centering
    \includegraphics[width=0.9\textwidth]{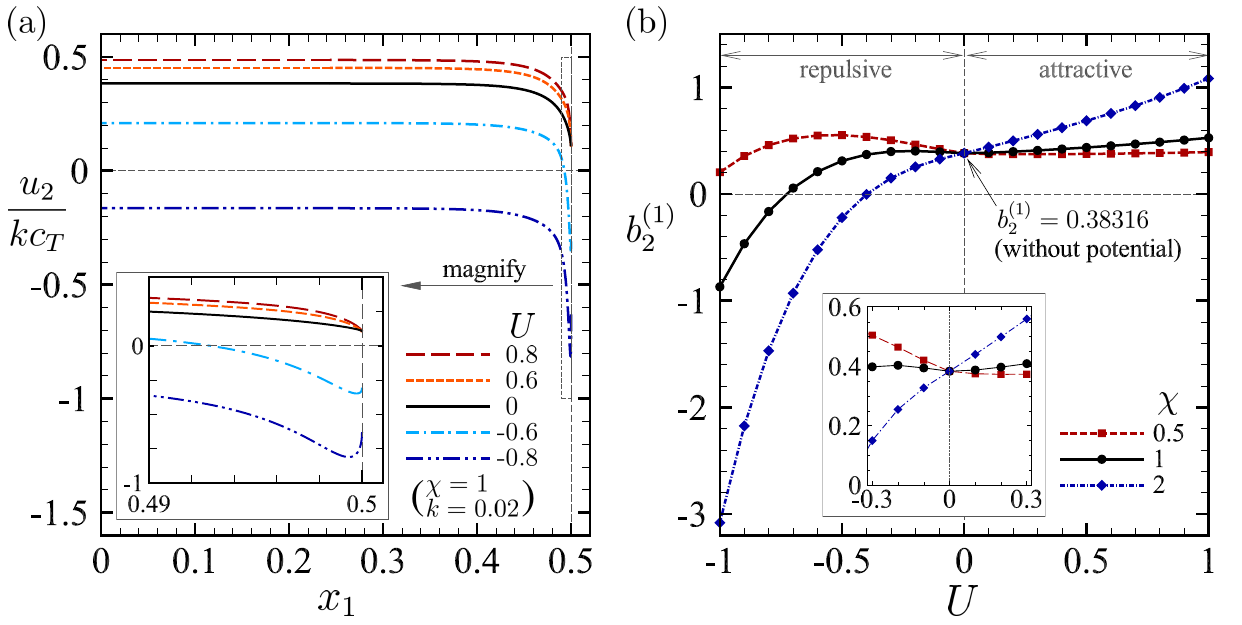}
    \caption{(a) Profiles of thermo-osmosis $u_2$ and (b) the thermal-slip coefficient $\b21$ for various magnitude of the potential $U$ obtained by the slip-flow theory. (a) $u_2(x_1)$ for $U=0.8$ (red-dash), $U=0.6$ (orange-dot), $U=0$ (black-slid), $U=-0.6$ (cyan-dash-dot), and $U=-0.8$ (blue-dash-dot-dot), where $\chi=1$ and $k=0.02$. The inset shows the magnification near $x_1=1/2$. (b) $\b21$ as a function of $U$ for $\chi=0.5$ (red-square), $\chi=1$ (black-circle), and $\chi=2$ (blue-diamond). The inset shows the magnification near $U=0$.
 }
    \label{fig:U-dependence}
\end{figure}

So far we have fixed $|U|=1$. In this section, we investigate the effect of $U$. Figure~\ref{fig:U-dependence}(a) shows the profiles of thermo-osmosis $u_2(x_1)$ for various values of $U=-0.8$, $-0.6$, $0$, $0.6$, and $0.8$ in the case of $\chi=1$ and $k=0.02$. For attractive potential $(U=0.6,\,0.8)$, $u_2$ is larger than that of the case without potential ($U=0$), indicating that the flow is enhanced due to the potential. In contrast, for repulsive potential $(U=-0.6,\,-0.8)$, the overall flow in the bulk is reduced $(U=-0.6)$ and eventually reversed $(U=-0.8)$. Near the plate, a more complex behavior is observed for $U=-0.6$, as shown in the inset. That is, $u_2$ is positive away from the plate but is negative near the plate. These observations suggest a general trend: attractive potentials tend to enhance thermo-osmosis from the cold to the hot region, whereas repulsive potentials tend to suppress or even reverse it.

However, the situation is more complex. Figure~\ref{fig:U-dependence}(b) shows the thermal-slip coefficient $b_2^{(1)}$, which determines the overall characteristics of the flow, as a function of $U$ for $\chi=0.5,\,1,\,2$. For large $\chi$ ($\chi=2$), the previously discussed trend appears to hold, that is, attractive (or repulsive) potential increases (or diminishes) thermo-osmosis. However, for smaller $\chi$ ($\chi=0.5$) and for small values of $|U|$, the inset shows a slight but opposite tendency. The case of $\chi=1$ appears to be a rough threshold between these two behaviors. For all values of $\chi$ shown in the figure, the rate of change of the thermal-slip coefficient with respect to $U$ (i.e., $|\partial \b21/\partial U|$) is larger for repulsive potentials than for attractive potentials. 
Phenomenologically, the parameter $U$ models the ``affinity" of the interaction between fluid molecules and solid surface.
The present model indicates that the thermal-slip coefficient $\b21$ is sensitive to the fluid-solid interaction, especially under repulsive conditions. 

\begin{figure}
    \centering
    \includegraphics[width=1\textwidth]{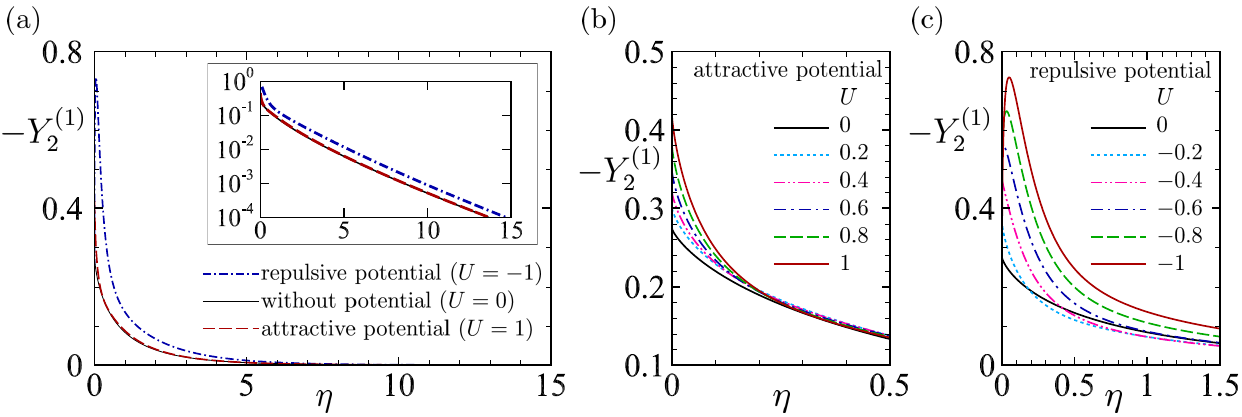}
    \caption{Profiles of the Knudsen-layer function $Y_2^{(1)}(\eta)$ for $\chi=1$. (a) The cases of repulsive potential ($U=-1$; blue-dash-dot), without potential ($U=0$; black-solid), and attractive potential ($U=1$; red-dash), where the inset shows the semi-logarithmic plot. (b) and (c) shows $Y_2^{(1)}(\eta)$ for the cases of attractive potential and repulsive potential, respectively, near $\eta =0$ with various $U$.}
    \label{fig:KL}
\end{figure}

Finally, Fig.~\ref{fig:KL} shows the Knudsen-layer correction $Y_2^{(1)}(\eta)$ for $\chi=1$, which represents the behavior of the macroscopic quantity inside the Knudsen layer [see Eq.~\eqref{eq:KLproblem-decomposed21-maintext}]. The overall behavior of $Y_2^{(1)}(\eta)$ is shown in Fig.~\ref{fig:KL}(a) for the cases of repulsive potential $(U=-1)$, no potential $(U=0)$, and attractive potential $(U=1)$. The inset, shown on a logarithmic scale, confirms that $Y_2^{(1)}$ decays exponentially as $\eta$ increases. 
Figure~\ref{fig:KL}(b) and Fig.~\ref{fig:KL}(c) show close-up views near the boundary ($\eta\approx0$) for the attractive and repulsive potential cases, respectively, for various $U$. For the attractive potential, the influence of the potential is rather localized near the plate. In contrast, for the repulsive potential, the decay of $Y_2^{(1)}$ appears to be slower, though it remains exponentially fast. It is worth noting that while exponential decay is well established in the absence of potential, its persistence in the present case is nontrivial, as the potential considered in this study follows a power-law function. 

\subsubsection{Adequateness of the scaling assumption}\label{sec:adequateness}

\begin{figure}[bt]
    \centering
    \includegraphics[width=0.6\textwidth]{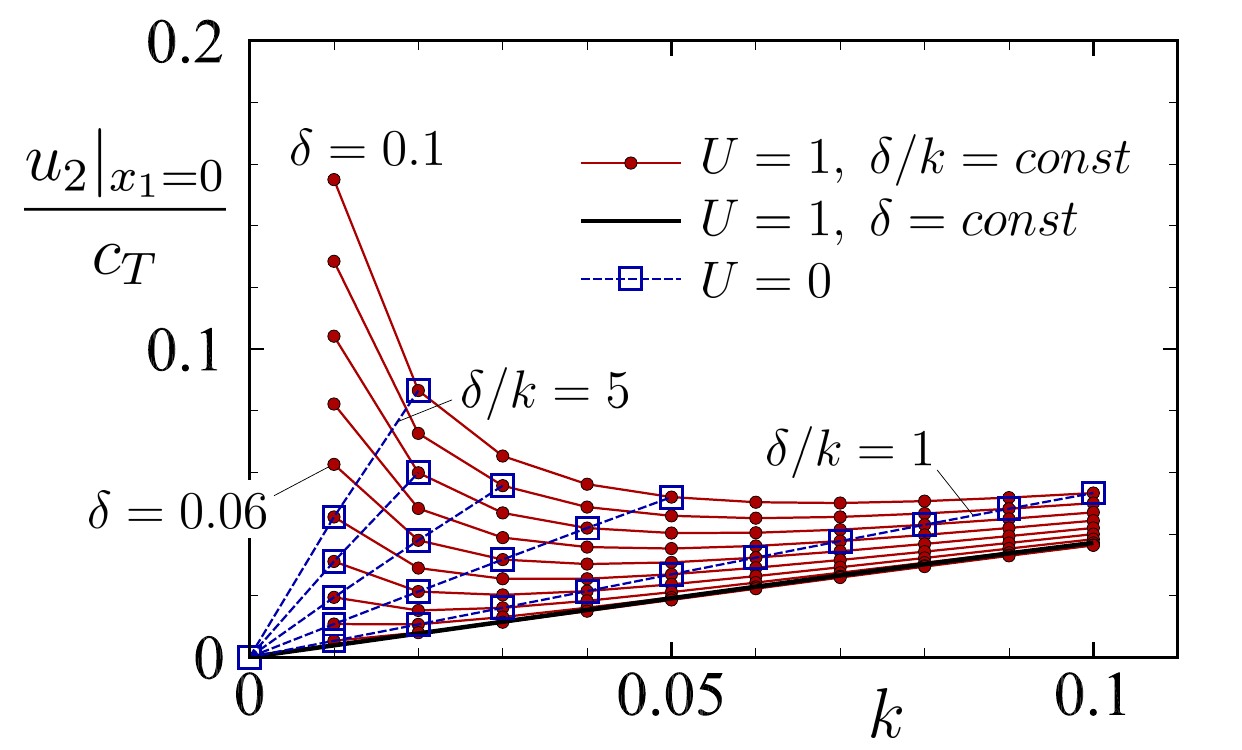}
    \caption{Thermo-osmosis at the channel center, $u_2|_{x_1=0}/c_T$, as a function of $k$ for various $\delta=0.01$, $0.02$, ..., $0.09$, $0.1$, obtained by the direct numerical analysis. The case without the potential $(U=0)$ is shown by the black-solid curve. The cases with $U=1$ are shown in two different manners: red-circle symbols (or blue-square symbols) connect the data point under the condition $\delta=0.01$, $0.02$, ..., $0.09$, $0.1$ (or $\delta/k=\chi=1$, ..., $5$). }
    \label{fig:adeq}
\end{figure}

In this section, we demonstrate the adequacy of the scaling relation \eqref{eq:same-order} in the slip-flow theory. Figure~\ref{fig:adeq} shows the thermo-osmosis velocity at the channel center $u_2|_{x_1=0}$ for various values of $k$ and $\delta$, obtained by direct numerical analysis. For the case without a potential ($U=0$; black-solid curve), the flow magnitude is proportional to $k$, being consistent with the prediction of the generalized slip-flow theory. Note that the slope of the curve corresponds to the slip coefficient [Eq.~\eqref{eq:bnum}].
However, this linear relation breaks down when $k$ is varied while $\delta$ is fixed (red-circle symbols), indicating that an asymptotic analysis based on an expansion with respect to $k$ is inadequate to describe the gas behavior for $k\ll1$. In contrast, when $\chi=\delta/k$ is kept constant (blue-square symbols), a clear linear relationship $u_2\propto k$ is observed. This result validates the appropriateness of the scaling in Eq.~\eqref{eq:same-order} and illustrates the reasoning behind its adoption in the present theory.  

\section{Discussion}\label{sec:discussion}
In the dimensional form, the thermal-slip boundary condition [Eq.~\eqref{eq:slip-coef} or \eqref{eq:slip-coef3}] is written as, up to the first order of $k$, 
\begin{align}
\bm{v}\cdot \bm{t} 
= -\slip \,\bm{t}\cdot\nabla T, \quad \slip = - \frac{\sqrt{\pi}}{2} \frac{v_0 \ell_0}{T_0}\b21 \quad (\text{on boundaries}), \label{eq:slip-discussion}
\end{align}
where 
$\slip$ [the unit m$^2$/(s K)] is the thermal-slip coefficient, and $\b21$ is the corresponding dimensionless thermal-slip coefficient of order unity, as
obtained in Secs.~\ref{sec:slip-flow-theory} and \ref{sec:thermo-osmosis}. Here, $v_0=\sqrt{2RT_0}$ is the thermal speed at the reference state, $\ell_0$ is the mean free path at the reference state, and $T_0$ is the reference temperature. Note that $\b21$ can be either positive or negative signs, although positive values $\b21>0$ (corresponding to flow from cold to hot) have been more commonly observed. 
\red{In the following, we discuss the validity of Eq.~\eqref{eq:slip-discussion} for gas and liquid phases separately.}

\subsection{Gases}

In the case of gases, which we have considered throughout this study based on the BGK model Eq.~\eqref{eq:bgk}, the viscosity $\mu$ is related to the mean free path $\ell_0$ as \cite{Sone2007}
\begin{align}
\mu = \frac{\spi}{2} \frac{p_0 \ell_0}{v_0}, \label{eq:viscosity}
\end{align} 
where $p_0$ is the reference pressure.
Consequently, the dimensionless viscosity is given by $\mu/(p_0 D/v_0)=(\spi/2)(\ell_0/D)=k$, as shown in the Stokes equation Eq.~\eqref{eq:Stokes-maintext}.
Using this relation of $\mu$ and $p_0=\rho_0RT_0$, the thermal-slip coefficient $\slip$ can be rewritten as, 
\begin{align}
\slip 
&= -\frac{\mu}{\rho_0 T_0} \sigma_T\quad (\sigma_T=2\b21).
\end{align}
In the experiment by Yamaguchi et al. \cite{Yamaguchi2016}, $\sigma_T$ was measured to be approximately $\sigma_T\approx 0.9$--$1$ for several monatomic gases, which is of the same order of magnitude as the present results [i.e., $\b21=O(1)$; see \red{Tables~\ref{tab:U=pm1}}]. 
This agreement is reasonable, since the experimental estimation of $\sigma_T$ in \cite{Yamaguchi2016} is also based on fitting the thermal-slip boundary condition to measured mass flow rates to obtain the thermal-slip coefficient. 
It is worth noting that the Stokes equations for incompressible viscous fluid, i.e., 
$\trnd{p}{X_i} = \mu (\trnd{^2 v_{i}}{X_j^2})$ can be recovered for the Hilbert part when $k \ll 1$ \cite{Sone2007}. 

\red{Purely-attractive $(U>0)$ and purely-repulsive $(U<0)$ potentials considered here represent two limiting cases of typical favorable and unfavorable solid-gas molecular interactions, respectively. Therefore, if we consider more realistic LJ-type potentials, which may lie between these two limiting cases, the values of $\b21$ is expected to fall well within the range of $\b21$ presented here in Table~\ref{tab:U=pm1}. In Ref.~\cite{Yamaguchi2021}, molecular dynamics simulations for an LJ gaseous system were carried out to investigate thermo-osmosis in a nanochannel. The estimated values of $\slip$ are in qualitative agreement between the molecular simulation and Eq.~\eqref{eq:slip-discussion}. That is, using $\b21=0.52706$ for $U=1$ and $\chi=1$ in Eq.~\eqref{eq:slip-discussion}, we obtain $\slip\approx -4.9\times10^{-10}$~m$^2$/(s~K) (Ref.~\cite{Yamaguchi2021}) and $\slip\approx -1.6\times10^{-10}$~m$^2$/(s~K) [Eq.~\eqref{eq:slip-discussion}] for $\Kn\approx0.1$ and 
$\slip\approx -7.2\times10^{-10}$~m$^2$/(s~K) (Ref.~\cite{Yamaguchi2021}) and $\slip\approx -4.9\times10^{-10}$~m$^2$/(s~K) [Eq.~\eqref{eq:slip-discussion}] for $\Kn\approx0.3$ (see \SM~\ref{sec:yamaguchi2021} for the detail of the estimation). 
However, it should be remarked that a potential including both attractive and repulsive parts (such as a LJ-type potential) complicates the numerical analysis of Knudsen-layer problems, since the external-force term $(\trnd{\Psi}{X_1})(\trnd{f}{\xi_1})$ [see Eq.~\eqref{eq:bgk}] is locally positive or negative for such cases, making it difficult to apply an implicit upwind finite-difference method widely used in the literature as well as in the present paper.}

\red{We also make some comparisons with a recent study \cite{Anzini2025}, which applied the framework of linear response theory to thermo-osmosis of gases and liquids in a closed channel consisting of two parallel confinements in both $X_1$ and $X_2$ directions. 
The major difference between the present study and Ref.~\cite{Anzini2025} is that here we consider an open channel, i.e., the longitudinal pressure gradient in the bulk is zero [Eq.~\eqref{eq:similarity-macro-2}], whereas in Ref.~\cite{Anzini2025} the effect of the backflow caused by a pressure gradient in the closed channel is present (no net mass flux condition is imposed in the longitudinal direction). The details of fluid-solid molecular interactions and the range of parameters ($\Kn$ and volume fraction) are also different. Despite these differences, there are some qualitative agreements as follows. (i) The velocity reversal between attractive-type and repulsive-type fluid-solid inter-molecular interaction potentials is observed in both studies. In particular, gas flow is driven from the cold to hot sides (or the hot to cold sides) for attractive (or repulsive) potentials near the wall. (In Ref.~\cite{Anzini2025}, the backflow in the opposite direction occurs away from the wall.) 
(ii) The velocity profile is flat (i.e., plug-like) away from the wall in both studies. 
(iii) For a hard wall, i.e., the specular reflection boundary condition with no potential, the flow vanishes (no thermo-osmosis occurs). The point (iii) can be seen in our problem by replacing Eq.~\eqref{eq:bc-nd} with a specular reflection condition and by neglecting the external-force term. For this case, 
\begin{align}
\phi=c_T(\z^2-\frac{5}{2})x_2 - c_T k \z_2(\z^2-\frac{5}{2}), \label{eq:specular}
\end{align}
is a solution that indeed realizes the thermo-osmosis setting $\tau=c_T x_2$ (linear temperature increase) and $P=0$ (uniform pressure) but results in $u_i=0$ (no flow), regardless of the value of $\Kn$. 
To further comparisons, we will need to find a solution that induces backflows due to a pressure gradient, for which the shear-slip coefficient $\b11$ will be necessary.
}

\red{\subsection{Liquids}}

The slip-flow theory can be basically applied to gases, since it is based on the Boltzmann equation \red{(or its model equations)}, which describes the evolution of the one-particle velocity distribution functions. \red{The Boltzmann equation is valid for the entire range of $0<\Kn<\infty$, provided that the volume fraction $\mathcal{V}$ $(\propto nd^3)$ is negligibly small, where $n$ is the number density of fluid molecules and $d$ the molecular diameter. Due to the smallness of volume fraction $\mathcal{V}$, long-range inter-molecular potentials, which are important in liquids [i.e., $\mathcal{V}=O(1)$], and finite-size effects of molecules are not incorporated into the Boltzmann equation}, and macroscopic quantities such as pressure are defined differently from those of liquids. Nevertheless, the present BGK model of the Boltzmann equation can be interpreted as describing a simplified ``fluid" (say, a BGK fluid), in which fluid molecules interact locally in space to drive the system toward local thermal equilibrium. From this perspective, the BGK fluid can be viewed as a toy model for studying certain phenomena in liquids.

In the context of liquids, 
the quantity $\slip T_0$, also denoted by $M_{21}$ or $\beta_{21}$, is often used \cite{Fu2017,Ganti2017} as a measure of thermal-slip coefficients. Accordingly, we consider the quantity $M_{21}=\slip T_0 = -(\spi/2) v_0 \ell_0 \b21$ for comparison and provide rough estimates of its magnitude. 
In Eq.~\eqref{eq:slip-discussion}, we may estimate each parameter as follows: 
\begin{itemize}
\item the average molecular speed $v_0\approx 10^2$--$10^3$ m/s,
\item the ``mean free path" $\ell_0\approx$ effective range of the potential $\delta D \approx0.1$--$1$~nm,
\item the (dimensionless) thermal-slip coefficient $|\b21|\approx 0.1$--$1$.
\end{itemize}
In the second estimate, we have used the assumption $\chi=\delta/k=\delta/(\ell_0/D)=O(1)$, implying that $\ell_0$ is on the order of $\delta D$. Note that the mean free path is not well defined for liquids, and $\ell_0$ here is interpreted as a characteristic length scale associated with the interaction potential.
With the above estimates, the (dimensional) thermal-slip coefficient $M_{21}$ can be evaluated as $|M_{21}|=|\slip T_0|\approx 10^{-9}$--$10^{-6}$ m$^2$/s. 
In literature, using the MD simulations, Ref.~\cite{Fu2017} reported $|M_{21}|\approx 1$--$400\times 10^{-8}$ m$^2$/s for LJ-potential-based liquid-solid interfaces and 
$|M_{21}|\approx 250\times 10^{-8}$ m$^2$/s for water-graphene surfaces, Ref.~\cite{Ganti2017} reported $|M_{21}|\approx 0.85$--$3.8\times 10^{-8}$ m$^2$/s for less-attractive LJ potential interaction and $|M_{21}|\approx 4.2$--$5.6\times10^{-6}$ m$^2$/s for purely repulsive potential interaction at liquid-solid interfaces, and Ref.~\cite{Qi2024} reported $|M_{12}|\approx1$--$5\times10^{-7}$~m$^2$/s for LJ-potential-based liquid-solid interfaces, while experimental studies \cite{Bregulla2016,Tsuji2023} reported $|M_{21}|\approx10^{-10}$--$10^{-9}$ m$^2$/s. 
It should be also noted that MD simulations have shown that the sign of $M_{21}$ (i.e., the direction of thermo-osmosis) can  reverse depending on the degree of hydrophilicity \cite{Fu2017,Wang2020a,Wang2021,Fan2024,Qi2024}. Our present results qualitatively reproduce this sign-reversal trend, as shown in Fig.~\ref{fig:U-dependence}(b), where the thermal-slip coefficient $\b21$ changes sign with the variation of the potential strength $U$. It should also be noted that the ``viscosity" in our BGK fluid is given by Eq.~\eqref{eq:viscosity}, which  is different from the standard microscopic definition of liquid viscosity based on the time correlation of the stress tensor \cite{Allen1987}. This is the why we regard the BGK fluid as a toy model when applied to liquids. 

Despite the oversimplified treatment from a molecular-physics perspective (e.g., long-range interaction, finite-volume effects, layering near interfaces, electric forces), the present approach based on a kinetic model exhibits some qualitative consistency with MD studies for liquids. Specifically, it captures (i) the correct order of magnitude and (ii) the trend with respect to the surface hydrophilic properties (i.e., attractive or repulsive potential). 
A notable strength of the present approach lies in its ability to systematically compute slip coefficients within the framework of the generalized slip-flow theory. In this framework, the slip coefficients are obtained by solving the Knudsen-layer problems [Eq.~\eqref{eq:KLproblem-decomposed21-maintext}], which are spatially one-dimensional boundary-value problems for integro-differential kinetic equations. These problems are numerically tractable, and the well-established solution procedure exist (e.g., Ref.~\cite{Ohwada1989}). 
In particular, the computational cost for the BGK model is relatively low. For example, a standard single CPU is sufficient to compute the thermal-slip coefficient with decent accuracy within a few minutes, using approximately $O(10^4)$ lattice points in the two-dimensional phase space. 
The main limitation of the present model is its omission of inter-molecular potential effects and finite-molecular-volume effects. Nevertheless, it may be possible to extend the model to include such effects by using kinetic models for nonideal gases, such as those developed for the van der Waals fluids \cite{Takata2021}. 


\section{Conclusion}\label{sec:conclusion}
In this paper, we have developed a slip-flow theory to predict slip coefficients in the slip-boundary condition of the Stokes equation, which describes micro- and nanoscale gas flows between two parallel plates in the presence of a near-wall potential. The approach is based on the generalized slip-flow theory in kinetic theory of gases, whereby the determination of slip coefficients is reduced to solving the Knudsen-layer problems (i.e.,  kinetic boundary-layer problems), namely, Eq.~\eqref{eq:KLproblem-decomposed21-maintext}. 
This framework offers a pathway to explore systematically
the influence of near-wall potentials on slip coefficients, while avoiding computationally costly 
molecular simulations. 

Using the developed slip-flow theory, we investigated thermo-osmosis between two parallel plates with a uniform temperature gradient, assuming the presence of a pseudo-Sutherland-type near-wall potential. A comparison between the predictions of the slip-flow theory and the results of numerical analysis showed good agreement, validating the proposed theoretical framework. We also systematically examined the influence of the near-wall potential on thermo-osmosis, thermal-slip coefficients, and Knudsen-layer corrections for various parameter sets corresponding to purely-attractive and purely-repulsive potentials. The results showed that a reversal in the flow direction of thermo-osmosis can occur when the near-wall potential becomes sufficiently repulsive. To be more specific, under a purely repulsive potential of decent strength, the thermo-osmosis flow is directed from hot to cold, which is opposite to the direction typically observed in conventional cases. 

There are several directions for extending the present slip-flow theory. 
\begin{enumerate}
\item The influence of the specific shape of a potential function (i.e., $\psid$) has not been investigated in this study. Physically realistic potential profiles, such as the LJ type that includes both attractive and repulsive parts, should be investigated. It is worth noting that the potential considered in this study [Eq.~\eqref{eq:potential-dim}] is a power-law type. When the potential decays more slowly, say $\Psi^\pm \propto |x_1\mp1/2|^{-1}$, a straightforward application of the slip-flow theory may fail (or may be limited) because of the slow decay of the Knudsen-layer solutions. In such cases, an asymptotic matching procedure may be required, as in the case of a two-dimensional Knudsen layer (or Knudsen zone) \cite{Taguchi2020}. Therefore, the applicability of the present slip-flow theory to more general classes of potentials remains an open question. 
\item \red{The influence of kinetic boundary conditions, which is diffuse reflection here, is considered important. It is reported that, in the case of dense fluids, the temperature gradient can induce thermo-osmosis even with a smooth hard wall \cite{Anzini2025}. This contrasts with a rarefied-gas case, where the specular reflection (without near-wall potential) does not allow thermo-osmosis [see Eq.~\eqref{eq:specular} and description below it], and therefore, is worth further investigation.}
\item The analysis of other slip and jump coefficients under near-wall potentials [Eqs.~\eqref{eq:KLproblem-decomposed10-maintext} and \eqref{eq:KLproblem-decomposed11-maintext}] is also important. In particular, the shear-slip coefficient has been the subject of longstanding interest and deserves a detailed examination in the context of near-wall interactions. 
\item The present analysis is limited to planar geometries and any effects from curved surfaces were not included. As the Knudsen-layer correction begins at the zeroth order in the Knudsen number, the curvature effect is expected to manifest in the first-order Knudsen-layer corrections when a curved surface is considered. Thus, to investigate thermo-osmosis around curved surfaces, including a spherical particle, an extension of the slip-flow theory to curved geometries is required. This is particularly relevant for understanding the mechanism of thermophoresis, the motion of suspended particles driven by temperature gradients in fluids (see, e.g., Ref.~\cite{Takata1995a} for gases and Ref.~\cite{Wuerger2010} for liquids).  
\item The inclusion of long-range inter-molecular interactions and finite-molecular-volume effects is important. Assessing the potential and limitations of the framework of the slip-flow theory in describing not only rarefied gases but also more complex fluids will be addressed in future work.  
\end{enumerate}

\begin{acknowledgments}
This work was supported by the Japan Society for the Promotion of Science KAKENHI Grants No. 22K18770, No. 22K03924, No. 24K00803, No.~25K01156 and also by the Japan Science and Technology Agency PRESTO Grant No. JPMJPR22O7. 
\end{acknowledgments}


\appendix

\setcounter{figure}{0}
\renewcommand\thefigure{A\arabic{figure}}   
\setcounter{table}{0}
\renewcommand\thetable{A\arabic{table}} 

\section{Knudsen-layer problems}\label{sec:KLother}

In this appendix, we explain the details of the Knudsen-layer problems that determine the slip and jump coefficients $\b11$, $\b21$, and $\c10$ and the Knudsen-layer corrections $Y_1^{(1)}$, $Y_2^{(1)}$, $\Omega_1^{(0)}$, and $\Theta_1^{(0)}$. We note that the potentials are localized near the plates, and thus the boundary layer of the one side is not affected by the potential of the other side. Therefore, only the boundary layer near the plate at $x_1=1/2$ is described here. 

To obtain $\b21$ and $Y_2^{(1)}$, which are necessary in the thermo-osmosis problem (Sec.~\ref{sec:thermo-osmosis}), we consider the following boundary-value problem:
\begin{subequations}\label{eq:KLproblem-decomposed21-maintext}
\begin{align}
&\zn\rnd{\p21}{\eta}-U\ddif{\tpsi}{\eta}\rnd{\p21}{\zn} = 
\exp(-2U\tpsi)\left(2Y_{2}^{(1)}-\p21\right) + \ITS , \\
&Y_2^{(1)}(\eta) =\frac{1}{2} \int_{\RR^3} \zrho^2 \p21 E\;\dd \bm{\z}, \\
&\ITS = -\left(2U\tpsi+2\zn U\ddif{\tpsi}{\eta}\right)
+[\exp(-2U\tpsi)-1]\left(\zn^2+\zrho^2-\frac{5}{2}\right), \\
&\p21 = -2\b21+\zn^2+\zrho^2-\frac{5}{2}\quad (\eta=0,\;\zn>0), \\
&\p21 \to 0 \quad (\eta\to\infty),
\end{align}
\end{subequations}
where the notations follow those in Ref.~\cite{Takata2012} and explained below one by one. First, the molecular velocity $\z_i$ is decomposed into $\zn=\z_in_i$ with $n_i$ the unit normal on the boundary at $x_1=1/2$ pointing to the gas (i.e., $\zn=-\z_1$) and $\zrho=\sqrt{\z_2^2+\z_3^2}$. The unknown function $\p21=\p21(\eta,\,\zn,\,\zrho)$ is the portion of the molecular velocity distribution function for the Knudsen-layer correction, and is related with the temperature gradient of the Hilbert part (see \SM~\ref{sec:KL} for detail). 
In short, the terms related with $\tpsi$ are the contributions found in this paper. Substituting $\tpsi=0$ into Eq.~\eqref{eq:KLproblem-decomposed21-maintext} formally reproduces the known result. 
This problem corresponds to the {\it thermal slip} under the presence of a near-wall potential field. 

The Knudsen-layer problem that determines $\c10$, $\Omega_1^{(0)}$, and $\Theta_1^{(0)}$ is summarized as 
\begin{subequations}\label{eq:KLproblem-decomposed10-maintext}
\begin{align}
&\zn\rnd{\p10}{\eta}-U\ddif{\tpsi}{\eta}\rnd{\p10}{\zn} = 
\exp(-2U\tpsi)\left[\Omega_1^{(0)}+\left(\z_1^2+\zrho^2-\frac{3}{2}\right)\Theta_1^{(0)}-\p10\right] +\ITJ, \\
&\Omega_1^{(0)}(\eta) = \int_{\RR^3} \p10 E\;\dd \bm{\z}, \quad 
\Theta_1^{(0)}(\eta)=\int_{\RR^3} \frac{2}{3}\left(\z_1^2+\zrho^2-\frac{3}{2}\right)\p10 E\;\dd \bm{\z}, \\
&\ITJ = -U\ddif{\tpsi}{\eta}\left(3\zn^2 + \zrho^2 -\frac{5}{2} - 2 \eta \zn\right)
+[\exp(-2U\tpsi)-1]\zn\left(\zn^2+\zrho^2-\frac{5}{2}\right),
\\
&\p10 = -2\sqrt{\pi}\int_{\zn<0} \zn \p10 E \;\dd \bm{\z} -(\zn^2+\zrho^2-2)\c10 + \zn\left(\zn^2+\zrho^2-\frac{5}{2}\right)\quad (\eta=0,\;\zn>0), \\
&\p10 \to 0 \quad (\eta\to\infty),
\end{align}
\end{subequations}
where $\p10=\p10(\eta,\,\zn,\,\zrho)$ is the unknown function. 
This problem corresponds to the {\it temperature jump} under the presence of a near-wall potential field. 

The Knudsen-layer problem that determines $\b11$ and $Y_1^{(1)}$ is summarized as 
\begin{subequations}\label{eq:KLproblem-decomposed11-maintext}
\begin{align}
&\zn\rnd{\p11}{\eta}-U\ddif{\tpsi}{\eta}\rnd{\p11}{\zn} 
= 
\exp(-2U\tpsi)\left(2Y_{1}^{(1)}-\p11\right) +\ISS, \\
&Y_1^{(1)}(\eta) =\frac{1}{2} \int_{\RR^3} \zrho^2 \p11 E\;\dd \bm{\z}, \\
&\ISS = -2U\ddif{\tpsi}{\eta}+[\exp(-2U\tpsi)-1]2\zn, \\
&\p11 = -2\b11+2\zn\quad (\eta=0,\;\zn>0), \\
&\p11 \to 0\quad (\eta\to\infty),
\end{align}
\end{subequations}
where $\p11=\p11(\eta,\,\zn,\,\zrho)$ is the unknown function. 
This problem corresponds to the {\it shear slip} under the presence of a near-wall potential field. 

We expect that solutions to Eqs.~\eqref{eq:KLproblem-decomposed21-maintext}, \eqref{eq:KLproblem-decomposed10-maintext}, and \eqref{eq:KLproblem-decomposed11-maintext} can be found only when the unknown coefficients $\c10$, $\b11$, and $\b12$ take particular values. 
Therefore, if we could find the the solutions, $\c10$, $\b11$, and $\b12$ are determined simultaneously. 
It should be noted that the existence and uniqueness of the solution for the Knudsen layer problem, formulated for the Boltzmann equation, is mathematically proved for the cases without the potential \cite{Bardos1986,Coron1988}.

\clearpage

\appendix
\onecolumngrid

\setcounter{page}{1}
\renewcommand{\appendixname}{Supplemental Materials}

\begin{center}
Supplemental Materials on
{\large
\\
{\bf \titleB}}
\\[1em]
{
Tetsuro Tsuji$^\ast$, Koichiro Takita, and Satoshi Taguchi\\[0.5em]
Graduate School of Informatics, Kyoto University, Kyoto 606-8501, Japan\\[0.5em]
$^\ast$tsuji.tetsuro.7x@kyoto-u.ac.jp
}
\end{center}

\setcounter{figure}{0}
\renewcommand\thefigure{S\arabic{figure}}   
\setcounter{table}{0}
\renewcommand\thetable{S\arabic{table}} 
\renewcommand\theequation{\thesection.\arabic{equation}} 
\renewcommand{\thesection}{S\arabic{section}}
\renewcommand{\thesubsection}{\Alph{subsection}}
\renewcommand{\thesubsubsection}{\Roman{subsubsection}}


\section{Slip-flow theory: the derivation of the slip and jump boundary conditions}\label{sec:slip-flow-theory-detail}

For convenience, we repeat the system considered in the main text, namely, Eqs.~\eqref{eq:bgk-nd}--\eqref{eq:macro}: 
\begin{subequations}\label{eq:sm-bgk}
\begin{align}
&
\z_i\rnd{\phi}{x_i} - U 
\dif{\psid}{x_1}
\rnd{\phi}{\z_1} = \frac{\alpha}{k} (\phi_\e - \phi), \label{eq:sm-bgk-a}\\
&
\phi_\e = \omega + 2 \z_i u_i + \left(\z^2-\frac{3}{2}\right)\tau, 
\label{eq:sm-bgk-b}\\
&\alpha = \bpsid^{-1} \exp(-2U\psid) \quad 
\left(\bpsid = \int_{-1/2}^{1/2} \exp(-2U\psid(x_1)) \;\dd x_1\right),
\label{eq:sm-alpha}\\
&
\begin{bmatrix}
\omega\\ u_i\\
\tau
\end{bmatrix}
=
\int_{\RR^3} 
\begin{bmatrix}
1\\ \z_i\\
\frac{2}{3}\left(\z^2-\frac{3}{2}\right)
\end{bmatrix} \phi E\; \dd \bm{\z}, \quad 
E=\pi^{-1/2}\exp(-\z^2), 
\label{eq:sm-bgk-c}\\
&
\text{(b.c.)}
\begin{dcases}
\phi = \phi_{\w}^\pm \quad \left(x_1=\pm \frac{1}{2},\;\z_1\lessgtr0\right),
\\
\phi_{\w}^\pm = 2\z_iu_{\w i}^\pm + (\z^2-2)\tau_\w^\pm  \pm 2\spi\int_{\z_1\gtrless 0} 
\z_1 \phi E \; \dd \bm{\z}, 
\end{dcases}\label{eq:sm-bc}\\
&P=\omega + \tau, \quad 
P_{ij}
=
2\int_{\RR^3} 
\z_i \z_j\phi E\; \dd \bm{\z}, 
\quad 
Q_i
=
\int_{\RR^3} 
\z_i \z^2
\phi E\; \dd \bm{\z},
\end{align}
\end{subequations}
where $\z=|\z_i| = (\zeta_j^2)^{1/2}$ and $u_{\w 1}^\pm = 0$.
In the following, we show the procedure of the generalized slip-flow theory, the asymptotic analysis with respect to small $k\sim \delta \ll1$.

\subsection{Fluid-dynamic part}\label{sec:fluid-dynamic-part}


\subsubsection{Hilbert solution}
We first consider the overall flow behavior, leaving aside the boundary condition \eqref{eq:sm-bc}. Accordingly, we neglect the effect of the near-wall potential in Eq.~\eqref{eq:sm-bgk-a}. To be more precise, we rewrite Eq.~\eqref{eq:sm-bgk-a} in the form
\begin{align}
\z_i\rnd{\phi}{x_i} = \frac{1}{k} (\phi_\e - \phi) 
+
\underbrace{\frac{\alpha-1}{k} (\phi_\e - \phi)  
+ U \dif{\psid}{x_1}
\rnd{\phi}{\z_1}}_{\equiv\; \I[\phi]}, \label{eq:bgk-nd-discard}
\end{align}
where $\I[\phi]=k^{-1}(\alpha-1)(\phi_\e - \phi)  
+ U (\tdif{\psid}{x_1})(\trnd{\phi}{\z_1})$, and neglect $\I[\phi]$ that is significant only near the boundaries. Note that $\I[\phi]$ depends linearly on $\phi$, since $\phi_\e$ is a linear functional of $\phi$ through $\omega$, $u_i$, and $\tau$ [see Eq.~\eqref{eq:sm-bgk-c}].
The contribution of $\I[\phi]$ will be discussed in Sec.~\ref{sec:KL}, where boundary effects are taken into account. Once we neglect $\I[\phi]$, Eq.~\eqref{eq:bgk-nd-discard} reduces to the conventional linearized BGK equation, which can be analyzed in a straightforward manner using a standard Hilbert expansion method (see, e.g., Chapter 3 of the textbook \cite{Sone2007}). In the following, we briefly summarize the results of this analysis. 

We seek a solution of Eq.~\eqref{eq:bgk-nd-discard} (neglecting the term $\I[\phi]$) that varies moderately in space variables, that is, we assume $|\partial\phi/\partial x_i|=O(\phi)$. This type of solution is called the Hilbert solution, and will be denoted by $\phi_\G$ in the following. 
The associated macroscopic quantities are also expressed with the subscript $\G$. 
The velocity distribution function $\phi_\G$ and the macroscopic quantities $h_\G$ ($h=\omega$, $u_i$, etc.) are expanded in a power series of $k$ as 
\begin{equation}
\begin{split}
&
\phi_{\G} =
\phi_{\G0}
+\phi_{\G1} k
+ O(k^2), \quad 
h_{\G} = 
h_{\G0}
+h_{\G1} k
+ O(k^2), 
\label{eq:hilbert-expansion}
\end{split}
\end{equation}
where the relations between $\phi_{\G m}$ and $h_{\G m}$ $(m=0,\,1,\,...)$ are given by 
\begin{equation}
\begin{split}
&
\omega_{\G m}
=
\int_{\RR^3} 
\phi_{\G m} E\; \dd \bm{\z}, \quad 
u_{i\G m}
=
\int_{\RR^3} 
\z_i\phi_{\G m} E\; \dd \bm{\z}, \\
&
\tau_{\G m}
=
\int_{\RR^3} 
\frac{2}{3}\left(\z^2-\frac{3}{2}\right)\phi_{\G m} E\; \dd \bm{\z}, 
\quad 
P_{\G m}=\omega_{\G m} + \tau_{\G m}, \\
&P_{ij\G m}
=
2\int_{\RR^3} 
\z_i \z_j\phi_{\G m} E\; \dd \bm{\z}, 
\quad 
Q_{i\G m}
=
\int_{\RR^3} 
\z_i \z^2
\phi_{\G m} E\; \dd \bm{\z} - \frac{5}{2} u_{i\G m}.
\label{eq:hGm}
\end{split}
\end{equation}

Substituting Eq.~\eqref{eq:hilbert-expansion}
into the BGK equation Eq.~\eqref{eq:bgk-nd-discard} with $\phi=\phi_\G$ (neglecting the term $\I[\phi_\G]$), we obtain
a series of equations for $\phi_{\G m}$ $(m=0,\,1,...)$:
\begin{align}
\phi_{\e\G 0} - \phi_{\G 0} = 0, \quad
\phi_{\e\G m} - \phi_{\G m} = \z_i\rnd{\phi_{\G m-1}}{x_i}
\quad(m=1,2,...), \label{eq:bgk-nd-a-fluid}
\end{align}
where $\phi_{\e\G m} = \omega_{\G m}
+2 \z_i u_{i\G m}  +
(\z^2-3/2) \tau_{\G m}$ (i.e., $m$-th order linearized local equilibrium distributions). 
Equation~\eqref{eq:bgk-nd-a-fluid} give the following expressions for $\phi_{\G m}$:
\begin{subequations}\label{eq:phiG}
\begin{align}
&\phi_{\G 0}=\omega_{\G 0}
+2 \z_i u_{i\G 0} +
\left(\z^2-\frac{3}{2}\right) \tau_{\G 0}, \label{e:Hilbert_0}\\
&
\phi_{\G1}
=\omega_{\G 1}
+2 \z_i u_{i\G 1} +
\left(\z^2-\frac{3}{2}\right) \tau_{\G 1} - \z_i\left(\z^2-\frac{5}{2}\right)\rnd{\tau_{\G 0}}{x_i} - 
2\z_i\z_j\rnd{u_{j\G0}}{x_i},  \label{e:Hilbert_1} \\
&\phi_{\G2} = 
\omega_{\G2}
+2\z_i u_{i\G2} 
+\left(\z^2-\frac{3}{2}\right)\tau_{\G2}
- \z_i\left(\z^2-\frac{5}{2}\right)\rnd{\tau_{\G 1} }{x_i} - 
2\z_i\z_j\rnd{u_{j\G1} }{x_i} \notag\\ 
&\hspace{2.5em} 
+\z_i\z_j\left(\z^2-\frac{5}{2}\right)\rnd{^2\tau_{\G0}}{x_i\partial x_j}
+2\z_k\z_j\z_i\rnd{^2u_{i\G0}}{x_k\partial x_j}-\z_j\rnd{P_{\G1}}{x_i}.
\end{align}
\end{subequations}
Multiplying $E$, $\z_j E$, and $\z^2 E$ to Eqs.~\eqref{eq:bgk-nd-a-fluid} and integrating over the whole $\z_i$ for $m=1,\,2,\,...$, the left-hand side vanishes, yielding the expanded conservation laws (i.e., the fluid-dynamic type equations). In the present case, the resulting system corresponds to the Stokes equations: 
\begin{subequations}\label{eq:stokes}
\begin{align}
&\rnd{P_{\G0}}{x_i}=0, \\
&\rnd{u_{i\G0}}{x_i}=0, \quad 
\rnd{P_{\G1}}{x_i}-\rnd{^2u_{i\G0}}{x_j^2}=0, \quad 
\rnd{^2\tau_{\G0}}{x_j^2} = 0,\\
&\rnd{u_{i\G1}}{x_i}=0, \quad 
\rnd{P_{\G2} }{x_i}-\rnd{^2u_{i\G1}}{x_j^2}=0, \quad 
\rnd{^2\tau_{\G1}}{x_j^2} = 0, 
\end{align}
\end{subequations}
supplemented by the equation of state $P_{\G m} = \omega_{\G m} + \tau_{\G m}$ ($m=0,\,1,\,2$).
The expressions of the stress tensor $P_{ij\G m}$ and the heat flux $Q_{i\G m}$ are given by
\begin{subequations}\label{eq:constitutive-equation}
\begin{align}
&P_{ij\G0} = P_{\G0}\delta_{ij}, \quad 
Q_{i\G0} = 0, \\
&P_{ij\G1} = P_{\G1}\delta_{ij} - \left(\rnd{u_{i\G0}}{x_j}+\rnd{u_{j\G0}}{x_i}\right), \quad 
Q_{i\G1} = -\frac{5}{4}\rnd{\tau_{\G0}}{x_i},\\
&P_{ij\G2} = P_{\G2}\delta_{ij} - \left(\rnd{u_{i\G1} }{x_j}+\rnd{u_{j\G1} }{x_i}\right)+\rnd{\tau_{\G0}}{x_ix_j}, \quad 
Q_{i\G2} = -\frac{5}{4}\rnd{\tau_{\G1} }{x_i} + \frac{1}{2}\rnd{P_{\G1}}{x_i}.
\end{align}
\end{subequations}
As expected from the absence of $\I[\phi_\G]$ in Eq.~\eqref{eq:bgk-nd-discard}, the Stokes system \eqref{eq:stokes} remains unchanged from the case without the near-wall potential.
By reconstructing $h_\G$ from the expansion in Eq.~\eqref{eq:hilbert-expansion}, the (expanded) Stokes equations Eq.~\eqref{eq:stokes} yield the fluid-dynamic equations Eq.~\eqref{eq:Stokeseqs} discussed in the main text.

\subsubsection{Consistency of the boundary conditions}\label{sec:bc-check}

We now consider the boundary condition that was set aside and examine whether the Hilbert solution in Eq.~\eqref{eq:phiG} satisfies the condition \eqref{eq:sm-bc}. To this end, we expand the boundary values $\tau_\w^\pm$ and $u_{\w i}^\pm$ in \eqref{eq:sm-bc} in powers of $k$ as follows:
\begin{align}
\tau_\w^\pm=\tau_{\w 0}^\pm + \tau_{\w 1}^\pm k+O(k^2), \quad
u_{\w i}^\pm=u_{\w i0}^\pm + u_{\w i1}^\pm k + O(k^2).
\end{align}
This yields the following boundary conditions for the zeroth- and first-order terms in the Hilbert expansion:
\begin{align}
\phi_{\G m} = \phi_{\w m}^\pm, \quad m=0,\,1, \quad \left(x_1=\pm\hf,\;\z_1\lessgtr 0\right),
\end{align}
where
\begin{subequations}\label{eq:bc-expand}
\begin{align}
&\phi_{\w0}^\pm = 2\z_i u_{\w i0}^\pm  + (\z^2-2)\tau_{\w 0}^\pm \pm 2\sqrt{\pi} \int_{\z_1\gtrless 0} \z_1 \phi_{\G0}E \; \dd \bm{\z}, \label{eq:bc-nd-k0}\\
&\phi_{\w1}^\pm = 2\z_i u_{\w i1}^\pm + (\z^2-2)\tau_{\w 1}^\pm \pm 2\sqrt{\pi} \int_{\z_1\gtrless 0} \z_1 \phi_{\G1}E \; \dd \bm{\z}, \label{eq:bc-nd-k1}
\end{align}
\end{subequations}
with $u_{\w 1 0}^\pm= u_{\w 1 1}^\pm=0$.
At the leading order, substituting \eqref{e:Hilbert_0} into \eqref{eq:bc-nd-k0}, $\phi_{\w0}^\pm$ is computed as
\begin{align}
&\phi_{\w0}^\pm=
2\z_iu_{\w i 0}^\pm + (\z^2-2)\tau_{\w0}^\pm + \omega_{\G0} \pm \sqrt{\pi}u_{1\G0} + \frac{1}{2}\tau_{\G0}  .
\end{align}
Therefore, to satisfy the zeroth-order boundary conditions,
we need to impose the following conditions on the boundaries:
\begin{align}
u_{i\G0} = u_{\w i 0}^\pm,\quad \tau_{\G0} = \tau_{\w 0}^\pm \quad \left(x_1 =\pm\hf\right). \label{eq:0th-order-bc}
\end{align}
The conditions in Eq.~\eqref{eq:0th-order-bc} are nothing but the no-slip and no-jump boundary conditions. 

Next, we consider the first-order boundary condition, which is given by 
$\phi_{\G 1} = \phi_{\w1}^+$ for $\z_1 < 0$ at $x_1= 1/2$ and $\phi_{\G 1} = \phi_{\w1}^-$ for $\z_1 > 0$ at $x_1= -1/2$ (with $u_{\w 1 1}= 0$).
Using the form of $\phi_{\G1}$ given in \eqref{e:Hilbert_1}, we observe that the boundary conditions cannot generally be satisfied due to a mismatch in the functional dependence on $\z_i$.
This inconsistency 
arises from assumption of moderate spatial variation of the Hilbert solution, i.e., $\partial\phi_\G/\partial x_i=O(\phi_\G)$.
Thus, a boundary-layer correction is necessary to satisfy the boundary condition at the first order in $k$.
This correction is referred to as the Knudsen-layer correction \cite{Sone2007}.


\subsection{Knudsen-layer correction}\label{sec:KL}


\subsubsection{Preliminaries}
We now consider the full equation \eqref{eq:bgk-nd-discard} [or \eqref{eq:sm-bgk-a}], retaining the term $\I[\phi]$, in order to account for the near-wall potential.
The solution $\phi$ then expressed in the form
\begin{align}
\phi = \phi_\G + \phi_\K, 
\end{align}
where $\phi_\K$ denotes the Knudsen-layer correction, which is assumed to decay with the distance from the boundaries so as to ensure matching with the Hilbert solution.
Accordingly, the macroscopic variables are expressed as
\begin{align}
    h = h_\G + h_\K.
\end{align}
Here, $h_\K$ is defined by the same expression as $h$ with $\phi$ replaced by $\phi_\K$.
The following discussion focuses on the Knudsen-layer correction associated with the boundary at  $x_1=1/2$, since the analysis for the correction associated with the boundary $x_1=-1/2$ can be carried out in a similar way.
We introduce the unit normal vector $\bm{n}=(-1,\,0,\,0)$ to the boundary, which is chosen to point into the gas region. We also
introduce the stretched coordinate variable $\eta$, the normal velocity component $\zn$, the tangential velocity component $\bz_i$, defined as
\begin{align}
\eta = \frac{1}{k}\left(\frac{1}{2}-x_1\right),\quad \zn=\z_in_i=-\z_1, \quad 
\bz_i=\z_i-\zn n_i.
\end{align}
Using $\eta$ and neglecting the effect of the potential from the opposite plate at $x_1=-1/2$, the potential $\psid$ is approximated as
\begin{align}
&
\psid(x_1)\approx\tpsi(\eta), \quad 
\tpsi \to 0\quad (\eta\to\infty), \quad 
\dif{\psid}{x_1} \approx - \frac{1}{k}\dif{\tpsi}{\eta}. 
\end{align}
Introducing $\phi=\phi_{\K}(\eta,x_2,x_3,\zn,\bz_i)$, the BGK equation \eqref{eq:sm-bgk-a} is recast as
\begin{subequations}\label{eq:bgk-nd-KL}
\begin{align}
&\z_n\rnd{\phi_\K}{\eta} 
+k \bz_i \rnd{\phi_{\K}}{x_i}
- U 
\dif{\tpsi}{\eta}
\rnd{\phi_\K}{\zn}
= 
\talpha (\phi_{\e \K} - \phi_\K) 
+k \I[\phi_\G], 
\label{eq:bgk-nd-KL-a}
\\
&\talpha = \talpha(\eta) = \bpsid^{-1}\exp(-2U\tpsi) [\approx \alpha(x_1)], \\
&
\phi_{\e\K} = \omega_\K + 2\z_iu_{i\K} + \left(\z^2-\frac{3}{2}\right)\tau_\K, \quad 
\begin{bmatrix}
\omega_\K\\ u_{i\K}\\
\tau_\K
\end{bmatrix}
=
\int_{\RR^3} 
\begin{bmatrix}
1\\ \z_i\\
\frac{2}{3}\left(\z^2-\frac{3}{2}\right)
\end{bmatrix} \phi_\K E\; \dd \bm{\z},
\end{align}
\end{subequations}
where $\I$ is defined in the sentence containing \eqref{eq:bgk-nd-discard}.
Note that the contribution from the Hilbert solution appears as an inhomogeneous term, i.e., $\I[\phi_\G]$ in \eqref{eq:bgk-nd-KL-a}.
Since $\phi_\K$ is a correction,
we assume that 
\begin{align}
\phi_\K\to0 \quad (\eta\to\infty).
\end{align}

In Sec.~\ref{sec:bc-check}, we have seen that the Knudsen layer correction is necessary at first order in $k$.
This observation might suggest that $\phi_\K=O(k)$. 
However, due to the presence of the inhomogeneous term $\I[\phi_G]$
in Eq.~\eqref{eq:bgk-nd-KL-a}, we must assume $\phi_\K=O(1)$, as will be demonstrated below. Accordingly, we expand $\phi_\K$ and the associated macroscopic quantities $h_\K(=\omega_\K,\,u_{i\K},...)$ as follows:
\begin{subequations}\label{eq:KL-expansion}
\begin{align}
&\phi_{\K} = \phi_{\K0} + \phi_{\K1} k + O(k^2), \quad 
h_{\K} = h_{\K0}+  h_{\K1} k + O(k^2), \\
&
\begin{cases}
\omega_{\K m}
=
\displaystyle\int_{\RR^3} 
\phi_{\K m} E\; \dd \bm{\z}, \quad 
u_{i\K m}
=
\displaystyle\int_{\RR^3} 
\z_i\phi_{\K m} E\; \dd \bm{\z}, \label{eq:KL-macro}\\[1em]
\tau_{\K m}
=
\displaystyle\int_{\RR^3} 
\frac{2}{3}\left(\z^2-\frac{3}{2}\right)\phi_{\K m} E\; \dd \bm{\z}, 
\quad 
P_{\K m}=\omega_{\K m} + \tau_{\K m}, \\[1em]
P_{ij\K m} = 2 \displaystyle\int_{\RR^3} 
\z_i\z_j\phi_{\K m} E\; \dd \bm{\z}, \quad 
Q_{i\K m} =  \displaystyle\int_{\RR^3} 
\z_i\z_j^2\phi_{\K m} E\; \dd \bm{\z} - \frac{5}{2} u_{i\K m}.
\end{cases}
\end{align}
\end{subequations}

Suppose that we can approximate $\bpsid$ [see Eq.~\eqref{eq:alpha} in the main text] as 
\begin{align}
\bpsid \approx 1 + \bpsid^{(1)} \delta + O(\delta^2) =
1 + c^{(1)} k + O(k^2), \quad c^{(1)} = \bpsid^{(1)}/\chi, 
\label{eq:alpha-expansion}
\end{align}
with $\bpsid^{(1)}$ being a constant depending on $U$ but not on $\delta$. 
In fact, we have $\displaystyle\lim_{\delta\to 0}\bpsid=1$ from Eqs.~\eqref{eq:alpha} and \eqref{eq:psi-nd}, and it is possible to explicitly estimate the validity of Eq.~\eqref{eq:alpha-expansion} for the specific potential considered in this paper [see Eq.~\eqref{eq:potential-dim} in the main text]. The detail of the estimate is provided in \SM~\ref{sec:potential}. Accordingly, $\talpha$ is expressed as 
\begin{align}
\talpha(\eta) = [1 - c^{(1)} k + O(k^2) ]\exp(-2U\tpsi(\eta)). 
\end{align}  

The macroscopic quantities $h_{\G m}$ and their spatial derivatives in the fluid-dynamic part are expanded around $x_1=1/2$ (or $\eta = 0$) as
\begin{subequations}\label{eq:h-expansion}
\begin{align}
&h_{\G m} =\zero{h_{\G m}} + (- k \eta) \zero{\rnd{h_{\G m}}{x_1}} +\hf (-k\eta)^2 \zero{\rnd{^2h_{\G m}}{x_1^2}} + \cdots,\\
&\rnd{^nh_{\G m}}{x_1^n} 
=\zero{\rnd{^nh_{\G m}}{x_1^n} } + (- k \eta) \zero{\rnd{^{n+1}h_{\G m}}{x_1^{n+1}}} +\hf (-k\eta)^2 \zero{\rnd{^{n+2}h_{\G m}}{x_1^{n+2}}} + \cdots,
\end{align}
\end{subequations}
where $n=1,2,...$.
Here, $\partial^n h_{\G m}/\partial x_1^n$ are all assumed to be $O(1)$ since the fluid-dynamic parts are moderately-varying quantities; the quantity in the parenthesis, $\zero{a}$, means that the quantity $a$ is evaluated at $\eta=0$. It should be noted that $\zero{a}$ is not a function of $\eta$ but a function of $(x_2,x_3)$.


\subsubsection{Knudsen-layer correction at the leading order}
Inserting Eqs.~\eqref{eq:KL-expansion}--\eqref{eq:h-expansion} into Eq.~\eqref{eq:bgk-nd-KL}, we obtain, at the leading order of $k$, 
\begin{align}
&\zn\rnd{\phi_{\K0}}{\eta}-U\ddif{\tpsi}{\eta}\rnd{\phi_{\K0}}{\zn}
-2 \zn \zero{\tau_{\G0}}U\ddif{\tpsi}{\eta}
= 
\exp(-2U\tpsi)\left(\phi_{\e\K0}-\phi_{\K0}\right),  \label{eq:bgk-phi-K0}
\end{align}
where $\phi_{\e\K 0}$ is defined by $\phi_{\e\K 0} = \omega_{\K 0} + 2 \z_i u_{i \K 0} + \left(\z^2-\frac{3}{2}\right)\tau_{\K 0}$.
Since $\phi_{\G0}$ satisfies the boundary condition as long as Eq.~\eqref{eq:0th-order-bc} is met, the boundary condition of $\phi_{\K0}$ and the condition at infinity are given by
\begin{subequations}\label{eq:bc-phi-K0}
\begin{align}
&\phi_{\K0} = -2\sqrt{\pi} \int_{\zn < 0} \zn \phi_{\K0} E \;\dd \bm{\z}\quad (\zn > 0, \;\eta = 0), \\ 
&\phi_{\K0} \to 0 \quad (\eta \to \infty).
\end{align}
\end{subequations} 
The solution $\phi_{\K 0}$ to Eqs.~\eqref{eq:bgk-phi-K0} and \eqref{eq:bc-phi-K0}, and the resulting Knudsen-layer corrections for the macroscopic quantities, are given by
\begin{subequations}\label{eq:KLsolution-0}  
\begin{align}
&\phi_{\K0}= 2 U \tpsi \zero{\tau_{\G0}}, \quad 
\omega_{\K 0} = 2 U \tpsi \zero{\tau_{\G0}}, \quad 
u_{i\K 0} = 0, \quad 
\tau_{\K 0} = 0, \\
&P_{\K 0} = 2 U \tpsi \zero{\tau_{\G0}}, \quad 
P_{ij \K 0} = 0, \quad 
Q_{i \K 0} = 0,
\end{align}
\end{subequations}
where the functional dependency of the quantities on $\eta$ has been omitted.
The perturbation of the density induced by the near-wall potential is confined within the Knudsen layer and is represented by the correction term $\omega_{\K 0}$. 


\subsubsection{Knudsen-layer correction at the first order}
The first-order equation is 
\begin{align}
&\zn\rnd{\phi_{\K1}}{\eta}-U\ddif{\tpsi}{\eta}\rnd{\phi_{\K1}}{\zn}
+U\ddif{\tpsi}{\eta}\rnd{\phi_{\G1}}{\z_1} 
+ 2 U \tpsi \bar{\z}_i \zero{\rnd{\tau_{\G0}}{x_i}}
-
2 \eta \z_1 U\ddif{\tpsi}{\eta} \zero{\rnd{\tau_{\G0}}{x_1}}
\notag\\
&= 
\exp(-2U\tpsi)\left(\phi_{\e\K1}-\phi_{\K1}\right)
+
(\exp(-2U\tpsi)-1)\left(\phi_{\e\G1}-\phi_{\G1}\right),  \label{eq:bgk-phi-K1}
\end{align}
where $\phi_{\e\K 1}$ is defined by $\phi_{\e\K 1} = \omega_{\K 1} + 2 \z_i u_{i \K 1} + \left(\z^2-\frac{3}{2}\right)\tau_{\K 1}$. 
In the derivation, we have used the relations $\phi_{\e\G0}-\phi_{\G0}=0$, $\phi_{\K0}=2U\tpsi \zero{\tau_{\G 0}}$, $\phi_{\e\K0}-\phi_{\K0}=0$, and $\zero{\partial u_{1\G0}/\partial x_1}=0$. The relation $\zero{\partial u_{1\G0}/\partial x_1}=0$ holds because we have $\zero{\partial u_{i\G0}/\partial x_i}=0$ due to mass conservation and $\zero{u_{2\G0}}=\zero{u_{3\G0}}=0$ [i.e., $\zero{\partial u_{2\G0}/\partial x_2}=\zero{\partial u_{3\G0}/\partial x_3}=0$] due to no-slip boundary condition at zeroth order. 
Note that $c^{(1)}$ in Eq.~\eqref{eq:alpha-expansion} does not appear here. 
The Hilbert part $\partial \phi_{\G1}/\partial \z_1$ and $\phi_{\e\G1}-\phi_{\G1}$ in Eq.~\eqref{eq:bgk-phi-K1} are expanded around $\eta=0$ as follows:
\begin{subequations}
\begin{align}
&\rnd{\phi_{\G1}}{\z_1} \approx 
2\zero{ u_{1\G1}}+  
2\z_1 \zero{ \tau_{\G1}} \notag\\
&\phantom{\rnd{\phi_{\G1}}{\z_1} \approx }
-\left(2\z_1^2 + \z^2 -\frac{5}{2}\right)\zero{\rnd{\tau_{\G0}}{x_1}}
- 2\z_1\bar{\z}_i \zero{\rnd{\tau_{\G0}}{x_i}} 
-2\bar{\z_i}\zero{\rnd{u_{i\G0}}{x_1}}, \\
&\phi_{\e\G1}-\phi_{\G1} 
\approx
\z_1\left(\z^2-\frac{5}{2}\right)\zero{\rnd{\tau_{\G0}}{x_1}}
+\bar{\z}_j\left(\z^2-\frac{5}{2}\right)\zero{\rnd{\tau_{\G0}}{x_j}}
+2\bar{\z_j}\z_1\zero{\rnd{u_{j\G0}}{x_1}},
\end{align} 
\end{subequations} 
where 
terms of $O(k)$ have been ignored. 
Finally, by neglecting terms of $O(k)$, Eq.~\eqref{eq:bgk-phi-K1} reduces to
\begin{subequations}\label{eq:bgk-phi-K1-v4}
\begin{align}
&\zn\rnd{\phi_{\K1}}{\eta}-U\ddif{\tpsi}{\eta}\rnd{\phi_{\K1}}{\zn} 
  +2U\ddif{\tpsi}{\eta}\left[\zero{u_{1\G1}}-\zn \zero{\tau_{\G1}}\right] 
=
\exp(-2U\tpsi)\left(\phi_{\e\K1}-\phi_{\K1}\right) 
+ \I_{\K1}, \label{eq:bgk-phi-K1-v4a} \\
& \I_{\K1} = -\ITJ \zero{\rnd{\tau_{\G0}}{x_1}}
-\ISS \bar{\z_i}\zero{\rnd{u_{i\G0}}{x_1}}
+\ITS \bar{\z}_i \zero{\rnd{\tau_{\G0}}{x_i}}, \label{eq:bgk-phi-K1-v4-b}
\\
&\ITJ = -U\rnd{\tpsi}{\eta}\left(2\zn^2 + \z^2 -\frac{5}{2} - 2 \eta \zn\right)
+[\exp(-2U\tpsi)-1]\zn\left(\z^2-\frac{5}{2}\right),
\\
&\ISS = -2U\rnd{\tpsi}{\eta}
+[\exp(-2U\tpsi)-1]2\zn, \\
&\ITS = -\left(2U\tpsi+2\zn U\rnd{\tpsi}{\eta}\right)
+[\exp(-2U\tpsi)-1]\left(\z^2-\frac{5}{2}\right).
\end{align}
\end{subequations}
Here, $\ITJ$, $\ISS$, and $\ITS$ in the inhomogeneous term represent, respectively, the contributions to the temperature-jump, shear-slip, and thermal-slip problems introduced below. 

Before presenting the boundary condition for $\phi_{\K 1}$, we show that $\zero{u_{1\G1}}=0$ holds. This relation simplifies both Eq.~\eqref{eq:bgk-phi-K1-v4a} and the boundary condition. 
Multiplying Eq.~\eqref{eq:bgk-phi-K1-v4a} by $E$ and integrating over the whole $\bm{\z}$ space, we obtain the following relation (i.e., conservation of mass): 
\begin{align}
\rnd{u_{1\K1}}{\eta} =
2 U\dif{\tpsi}{\eta} \left[\zero{u_{1\G1}}+u_{1\K1} \right].
\end{align}
The solution to this equation that vanishes as $\eta \to \infty$ is given by 
\begin{align}
u_{1\K1}=\zero{u_{1\G1}}[\exp(2U\tpsi)-1]. \label{eq:masscons1}
\end{align}
However, 
the mass flux across the boundary must vanish due to the diffuse reflection, i.e., 
\begin{align}
\zero{u_{1\G1}}+u_{1\K1}=0\quad (\eta=0). \label{eq:masscons2}
\end{align} 
Equations~\eqref{eq:masscons1} and \eqref{eq:masscons2} imply $\zero{u_{1\G1}}=0$, and hence $u_{1\K1}=0$ for all $\eta$. 

The boundary condition on the plate $x_1=1/2$ (i.e., $\eta=0$) for $\phi_{\K1}$ is derived by requiring that the sum $\phi_{\G1}+\phi_{\K1}$ satisfies the diffuse reflection condition. In other words, we impose $\phi_{\G1}+\phi_{\K1} = \phi_{\w 1}^+$ for $\zeta_n>0$, where $\phi_{\w 1}^+$ is given by Eq.~\eqref{eq:bc-nd-k1} with $\phi_{\G1}$ in the integral changed to $\phi_{\G1}+\phi_{\K1}$. This yields
\begin{align}
\phi_{\K1} &= 2 \bz_i u_{\w i1}^+ + (\z^2-2)\tau_{\w1}^+-2\sqrt{\pi}\int_{\zn<0} \zn (\phi_{\G1}+\phi_{\K1}) E \;\dd \bm{\z} - \phi_{\G1} \notag\\
&=-2\sqrt{\pi}\int_{\zn<0} \zn \phi_{\K1} E \;\dd \bm{\z} - (\z^2-2)[\zero{\tau_{\G1}}-\tau_{\w1}^+] - \zn\left(\z^2-\frac{5}{2}\right)\zero{\rnd{\tau_{\G 0}}{x_1}}\notag\\
&\phantom{=}
+ \bar{\z_i}\left\{-2[\zero{u_{i\G 1}}-u_{\w i 1}^+] + \left(\z^2-\frac{5}{2}\right)
\zero{\rnd{\tau_{\G 0}}{x_i}}  
-2\zn\zero{\rnd{u_{i\G0}}{x_1}}
\right\} \quad (\eta=0,\;\zn>0). 
\label{eq:phiK1-bc}
\end{align}
Note that $u_{\w i 1} n_i = -u_{\w 1 1} =0$ and $\zero{u_{1\G1}}=0$ have been used. The condition at infinity reads 
\begin{align} \label{e:KLP_bc_infinity}
\phi_{\K1}\to0 \quad (\eta\to\infty).
\end{align}

Equations~\eqref{eq:bgk-phi-K1-v4}--\eqref{e:KLP_bc_infinity} constitute a one-dimensional boundary-value problem in a half-space.
In Eq.~\eqref{eq:bgk-phi-K1-v4} and the boundary condition~\eqref{eq:phiK1-bc}, the following quantities are involved: the normal derivative of the temperature $\zero{\partial\tau_{\G 0}/\partial x_1}$, the tangential derivatives of the temperature and flow velocity, $\zero{\trnd{\tau_{\G 0}}{x_i}}$ and $\zero{\trnd{u_{i\G0}}{x_1}}$ $(i=2,\,3)$, evaluated on the boundary. These quantities are regarded as given, since they are determined by solving the zeroth-order fluid-dynamic equations [Eq.~\eqref{eq:stokes}] subject to the boundary conditions \eqref{eq:0th-order-bc}.
On the other hand, $\zero{\tau_{\G1}}$ and $\zero{u_{i\G1}}$ $(i=2,\,3)$ in Eq.~\eqref{eq:phiK1-bc} are constants that are to be determined together with the solution of the problem, as described below. These constants serve as boundary conditions for the first-order Stokes equations.


\subsubsection{Decomposing the Knudsen-layer problem into elementary problems}

Considering the forms of the inhomogeneous terms and the linearity of the problem, we seek a solution $\phi_{\K1}$ in the following form:
\begin{align}
\phi_{\K1} = &
2U\tpsi \zero{\tau_{\G1}}
-\p10\zero{\rnd{\tau_{\G0}}{x_1}} 
-\p11\bar{\z}_j\zero{\rnd{u_{j\G0}}{x_1}}  
+\p21\bar{\z}_j\zero{\rnd{\tau_{\G0}}{x_j}}.\label{eq:decompose}
\end{align}
Here, the functions $\p10=\p10(\eta,\zn,\zrho)$, $\p11=\p11(\eta,\zn,\zrho)$, and $\p21=\p21(\eta,\zn,\zrho)$, where $\zn=\z_in_i(=-\z_1)$ and $\zrho=(\z_2^2+\z_3^2)^{1/2}$, are new unknowns to be determined. 
[The negative signs of the second and third terms on the right-hand side are chosen for consistency with previous studies (e.g., Ref.~\cite{Takata2012}). These mixed signs can be unified by rewriting the derivative using the relation  $n_i(\trnd{}{x_i})=-(\trnd{}{x_1})$.]
The first term on the right-hand side represents the inhomogeneous solution corresponding to the term $-2U(\dd\tpsi/\dd \eta)\zn \zero{\tau_{\G0}}$ in Eq.~\eqref{eq:bgk-phi-K1-v4a}, and satisfies the homogeneous boundary condition as $\phi_{\K0}$ does. 

We first introduce the Knudsen-layer corrections to the macroscopic quantities. By substituting Eq.~\eqref{eq:decompose} into Eq.~\eqref{eq:KL-macro}, the Knudsen-layer part of the macroscopic quantities are obtained as
\begin{subequations}\label{eq:KLsol-1st}
\begin{align}
&
\omega_{\K1}(\eta) = -\zero{\rnd{\tau_{\G0}}{x_1}}\Omega_1^{(0)}(\eta)
+2U\tpsi(\eta)\zero{\tau_{\G1}}, \\
&u_{1\K1} = 0\quad (\because \text{mass conservation}),\\
&
u_{j\K1}(\eta) = 
 -\zero{\rnd{u_{j\G0}}{x_1}}Y_1^{(1)}(\eta)+
\zero{\rnd{\tau_{\G0}}{x_j}}Y_2^{(1)}(\eta) \quad (j=2,\,3), \\
&\tau_{\K1}(\eta) =-\zero{\rnd{\tau_{\G0}}{x_1}}\Theta_1^{(0)}(\eta), 
\end{align}
\end{subequations}
where the dependence on $(x_2,x_3)$ is omitted for brevity, and the Knudsen-layer functions $\Omega_1^{(0)}$, $Y_j^{(1)}$, and $\Theta_1^{(0)}$ (i.e., the macroscopic quantities) are defined by
\begin{subequations} \label{eq:macro_KL}
\begin{align}
&\Omega_1^{(0)} = \int_{\RR^3} \p10 E\;\dd \bm{\z}, \\
&Y_j^{(1)} = \int_{\RR^3} \z_2^2 \pj1 E\;\dd \bm{\z}= \int_{\RR^3} \z_3^2 \pj1 E\;\dd \bm{\z}=\frac{1}{2} \int_{\RR^3} \zrho^2 \pj1 E\;\dd \bm{\z} \quad (j=2,\,3),\\ 
&
\Theta_1^{(0)}=\int_{\RR^3} \frac{2}{3}\left(\z_1^2+\zrho^2-\frac{3}{2}\right)\p10 E\;\dd \bm{\z}.
\end{align}
\end{subequations}
With these functions,
the local Maxwellian $\phi_{\e \K 1}$ is expressed as
\begin{align}
\phi_{\e\K1} = &
2U\tpsi \zero{\tau_{\G1}}
-\zero{\rnd{\tau_{\G0}}{x_1}}\left[\Omega_1^{(0)}+\left(\zn^2+\zrho^2-\frac{3}{2}\right)\Theta_1^{(0)}\right] \notag\\
&-2 \bar{\z}_i\zero{\rnd{u_{i\G0}}{x_1}}Y_1^{(1)}+
2 \bar{\z}_i\zero{\rnd{\tau_{\G0}}{x_i}}Y_2^{(1)}.
\end{align}


Now, corresponding to the decomposition \eqref{eq:decompose}, we set the constants
$\zero{\tau_{\G1}}-\tau_{\w1}^+$ and $\zero{u_{i\G1}}-u_{\w i 1}^+$ $(i=2,\,3)$ appearing in \eqref{eq:phiK1-bc} in the following forms:
\begin{subequations}\label{eq:slip-coef-general}
\begin{align} 
&\zero{\tau_{\G1}}-\tau_{\w1}^+=-\c10\zero{\rnd{\tau_{\G0}}{x_1}}, \\ 
& \zero{u_{i\G1}}-u_{\w i1}^+=-\b11 \zero{\rnd{u_{i\G0}}{x_1}}
+\b21
\zero{\rnd{\tau_{\G0}}{x_i}} \quad (i=2,\,3),
\end{align}
\end{subequations}
where $\c10$, $\b11$, and $\b21$ are constants to be determined. 
Then, the boundary conditions for $\p10$, $\p11$, and $\p21$ in the decomposed problems are obtained as, at $\eta = 0$ and for $\zn>0$, 
\begin{subequations}
\begin{align}
&\p10 = -2\sqrt{\pi}\int_{\zn<0} \zn \p10 E \;\dd \bm{\z} -(\z^2-2)\c10 + \zn\left(\z^2-\frac{5}{2}\right), \\
&\p11 = -2\b11 + 2\zn, \\
&\p21 = -2\b21+\z^2-\frac{5}{2}. 
\end{align}
\end{subequations}

To summarize, we have obtained the following three problems to analyze the Knudsen layer at first order in $k$:

\vspace{0.5em}
\noindent\underline{(i) temperature jump}
\begin{subequations}\label{eq:KLproblem-decomposed10}
\begin{align}
&\zn\rnd{\p10}{\eta}-U\ddif{\tpsi}{\eta}\rnd{\p10}{\zn} = 
\exp(-2U\tpsi)\left[\Omega_1^{(0)}+\left(\z_1^2+\zrho^2-\frac{3}{2}\right)\Theta_1^{(0)}-\p10\right] +\ITJ, \\
&\p10 = -2\sqrt{\pi}\int_{\zn<0} \zn \p10 E \;\dd \bm{\z} -(\zn^2+\zrho^2-2)\c10 + \zn\left(\zn^2+\zrho^2-\frac{5}{2}\right)\quad (\eta=0,\;\zn>0), \\
&\p10 \to 0 \quad (\eta\to\infty),
\end{align}
\end{subequations}

\noindent\underline{(ii) shear slip}
\begin{subequations}\label{eq:KLproblem-decomposed11}
\begin{align}
&\zn\rnd{\p11}{\eta}-U\ddif{\tpsi}{\eta}\rnd{\p11}{\zn} 
= 
\exp(-2U\tpsi)\left(2Y_{1}^{(1)}-\p11\right) +\ISS, \\
&\p11 = -2\b11+2\zn\quad (\eta=0,\;\zn>0), \\
&\p11 \to 0\quad (\eta\to\infty),
\end{align}
\end{subequations}

\noindent\underline{(iii) thermal slip}
\begin{subequations}\label{eq:KLproblem-decomposed21}
\begin{align}
&\zn\rnd{\p21}{\eta}-U\ddif{\tpsi}{\eta}\rnd{\p21}{\zn} = 
\exp(-2U\tpsi)\left(2Y_{2}^{(1)}-\p21\right) + \ITS , \\
&\p21 = -2\b21+\zn^2+\zrho^2-\frac{5}{2}\quad (\eta=0,\;\zn>0), \\
&\p21 \to 0 \quad (\eta\to\infty),
\end{align}
\end{subequations}
where $\Omega_1^{(0)}$, $\Theta_1^{(0)}$, and $Y_1^{(1)}$ are defined in Eq.~\eqref{eq:macro_KL}.
In these problems, the constants $\c10$, $\b11$, and $\b12$ are not arbitrary, but must be determined together with the solution. This property has been numerically confirmed through our numerical analysis of the thermal-slip problem Eq.~\eqref{eq:KLproblem-decomposed21}; the others will be examined in future work. Once these constants are obtained, the relation~\eqref{eq:slip-coef-general} provides jump and slip conditions at $x_1=1/2$ for the fluid-dynamic equations.

It should be noted that, unlike in the case of standard Knudsen-layer problems without a near-wall potential (see, e.g., Ref.~\cite{Takata2012}), the coefficients $\c10$, $\b11$, $\b12$, as well as the functions $\Omega_1^{(0)}$, $\Theta_1^{(0)}$, $Y_1^{(1)}$, $Y_2^{(1)}$, depend on the specific form of the near-wall potential, namely, $\tpsi(\eta)$ and $U$.
In this sense, the slip and jump coefficients are not as universal as those in the case without a potential.
Nevertheless, since the problems~\eqref{eq:KLproblem-decomposed10}--\eqref{eq:KLproblem-decomposed21} can be solved quickly on standard hardware (within minutes) when the BGK model (or similar kinetic models) is used, systematic investigations on the effects of $\tpsi$, $U$, and $\chi(=\delta/k)$ can be carried out with relative ease.

Now, we summarize the boundary conditions at both boundaries $x_1=\pm 1/2$. The preceding discussion in Sec.~\ref{sec:KL} also applies to the boundary layer near the other plate at $x_1=-1/2$, where the unit normal is given by $\bm{n}=(1,0,0)$. Summarizing the results at the plates $x_1=\pm1/2$, 
we obtain the following first-order boundary conditions for the Stokes equation Eq.~\eqref{eq:stokes}:
\begin{align}
\begin{cases}
\zero{\tau_{\G1}}-\tau_{\w1}^\pm=\c10n_i\zero{\drnd{\tau_{\G0}}{x_i}},
\quad
\zero{u_{1\G1}} = 0, \\[1.5em]
[\zero{u_{i\G1}}-u_{\w i 1}^\pm]t_i=\b11 n_j t_i \zero{\drnd{u_{i\G0}}{x_j}}+\b21t_i\zero{\drnd{\tau_{\G0}}{x_i}}, 
\end{cases}\quad 
\left(x_1=\pm\frac{1}{2}\right),
\label{eq:slip-coef2}
\end{align}
where $n_i$ and $t_i$ denote the unit normal and tangential vectors to the plates $x_1 = \pm 1/2$, respectively ($\bm{n}$ points to the gas region).

Finally, considering the expansions in Eqs.~\eqref{eq:bc-expand} and \eqref{eq:KL-expansion} and neglecting terms of $O(k^2)$, we obtain the Knudsen-layer corrections \eqref{eq:KLcorrections}, the slip boundary condition \eqref{eq:slip-coef}, and the jump boundary condition \eqref{eq:jump-coef} as shown in the main text.


\section{Estimates of the potential for small effective ranges}\label{sec:potential} 
\begin{figure}
    \centering
    \includegraphics[width=0.9\textwidth]{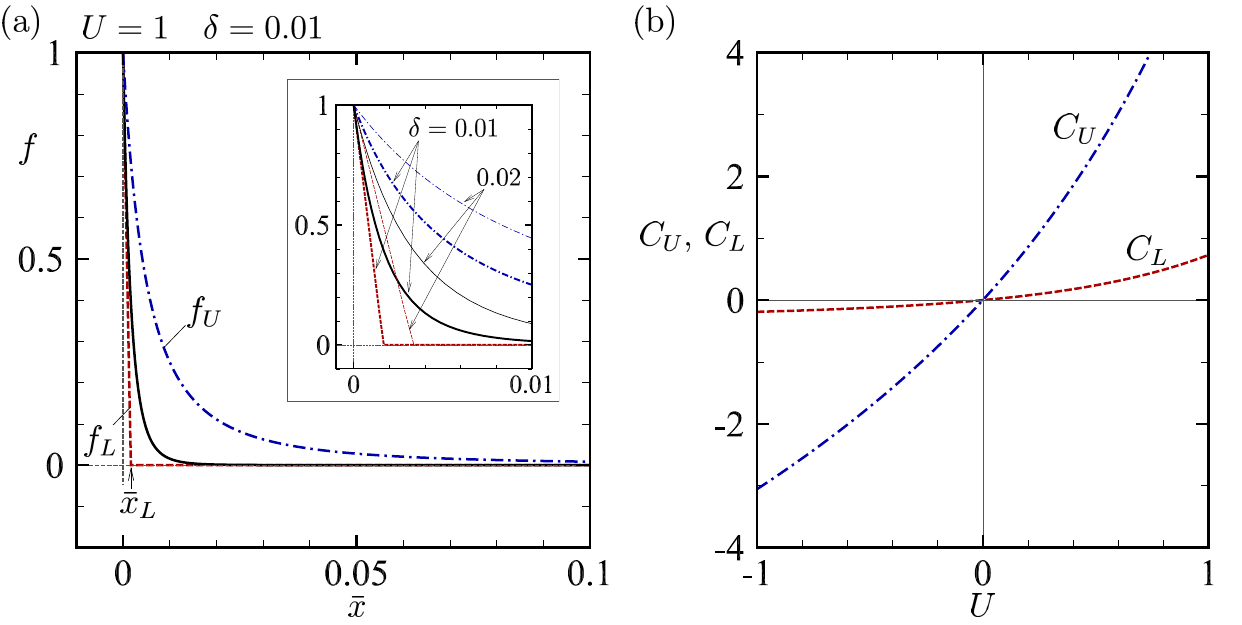}
    \caption{(a) Schematic explanation of $f_L\leq f\leq f_U$ for the case of $U=1$ and $\delta=0.01$, where $f$ (black-solid), $f_L$ (red-dash), and $f_U$ (blue-dash-dot) are shown. The inset shows the magnification near $\bar{x}=0$ and for $\delta=0.01$ (bold) and $\delta=0.02$ (thin). (b) Relation between $C_U$ (or $C_L$) and $U$.}
    \label{fig:f}
\end{figure}

We evaluate $\bpsid$ for the specific case Eq.~\eqref{eq:psi-nd} in the main text, when $\delta$ is small. 
Recalling that the potential has the form
$\psid=\psid^++\psid^-$, 
$\psid^\pm = -\left(\delta/[\pm (1/2 + \delta)-x_1]\right)^6$, 
we reduce $\bpsid$ as 
\begin{align}
\bpsid
= \int_{-1/2}^{1/2} \exp(-2U\psid(x_1)) \;\dd x_1 = 
2\int_{0}^{1/2} \exp(2 U f(\bar{x})) \;\dd \bar{x}, 
\end{align}
where the second equality follows from the change of integration variables $\bar{x}=1/2-x_1$ and $\bar{x}=1/2+x_1$ for the integration ranges $[0,1/2]$ and $[-1/2,0]$, respectively, and an auxiliary function
$f(\bar{x})=[\delta/(\bar{x} + \delta)]^6+[\delta/(\bar{x} - \delta -1)]^6$ has been introduced. 
The function $f(\bar{x})$ can be bounded as $f_L\leq f \leq f_U$ $(0\leq \bar{x}\leq 1/2)$, where 
\begin{align}
f_L=\max(-\frac{\bar{x}}{\bar{x}_L}+1,0), \quad 
f_U=
\frac{\delta^2}{(\bar{x}+\delta)^2}+\frac{\delta^6}{(\delta+1)^6},  
\end{align}    
with $\bar{x}_L = \left[6/\delta-(6\delta^6)/(\delta+1)^7\right]^{-1}$ [see Fig.~\ref{fig:f}(a)].
Then, due to the monotonicity of the exponential function, we can estimate explicitly $\bpsid$ as
\begin{align}
1+C_L \delta + O(\delta^2) \lessgtr
\bpsid \lessgtr
1+ C_U \delta + O(\delta^2) \quad (U \gtrless 0), 
\end{align}
where the constants $C_L$ and $C_U$ depending on $U$ are given by
\begin{align}
C_L= \frac{\exp(2U)-1}{6 U} -\frac{1}{3}, \quad 
C_U= 
4\int_0^{2U} \exp\left(\frac{t^2}{2U}\right) \dd t
+2[1-\exp(2U)],
\end{align}
[see Fig.~\ref{fig:f}(b)].
Therefore, for small $\delta$, we can assume that
$\bpsid = 1 + C_0 \delta + O(\delta^2)$ $(\delta \ll1)$
with $C_0$ being a constant depending only on $U$ (and not on $\delta$). 
The above estimate justifies Eq.~\eqref{eq:alpha-expansion} used in the Knudsen-layer analysis. 

\section{Some remarks on the numerical analyses}\label{sec:numerical-appendix}

As described in the maintext, the present paper uses two approaches, i.e., (i) the slip-flow theory (Sec.~\ref{sec:slip-flow-theory-TOproblem}) and (ii) the numerical analysis  (Sec.~\ref{sec:numerical}), to investigate the thermo-osmosis problem [Fig.~\ref{fig:problem}(b)]. 
The former approach requires also the preliminary numerical analysis of the Knudsen-layer problem Eq.~\eqref{eq:KLproblem-decomposed21-maintext}. Thus, we need to solve numerically the Knudsen-layer problem Eq.~\eqref{eq:KLproblem-decomposed21-maintext} and the thermo-osmosis problem Eq.~\eqref{eq:bgk-numerical-similarity}. In this \SM, some remarks on the numerical analyses of these two problems are summarized. 


\subsection{Numerical analysis of the Knudsen-layer problem}\label{sec:numerical-KL}

To solve numerically the Knudsen-layer problem for the thermal slip, i.e., Eq.~\eqref{eq:KLproblem-decomposed21-maintext}, we introduce the following marginal function:
\begin{subequations}\label{eq:gKL}
\begin{align}
&\gKL(\eta,\zn) = \hf E_n \iint_{-\infty}^{\infty}
\zrho^2 \p21(\eta,\,\zn,\zrho) E_{23} \; \dd\z_2\dd\z_3. \\ 
&E_n=\frac{\exp(-\zn^2)}{\pi^{1/2}},\;E_{23}=\frac{\exp(-\zrho^2)}{\pi}, \quad 
\zrho = \sqrt{\z_2^2+\z_3^2}.
\end{align}
\end{subequations}
By multiplying Eq.~\eqref{eq:KLproblem-decomposed21-maintext} by $(1/2) E_n E_{23} \zrho^2$ and integrating over the whole $\z_2$ and $\z_3$, we obtain the following system for $\gKL(\zn,\eta)$:
\begin{subequations}\label{eq:KLproblem-decomposed21-marginal}
\begin{align}
&
\begin{cases}
\zn\drnd{\gKL}{\eta}-U\ddif{\tpsi}{\eta}\left(\drnd{\gKL}{\zn} + 2\zn\gKL \right)
= \alphaKL\left(\uKL E_n-\gKL\right)
+ \IKL E_n, \\
\alphaKL = \exp(-2U\tpsi), \quad 
\uKL(=Y_2^{(1)}) =\displaystyle\int_{-\infty}^{\infty} \gKL \;\dd \zn, \\
\IKL = -U\tpsi -\zn U\ddif{\tpsi}{\eta} + [\exp(-2U\tpsi)-1]\left(\dfrac{1}{2}\zn^2-\dfrac{1}{4}\right), 
\end{cases}
\\
&\gKL = \left(-\b21+\hf\zn^2-\frac{1}{4}\right)E_n\quad (\eta=0,\;\zn>0), 
\\
&\gKL \to 0 \quad (\eta \to \infty).  
\end{align}
\end{subequations}
In the derivation, we have used $\iint \zrho^2 E_{23} \dd\z_2\dd\z_3=1$ and $\iint \zrho^4 E_{23} \dd\z_2\dd\z_3=2$, and the Knudsen-layer function $Y_{2}^{(1)}(\eta)$ has been renamed as $\uKL(\eta)$ for later convenience. 
We solve this boundary-value problem numerically to obtain the function $Y_2^{(1)}(=\uKL)$ and the constant $\b21$. 
A standard approach for the thermal transpiration problem \cite{Ohwada1989} can be applied here. Roughly speaking, we obtain $\b21$ by an iteration scheme based on the finite-difference method so that the solution $\gKL$ decays to zero at large $\eta$. 

\subsection{Numerical analysis of the thermo-osmosis problem}\label{sec:numerical-TO}

To solve numerically the thermo-osmosis problem Eq.~\eqref{eq:bgk-numerical-similarity}, we introduce the following marginal function:
\begin{subequations}
\begin{align}
\gTO (x_1,\z_1) = E_1 \iint_{-\infty}^\infty \z_2^2 \phi_T E_{23} \; \dd \z_2 \dd \z_3, \\
E_1=\frac{\exp(-\z_1^2)}{\pi^{1/2}},\;E_{23}=\frac{\exp(-\z_2^2-\z_3^2)}{\pi}.
\end{align}
\end{subequations}
Multiplying Eq.~\eqref{eq:bgk-numerical-similarity} by $\z_2^2 E_1 E_{23}$ and integrating with respect to $\z_2$ and $\z_3$, we obtain 
\begin{subequations}\label{eq:TOproblem-marginal}
\begin{align}
&
\begin{cases}
\z_1 \drnd{\gTO}{x_1}
-U\ddif{\psid}{x_1}
\left(\drnd{\gTO}{\z_1}+2\z_1 \gTO\right) = \dfrac{\alphaTO}{k}(\uTO E_1 -\gTO) + 
\ITO E_1, \\
\alphaTO = \alpha = \bpsid^{-1} \exp(-2U\psid) \quad 
\left(\bpsid = \displaystyle\int_{-1/2}^{1/2} \exp(-2U\psid(x_1)) \;\dd x_1\right). 
\\
\uTO(=u_T)=\displaystyle\int_{-\infty}^{\infty} \gTO \; \dd \z_1, \quad 
\ITO = -U\psid - \left(\dfrac{1}{2} \z_1^2 - \dfrac{1}{4}\right), 
\end{cases}\\
&
\gTO=0\quad (x_1=1/2,\; \z_1 < 0), \\
&
\gTO(x_1,\z_1) = \gTO(-x_1,-\z_1) \quad (x_1=0,\;\z_1 >0), \label{eq:reflection}
\end{align}
\end{subequations}
where $u_T(x_1)$ has been renamed as $\uTO(x_1)$ for the comparison with Eq.~\eqref{eq:KLproblem-decomposed21-marginal} and the symmetry with respect to $x_1=0$ [Eq.~\eqref{eq:reflection}] has been applied to reduce the computational domain from $x_1\in[-1/2,1/2]$ to $x_1\in[0,1/2]$. 

It should be noted that the reduced Knudsen-layer problem Eq.~\eqref{eq:KLproblem-decomposed21-marginal} and the reduced thermo-osmosis problem Eq.~\eqref{eq:TOproblem-marginal} have similar structures, which are highlighted by the differences of the subscripts ``$\mathrm{KL}$" and ``$\mathrm{TO}$". Therefore, similar numerical schemes can be applied as explained below. However, the reduced thermo-osmosis problem Eq.~\eqref{eq:TOproblem-marginal} includes the physical parameter $k$ in addition to the parameters related to the potential (i.e, $\psid$, $U$, and $\delta$) and therefore is more involved for a systematic parameter study. 


\subsection{Overview of the numerical scheme}\label{sec:numerical-overview}
Here, it is sufficient to describe the numerical scheme for Eq.~\eqref{eq:KLproblem-decomposed21-marginal}. The numerical scheme for Eq.~\eqref{eq:TOproblem-marginal} can be constructed in the same manner. 

We truncate the range of $\zn$ to $\zn\in[-\zmax,\zmax]$, where $\zmax=5$ is chosen so that $E_n \approx \exp(-\zmax^2)$ is negligibly small. 
The range of $\eta$ is also truncated to $\eta\in[0,\etamax]$, where $\etamax=25$ is chosen so that $\gKL$ becomes negligibly small there. 
The discrete independent variables are introduced as 
\begin{equation}
\begin{split}
&\eta = \eta^{(i)} \quad (i=0,1,...,N_{\eta}), \quad 
0=\eta^{(0)}<\cdots<\eta^{(N_\eta)} = \etamax, \\
&\zn = \pm\z^{(j)}\quad (\zn\gtrless 0;\,j=1,2,...,N_{\z}), \quad 
0<\epsilon =\z^{(1)}<\cdots<\z^{(N_\z)} = \zmax, \label{eq:lattice}
\end{split}
\end{equation}
where $\eta^{(i)}$ and $\z^{(j)}$ are the functions that determine a lattice in the $(\eta,\,\zn)$ space; $\epsilon\approx 0$ is a parameter to define the lattice points at $\zn=\pm 0$. (Practically, we set $\epsilon=10^{-16}$.) 
These two lattice points are introduced for convenience since the boundary condition at $\eta=0$ is defined for $\zn>0$, meaning that there is no guarantee that $\gKL$ is continuous at $\zn=0$ there. In fact, by considering the characteristic curves of the BGK equation, it is readily seen that $\gKL$ is discontinuous when $U\leq0$, i.e., the cases without the potential and with the repulsive potential (see, e.g., Ref.~\cite{Sone1992} for similar situations found in curved surfaces). 
The case of $U=0$ can be appropriately treated by introducing the two lattice points at $\zn=\pm 0$ as above. In the case of $U<0$, the discontinuity enters the computational domain along the characteristics, and thus the accuracy of the numerical scheme may deteriorate considerably. Therefore, in this paper, we use a relatively large number of lattice points ($N_\eta=1500$ and $N_\z=1500$; see Table~\ref{tab:numerical-parameter}). 
The lattices $\eta^{(i)}$ and $\z^{(j)}$ are constructed based on cubic polynomials so that fine lattice points are densely placed near $\eta=0$ and $\z=0$; the minimum intervals for $\eta^{(i)}$ and $\z^{(j)}$ are $1.0\times 10^{-3}$ and $1.5\times 10^{-9}$, respectively. 
The effects of the values of $N_\eta$ and $N_\z$ on the numerical results will also be investigated below. It should be remarked that for faithfully accurate numerical analysis, the discontinuities must be treated in appropriate manners using a hybrid scheme \cite{Sugimoto1992} or a method of characteristics \cite{Tsuji2013}. 

On the lattice points Eq.~\eqref{eq:lattice}, the derivatives with respect to $\eta$ and $\zn$ are approximated by using the second order upwind finite-difference methods. Here, to construct the finite-difference methods for the $\zn$-derivative, a condition $\trnd{\gKL}{\zn}=0$ $(|\zn|\to\infty)$, which is consistent with Eq.~\eqref{eq:gKL}, is imposed at $\zn=\pm\zmax$ for the cases of $U\gtrless 0$.
The numerical quadrature to obtain $\uKL$ is the Simpson's rule. The whole scheme is constructed in a semi-implicit manner, that is, $\gKL$ and $\uKL$ are treated in implicit and explicit manners, respectively. 
By iteration, we can obtain the numerical solution to Eq.~\eqref{eq:KLproblem-decomposed21-marginal}, together with the thermal-slip coefficient $\b21$ and the Knudsen-layer correction $Y_2^{(1)}(=\uKL)$. 


\subsection{Accuracy checks}\label{sec:accuracy}
We fix $\zmax=5$. Other numerical parameters are shown in Table~\ref{tab:numerical-parameter}. The physical parameters for the accuracy check we present here are $U=-1$ (repulsive potential) and $\chi=1$ and $4$. Note that the case of repulsive potential $(U=-1)$ is more difficult to compute compared to the case of attractive potential due to the discontinuities in the velocity distribution function, as mentioned above. In this study, we ignore the discontinuities for simplicity, but instead, we use rather fine lattices as shown in Table~\ref{tab:numerical-parameter}. 

The numerical values of the thermal-slip coefficients $\b21$ are listed in Table~\ref{tab:numerical-parameter}. The comparison between the reference A0 and B$x$ ($x=1,...,4$) (or C$x$) shows that the result of the reference case is accurate to three decimal places with respect to the $N_\eta$ (or $N_\z$) lattice. 
The comparison among the results of D$x$ with $\chi=4$, in which the widest effective range of the potential is used, shows that $\etamax=25$ is sufficiently large to obtain the thermal-slip coefficient accurate to three decimal places.

\begin{table}[bt]
    \centering
    \caption{Results of the accuracy check on the thermal-slip coefficient $\b21$. 
    Lattice A0 represents the reference case. B1--B4 represent the cases with different numbers of the $\eta$-lattice, $N_\eta$, C1--C4 represent the cases with different numbers of $\zn$-lattice, $N_\z$, and D1--D4 represent the cases with different computational domains $\etamax$. In D1--D4, the effect of $\etamax$ is examined by using a larger value of $\chi$, which represents the effective range of the potential.
     }
    \label{tab:numerical-parameter}
    {\tabcolsep = 1em 
    \begin{tabular}{llllll}
    \hline
         lattice name& $N_\eta$ & $\etamax$ & $N_\z$ & $\chi$ &$\b21$     \\ \hline
        A0 (reference)& $1500$   & $25$      & $1500$ & $1$    &$-0.87034$ \\ 
             B1      & $500$    & $25$      & $1500$ & $1$    &$-0.87462$ \\ 
             B2      & $1000$   & $25$      & $1500$ & $1$    &$-0.87100$ \\ 
             B3(=A0) & $1500$   & $25$      & $1500$ & $1$    &$-0.87034$ \\ 
             B4      & $2000$   & $25$      & $1500$ & $1$    &$-0.87011$ \\ 
             C1      & $1500$   & $25$      & $500$  & $1$    &$-0.87018$ \\ 
             C2      & $1500$   & $25$      & $1000$ & $1$    &$-0.87031$ \\ 
             C3(=A0) & $1500$   & $25$      & $1500$ & $1$    &$-0.87034$ \\ 
             C4      & $1500$   & $25$      & $2000$ & $1$    &$-0.87035$ \\ 
             D1      & $1300$   & $16.3$    & $1500$ & $4$    &$-7.79106$ \\ 
             D2      & $1500$   & $25$      & $1500$ & $4$    &$-7.80224$ \\ 
             D3      & $1700$   & $36.4$    & $1500$ & $4$    &$-7.80475$ \\ 
             D4      & $2000$   & $59.3$    & $1500$ & $4$    &$-7.80548$ \\ 
             \hline
    \end{tabular}
    }
\end{table}

\section{Values of thermal-slip coefficients}\label{sec:values-of-b21}

Table~\ref{tab:U=1} presents the thermal-slip coefficients $\bnum$ for various $k$ (column) and $\chi$ (row), together with $\b21$. 
The case of $\chi=0$ corresponds to the case without the potential, and thus can be considered as a reference case. Note that $\b21$ in the slip-flow theory is independent of $k$. 
The case of $\chi=0$ results in $\bnum\approx\b21=0.38316$ for small $k$; this value is identical with that presented in the textbook~\cite{Sone2007}. Note that $\b21$ is denoted by $-K_1$ in Ref.~\cite{Sone2007}. 
See also Fig.~\ref{fig:U=1} in the main text.

Next, Table~\ref{tab:U=-1} shows the thermal-slip coefficients $\bnum$ and $\b21$ for $U=-1$ (repulsive potential) with various $\chi$. 
As in the case of $U=1$, for a fixed $\chi>0$, it is seen from Table~\ref{tab:U=-1} that $\bnum$ approaches $\b21$ as $k$ decreases. 
As the range of the potential $\chi$ increases, $\bnum$ and $\b21$ increase first ($\chi=0.25$) and then decrease and eventually change sign.
See also Fig.~\ref{fig:U=-1} in the main text. 

\begin{table}[tb]
\begin{center}
\caption{Thermal-slip coefficients for $U=1$ (attractive potential). The values of $\bnum$ for $k=0.05$, ..., $0.001$ and $\b21$ with various $\chi$, i.e., the effective range of the potential.}
{\tabcolsep = 1em 
\renewcommand{\arraystretch}{1.0}
\begin{tabular}{ccccccccc}\hline
     & \multicolumn{6}{c}{$\bnum$} & $\b21$ \\ 
     \cline{2-7}
    $\chi\backslash k$ & $0.05$ & $0.02$ & $0.01$ & $0.005$ & $0.002$ & $0.001^\dagger$ & --- \\ \hline
    0\ss & $0.38213$& $0.38316 $& $0.38316 $& $0.38316 $& $0.38316 $& $0.38317 $& $0.38316 $\\
    0.25 & $0.37318 $& $0.37054 $& $0.36928 $& $0.36865 $& $0.36827 $& $0.36814 $& $0.36798 $\\
    0.5\s& $0.40254 $& $0.39717 $& $0.39498 $& $0.39388 $& $0.39324 $& $0.39301 $& $0.39275 $\\
    0.75 & $0.45865 $& $0.45207 $& $0.44946 $& $0.44818 $& $0.44742 $& $0.44714 $& $0.44685 $\\
    1\ss & $0.53745 $& $0.53163 $& $0.52933 $& $0.52822 $& $0.52757 $& $0.52733 $& $0.52706 $\\
    1.5\s& $0.75398 $& $0.75729 $& $0.75842 $& $0.75912 $& $0.75961 $& $0.75972 $& $0.75982 $\\
    2\ss & $1.0389\q$& $1.0639\q$& $1.0733\q$& $1.0784\q$& $1.0816\q$& $1.0826\q$& $1.0836\q$\\
    2.5\s& $1.3833\q$& $1.4452\q$& $1.4688\q$& $1.4815\q$& $1.4894\q$& $1.4920\q$& $1.4944\q$\\
    3\ss & $1.7799\q$& $1.8962\q$& $1.9412\q$& $1.9654\q$& $1.9806\q$& $1.9856\q$& $1.9901\q$\\
    3.5\s& $2.2230\q$& $2.4130\q$& $2.4876\q$& $2.5278\q$& $2.5532\q$& $2.5616\q$& $2.5688\q$\\
    4\ss & $2.7073\q$& $2.9924\q$& $3.1056\q$& $3.1671\q$& $3.2059\q$& $3.2188\q$& $3.2294\q$\\
    \hline
        \multicolumn{8}{l}{$^\dagger$ For $x_1$ variable, the lattice with $\times2$ finer points is used.} &
\end{tabular}
}
\label{tab:U=1}
\end{center}
\end{table}

\begin{table}[tb]
\begin{center}
\caption{Thermal-slip coefficients for $U=-1$ (repulsive potential). The values of $\bnum$ for $k=0.05$, ..., $0.001$ and $\b21$ with various $\chi$, i.e., the effective range of the potential.}
{\tabcolsep = 0.65em 
\renewcommand{\arraystretch}{1.0}
\begin{tabular}{ccccccccc}\hline
     & \multicolumn{6}{c}{$\bnum$} & $\b21$ \\ 
     \cline{2-7}
    $\chi\backslash k$ & $0.05$ & $0.02$ & $0.01^\dagger$ & $0.005^\dagger$ & $0.002^\ddagger$ & $0.001^\ddagger$ & ---$^\dagger$ \\ \hline
    0\ss & $\m0.38213 $ & $\m0.38316 $ & $\m0.38316 $ & $\m0.38316 $ & $\m0.38316 $ & $\m0.38317 $ & $\m0.38316 $\\
    0.25 & $\m0.72387 $ & $\m0.72980 $ & $\m0.73168 $ & $\m0.73249 $ & $\m0.73309 $ & $\m0.73323 $ & $\m0.73350 $\\ 
    0.5\s& $\m0.18523 $ & $\m0.19660 $ & $\m0.20032 $ & $\m0.20197 $ & $\m0.20315 $ & $\m0.20343 $ & $\m0.20388 $\\
    0.75 & $ -0.35839 $ & $ -0.34141 $ & $ -0.33585 $ & $ -0.33332 $ & $ -0.33158 $ & $ -0.33113 $ & $ -0.33051 $\\
    1\ss & $ -0.90738 $ & $ -0.88459 $ & $ -0.87713 $ & $ -0.87369 $ & $ -0.87137 $ & $ -0.87074 $ & $ -0.86994 $\\
    1.5\s& $ -2.0230\q$ & $ -1.9876\q$ & $ -1.9761\q$ & $ -1.9706\q$ & $ -1.9671\q$ & $ -1.9661\q$ & $ -1.9649\q$\\
    2\ss & $ -3.1643\q$ & $ -3.1147\q$ & $ -3.0984\q$ & $ -3.0907\q$ & $ -3.0857\q$ & $ -3.0842\q$ & $ -3.0825\q$\\
    2.5\s& $ -4.3335\q$ & $ -4.2676\q$ & $ -4.2457\q$ & $ -4.2352\q$ & $ -4.2286\q$ & $ -4.2266\q$ & $ -4.2242\q$\\
    3.0\s& $ -5.5326\q$ & $ -5.4479\q$ & $ -5.4194\q$ & $ -5.4056\q$ & $ -5.3969\q$ & $ -5.3942\q$ & $ -5.3908\q$\\
    3.5\s& $ -6.7628\q$ & $ -6.6572\q$ & $ -6.6206\q$ & $ -6.6028\q$ & $ -6.5917\q$ & $ -6.5882\q$ & $ -6.5831\q$\\
    4.0\s& $ -8.0250\q$ & $ -7.8968\q$ & $ -7.8506\q$ & $ -7.8278\q$ & $ -7.8138\q$ & $ -7.8094\q$ & $ -7.8015\q$\\
    \hline 
    \multicolumn{8}{l}{$^\dagger$ For $x_1$ variable, the lattice with $\times2$ finer points is used.} &\\
    \multicolumn{8}{l}{$^\ddagger$ For $x_1$ variable, the lattice with $\times4$ finer points is used.} &
\end{tabular}\\
}
\label{tab:U=-1}
\end{center}
\end{table}

\red{
\section{Estimation of thermal slip coefficient from molecular simulation in gases}\label{sec:yamaguchi2021}
}

\red{
Table~\ref{tab:yamaguchi2021} summarizes the parameter values to deduce thermal slip coefficients $\slip$ for gases from the molecular dynamics simulation data in Ref.~\cite{Yamaguchi2021}. The magnitude of the temperature gradient is computed from the temperature difference ($=100~$K) divided by the distance ($=0.4\times$channel length) between the hot and cold regions (see Fig.~1 of Ref.~\cite{Yamaguchi2021}).}

\red{
The thermo-osmotic speed in Table~\ref{tab:yamaguchi2021} is estimated from Fig.~6 of Ref.~\cite{Yamaguchi2021}. 
In Fig.~6 of Ref.~\cite{Yamaguchi2021}, the effect of Poiseuille-type backflow is included because the simulation was carried out in a periodic domain, in which a unidirectional thermal transpiration flow is not allowed due to mass conservation. The backflow may slightly underestimate the magnitude of thermo-osmosis in Table~\ref{tab:yamaguchi2021}.} 

\red{
The mean free path $\ell_0$ in Table~\ref{tab:yamaguchi2021} is computed based on a formula for hard-sphere molecules in the dilute regime using the physical parameters of cases A and B, although the simulation was carried out for a rather dense regime with volume fraction $\approx 0.1$. Regarding the value of $\b21$, we use the result for $U=1$ and $\chi=1$ as a typical value. The values of $\slip$ in the table are presented in the main text.
}

\begin{table}[t]
\centering
\caption{\red{Parameter values for the estimation of $\slip$ in Ref.~\cite{Yamaguchi2021}. 
Common parameters for case A and case B are summarized as follows: Argon mass $6.63\times 10^{-26}$~kg, Argon diameter $0.340$~nm, channel length $55.4$~nm, channel height $3$~nm, reference temperature $300$~K, the magnitude of temperature gradient $4.51\times10^9$~K/m, and thermal speed at the reference temperature $v_0=353$~m/s.}}
\label{tab:yamaguchi2021}
\red{
\begin{tabular}{lll}
\hline
{parameter} & {case A} & {case B} \\
\hline
&\multicolumn{2}{c}{data of Ref.~\cite{Yamaguchi2021}}\\
\cline{2-3}
channel width & $20.2$~nm & $50.4$~nm \\
number of atoms & $21950$   & $18275$ \\
Knudsen number & $0.099$ & $0.3$ \\
thermo-osmotic flow speed & $-2.21$~m/s & $-3.25$~m/s \\
thermal-slip coefficient $K_{\mathrm{TS}}$ from Ref.~\cite{Yamaguchi2021} & $-4.9\times10^{-10}$~m$^2$/(s~K) & $-7.2\times10^{-10}$~m$^2$/(s~K) \\
\hline
&\multicolumn{2}{c}{present study with $\b21=0.52706$ $(U=1,\,\chi=1)$}\\
\cline{2-3}
mean free path $\ell_0$ & $2.98\times10^{-10}$~m & $8.92\times10^{-10}$~m \\
thermal-slip coefficient $K_{\mathrm{TS}}$ of Eq.~\eqref{eq:slip-discussion} & $-1.6\times10^{-10}$~m$^2$/(s~K) & $-4.9\times10^{-10}$~m$^2$/(s~K) \\
\hline
\end{tabular}
}
\end{table}



\begin{thebibliography}{68}%
\makeatletter
\providecommand \@ifxundefined [1]{%
 \@ifx{#1\undefined}
}%
\providecommand \@ifnum [1]{%
 \ifnum #1\expandafter \@firstoftwo
 \else \expandafter \@secondoftwo
 \fi
}%
\providecommand \@ifx [1]{%
 \ifx #1\expandafter \@firstoftwo
 \else \expandafter \@secondoftwo
 \fi
}%
\providecommand \natexlab [1]{#1}%
\providecommand \enquote  [1]{``#1''}%
\providecommand \bibnamefont  [1]{#1}%
\providecommand \bibfnamefont [1]{#1}%
\providecommand \citenamefont [1]{#1}%
\providecommand \href@noop [0]{\@secondoftwo}%
\providecommand \href [0]{\begingroup \@sanitize@url \@href}%
\providecommand \@href[1]{\@@startlink{#1}\@@href}%
\providecommand \@@href[1]{\endgroup#1\@@endlink}%
\providecommand \@sanitize@url [0]{\catcode `\\12\catcode `\$12\catcode `\&12\catcode `\#12\catcode `\^12\catcode `\_12\catcode `\%12\relax}%
\providecommand \@@startlink[1]{}%
\providecommand \@@endlink[0]{}%
\providecommand \url  [0]{\begingroup\@sanitize@url \@url }%
\providecommand \@url [1]{\endgroup\@href {#1}{\urlprefix }}%
\providecommand \urlprefix  [0]{URL }%
\providecommand \Eprint [0]{\href }%
\providecommand \doibase [0]{https://doi.org/}%
\providecommand \selectlanguage [0]{\@gobble}%
\providecommand \bibinfo  [0]{\@secondoftwo}%
\providecommand \bibfield  [0]{\@secondoftwo}%
\providecommand \translation [1]{[#1]}%
\providecommand \BibitemOpen [0]{}%
\providecommand \bibitemStop [0]{}%
\providecommand \bibitemNoStop [0]{.\EOS\space}%
\providecommand \EOS [0]{\spacefactor3000\relax}%
\providecommand \BibitemShut  [1]{\csname bibitem#1\endcsname}%
\let\auto@bib@innerbib\@empty
\bibitem [{\citenamefont {Derjaguin}\ \emph {et~al.}(1987)\citenamefont {Derjaguin}, \citenamefont {Churaev},\ and\ \citenamefont {Muller}}]{Derjaguin1987}%
  \BibitemOpen
  \bibfield  {author} {\bibinfo {author} {\bibfnamefont {B.~V.}\ \bibnamefont {Derjaguin}}, \bibinfo {author} {\bibfnamefont {N.~V.}\ \bibnamefont {Churaev}},\ and\ \bibinfo {author} {\bibfnamefont {M.~M.}\ \bibnamefont {Muller}},\ }\href@noop {} {\emph {\bibinfo {title} {Surface {F}orces}}}\ (\bibinfo  {publisher} {Springer},\ \bibinfo {year} {1987})\BibitemShut {NoStop}%
\bibitem [{\citenamefont {Bregulla}\ \emph {et~al.}(2016)\citenamefont {Bregulla}, \citenamefont {W{\"u}rger}, \citenamefont {G{\"u}nther}, \citenamefont {Mertig},\ and\ \citenamefont {Cichos}}]{Bregulla2016}%
  \BibitemOpen
  \bibfield  {author} {\bibinfo {author} {\bibfnamefont {A.~P.}\ \bibnamefont {Bregulla}}, \bibinfo {author} {\bibfnamefont {A.}~\bibnamefont {W{\"u}rger}}, \bibinfo {author} {\bibfnamefont {K.}~\bibnamefont {G{\"u}nther}}, \bibinfo {author} {\bibfnamefont {M.}~\bibnamefont {Mertig}},\ and\ \bibinfo {author} {\bibfnamefont {F.}~\bibnamefont {Cichos}},\ }\bibfield  {title} {\bibinfo {title} {Thermo-osmotic flow in thin films},\ }\href {https://doi.org/10.1103/physrevlett.116.188303} {\bibfield  {journal} {\bibinfo  {journal} {Physical Review Letters}\ }\textbf {\bibinfo {volume} {116}},\ \bibinfo {pages} {188303} (\bibinfo {year} {2016})}\BibitemShut {NoStop}%
\bibitem [{\citenamefont {Dietzel}\ and\ \citenamefont {Hardt}(2017)}]{Dietzel2017}%
  \BibitemOpen
  \bibfield  {author} {\bibinfo {author} {\bibfnamefont {M.}~\bibnamefont {Dietzel}}\ and\ \bibinfo {author} {\bibfnamefont {S.}~\bibnamefont {Hardt}},\ }\bibfield  {title} {\bibinfo {title} {Flow and streaming potential of an electrolyte in a channel with an axial temperature gradient},\ }\href {https://doi.org/10.1017/jfm.2016.844} {\bibfield  {journal} {\bibinfo  {journal} {Journal of Fluid Mechanics}\ }\textbf {\bibinfo {volume} {813}},\ \bibinfo {pages} {1060} (\bibinfo {year} {2017})}\BibitemShut {NoStop}%
\bibitem [{\citenamefont {Wang}\ \emph {et~al.}(2020{\natexlab{a}})\citenamefont {Wang}, \citenamefont {Su}, \citenamefont {Zhang}, \citenamefont {Zhang},\ and\ \citenamefont {Zhang}}]{Wang2020}%
  \BibitemOpen
  \bibfield  {author} {\bibinfo {author} {\bibfnamefont {X.}~\bibnamefont {Wang}}, \bibinfo {author} {\bibfnamefont {T.}~\bibnamefont {Su}}, \bibinfo {author} {\bibfnamefont {W.}~\bibnamefont {Zhang}}, \bibinfo {author} {\bibfnamefont {Z.}~\bibnamefont {Zhang}},\ and\ \bibinfo {author} {\bibfnamefont {S.}~\bibnamefont {Zhang}},\ }\bibfield  {title} {\bibinfo {title} {Knudsen pumps: a review},\ }\href {https://doi.org/10.1038/s41378-020-0135-5} {\bibfield  {journal} {\bibinfo  {journal} {Microsystems {\&} Nanoengineering}\ }\textbf {\bibinfo {volume} {6}},\ \bibinfo {pages} {26} (\bibinfo {year} {2020}{\natexlab{a}})}\BibitemShut {NoStop}%
\bibitem [{\citenamefont {An}\ \emph {et~al.}(2014)\citenamefont {An}, \citenamefont {Gupta},\ and\ \citenamefont {Gianchandani}}]{An2014}%
  \BibitemOpen
  \bibfield  {author} {\bibinfo {author} {\bibfnamefont {S.}~\bibnamefont {An}}, \bibinfo {author} {\bibfnamefont {N.~K.}\ \bibnamefont {Gupta}},\ and\ \bibinfo {author} {\bibfnamefont {Y.~B.}\ \bibnamefont {Gianchandani}},\ }\bibfield  {title} {\bibinfo {title} {A {S}i-micromachined 162-stage two-part {K}nudsen pump for on-chip vacuum},\ }\href {https://doi.org/10.1109/JMEMS.2013.2281316} {\bibfield  {journal} {\bibinfo  {journal} {Journal of Microelectromechanical Systems}\ }\textbf {\bibinfo {volume} {23}},\ \bibinfo {pages} {406} (\bibinfo {year} {2014})}\BibitemShut {NoStop}%
\bibitem [{\citenamefont {Fr\"{a}nzl}\ and\ \citenamefont {Cichos}(2022)}]{Fraenzl2022}%
  \BibitemOpen
  \bibfield  {author} {\bibinfo {author} {\bibfnamefont {M.}~\bibnamefont {Fr\"{a}nzl}}\ and\ \bibinfo {author} {\bibfnamefont {F.}~\bibnamefont {Cichos}},\ }\bibfield  {title} {\bibinfo {title} {Hydrodynamic manipulation of nano-objects by optically induced thermo-osmotic flows},\ }\href {https://doi.org/10.1038/s41467-022-28212-z} {\bibfield  {journal} {\bibinfo  {journal} {Nature Communications}\ }\textbf {\bibinfo {volume} {13}},\ \bibinfo {pages} {656} (\bibinfo {year} {2022})}\BibitemShut {NoStop}%
\bibitem [{\citenamefont {Ganti}\ \emph {et~al.}(2017)\citenamefont {Ganti}, \citenamefont {Liu},\ and\ \citenamefont {Frenkel}}]{Ganti2017}%
  \BibitemOpen
  \bibfield  {author} {\bibinfo {author} {\bibfnamefont {R.}~\bibnamefont {Ganti}}, \bibinfo {author} {\bibfnamefont {Y.}~\bibnamefont {Liu}},\ and\ \bibinfo {author} {\bibfnamefont {D.}~\bibnamefont {Frenkel}},\ }\bibfield  {title} {\bibinfo {title} {Molecular simulation of thermo-osmotic slip},\ }\href {https://doi.org/10.1103/physrevlett.119.038002} {\bibfield  {journal} {\bibinfo  {journal} {Physical Review Letters}\ }\textbf {\bibinfo {volume} {119}},\ \bibinfo {pages} {038002} (\bibinfo {year} {2017})}\BibitemShut {NoStop}%
\bibitem [{\citenamefont {Anzini}\ \emph {et~al.}(2019)\citenamefont {Anzini}, \citenamefont {Colombo}, \citenamefont {Filiberti},\ and\ \citenamefont {Parola}}]{Anzini2019}%
  \BibitemOpen
  \bibfield  {author} {\bibinfo {author} {\bibfnamefont {P.}~\bibnamefont {Anzini}}, \bibinfo {author} {\bibfnamefont {G.~M.}\ \bibnamefont {Colombo}}, \bibinfo {author} {\bibfnamefont {Z.}~\bibnamefont {Filiberti}},\ and\ \bibinfo {author} {\bibfnamefont {A.}~\bibnamefont {Parola}},\ }\bibfield  {title} {\bibinfo {title} {Thermal forces from a microscopic perspective},\ }\href {https://doi.org/10.1103/physrevlett.123.028002} {\bibfield  {journal} {\bibinfo  {journal} {Physical Review Letters}\ }\textbf {\bibinfo {volume} {123}},\ \bibinfo {pages} {028002} (\bibinfo {year} {2019})}\BibitemShut {NoStop}%
\bibitem [{\citenamefont {Tsuji}\ \emph {et~al.}(2023)\citenamefont {Tsuji}, \citenamefont {Mei},\ and\ \citenamefont {Taguchi}}]{Tsuji2023}%
  \BibitemOpen
  \bibfield  {author} {\bibinfo {author} {\bibfnamefont {T.}~\bibnamefont {Tsuji}}, \bibinfo {author} {\bibfnamefont {S.}~\bibnamefont {Mei}},\ and\ \bibinfo {author} {\bibfnamefont {S.}~\bibnamefont {Taguchi}},\ }\bibfield  {title} {\bibinfo {title} {Thermo-osmotic slip flows around a thermophoretic microparticle characterized by optical trapping of tracers},\ }\href {https://doi.org/10.1103/physrevapplied.20.054061} {\bibfield  {journal} {\bibinfo  {journal} {Physical Review Applied}\ }\textbf {\bibinfo {volume} {20}},\ \bibinfo {pages} {054061} (\bibinfo {year} {2023})}\BibitemShut {NoStop}%
\bibitem [{\citenamefont {Sone}(1991)}]{Sone1991}%
  \BibitemOpen
  \bibfield  {author} {\bibinfo {author} {\bibfnamefont {Y.}~\bibnamefont {Sone}},\ }\bibfield  {title} {\bibinfo {title} {A simple demonstration of a rarefied gas flow induced over a plane wall with a temperature gradient},\ }\href {https://doi.org/10.1063/1.857981} {\bibfield  {journal} {\bibinfo  {journal} {Physics of Fluids}\ }\textbf {\bibinfo {volume} {3}},\ \bibinfo {pages} {997} (\bibinfo {year} {1991})}\BibitemShut {NoStop}%
\bibitem [{\citenamefont {Cercignani}(2000)}]{Cercignani2000}%
  \BibitemOpen
  \bibfield  {author} {\bibinfo {author} {\bibfnamefont {C.}~\bibnamefont {Cercignani}},\ }\href@noop {} {\emph {\bibinfo {title} {Rarefied {G}as {D}ynamics: from {B}asic {C}oncepts to {A}ctual {C}alculations}}}\ (\bibinfo  {publisher} {Cambridge University Press},\ \bibinfo {year} {2000})\BibitemShut {NoStop}%
\bibitem [{\citenamefont {Sone}(2007)}]{Sone2007}%
  \BibitemOpen
  \bibfield  {author} {\bibinfo {author} {\bibfnamefont {Y.}~\bibnamefont {Sone}},\ }\href@noop {} {\emph {\bibinfo {title} {Molecular Gas Dynamics: Theory, Techniques, and Applications}}}\ (\bibinfo  {publisher} {Birkh\"{a}user Boston},\ \bibinfo {year} {2007})\BibitemShut {NoStop}%
\bibitem [{\citenamefont {Takata}\ and\ \citenamefont {Hattori}(2012)}]{Takata2012}%
  \BibitemOpen
  \bibfield  {author} {\bibinfo {author} {\bibfnamefont {S.}~\bibnamefont {Takata}}\ and\ \bibinfo {author} {\bibfnamefont {M.}~\bibnamefont {Hattori}},\ }\bibfield  {title} {\bibinfo {title} {Asymptotic theory for the time-dependent behavior of a slightly rarefied gas over a smooth solid boundary},\ }\href {https://doi.org/10.1007/s10955-012-0512-z} {\bibfield  {journal} {\bibinfo  {journal} {Journal of Statistical Physics}\ }\textbf {\bibinfo {volume} {147}},\ \bibinfo {pages} {1182} (\bibinfo {year} {2012})}\BibitemShut {NoStop}%
\bibitem [{\citenamefont {Ohwada}\ \emph {et~al.}(1989{\natexlab{a}})\citenamefont {Ohwada}, \citenamefont {Sone},\ and\ \citenamefont {Aoki}}]{Ohwada1989a}%
  \BibitemOpen
  \bibfield  {author} {\bibinfo {author} {\bibfnamefont {T.}~\bibnamefont {Ohwada}}, \bibinfo {author} {\bibfnamefont {Y.}~\bibnamefont {Sone}},\ and\ \bibinfo {author} {\bibfnamefont {K.}~\bibnamefont {Aoki}},\ }\bibfield  {title} {\bibinfo {title} {Numerical analysis of the {P}oiseuille and thermal transpiration flows between two parallel plates on the basis of the {B}oltzmann equation for hard-sphere molecules},\ }\href {https://doi.org/10.1063/1.857478} {\bibfield  {journal} {\bibinfo  {journal} {Physics of Fluids}\ }\textbf {\bibinfo {volume} {1}},\ \bibinfo {pages} {2042} (\bibinfo {year} {1989}{\natexlab{a}})}\BibitemShut {NoStop}%
\bibitem [{\citenamefont {Rojas-C\'ardenas}\ \emph {et~al.}(2011)\citenamefont {Rojas-C\'ardenas}, \citenamefont {Graur}, \citenamefont {Perrier},\ and\ \citenamefont {Meolans}}]{RojasCardenas2011}%
  \BibitemOpen
  \bibfield  {author} {\bibinfo {author} {\bibfnamefont {M.}~\bibnamefont {Rojas-C\'ardenas}}, \bibinfo {author} {\bibfnamefont {I.}~\bibnamefont {Graur}}, \bibinfo {author} {\bibfnamefont {P.}~\bibnamefont {Perrier}},\ and\ \bibinfo {author} {\bibfnamefont {J.~G.}\ \bibnamefont {Meolans}},\ }\bibfield  {title} {\bibinfo {title} {Thermal transpiration flow: A circular cross-section microtube submitted to a temperature gradient},\ }\href {https://doi.org/10.1063/1.3561744} {\bibfield  {journal} {\bibinfo  {journal} {Physics of Fluids}\ }\textbf {\bibinfo {volume} {23}},\ \bibinfo {pages} {031702} (\bibinfo {year} {2011})}\BibitemShut {NoStop}%
\bibitem [{\citenamefont {Yamaguchi}\ \emph {et~al.}(2016)\citenamefont {Yamaguchi}, \citenamefont {Perrier}, \citenamefont {Ho}, \citenamefont {M{\'e}olans}, \citenamefont {Niimi},\ and\ \citenamefont {Graur}}]{Yamaguchi2016}%
  \BibitemOpen
  \bibfield  {author} {\bibinfo {author} {\bibfnamefont {H.}~\bibnamefont {Yamaguchi}}, \bibinfo {author} {\bibfnamefont {P.}~\bibnamefont {Perrier}}, \bibinfo {author} {\bibfnamefont {M.~T.}\ \bibnamefont {Ho}}, \bibinfo {author} {\bibfnamefont {J.~G.}\ \bibnamefont {M{\'e}olans}}, \bibinfo {author} {\bibfnamefont {T.}~\bibnamefont {Niimi}},\ and\ \bibinfo {author} {\bibfnamefont {I.}~\bibnamefont {Graur}},\ }\bibfield  {title} {\bibinfo {title} {Mass flow rate measurement of thermal creep flow from transitional to slip flow regime},\ }\href {https://doi.org/10.1017/jfm.2016.234} {\bibfield  {journal} {\bibinfo  {journal} {Journal of Fluid Mechanics}\ }\textbf {\bibinfo {volume} {795}},\ \bibinfo {pages} {690} (\bibinfo {year} {2016})}\BibitemShut {NoStop}%
\bibitem [{\citenamefont {Anzini}\ \emph {et~al.}(2022)\citenamefont {Anzini}, \citenamefont {Filiberti},\ and\ \citenamefont {Parola}}]{Anzini2022}%
  \BibitemOpen
  \bibfield  {author} {\bibinfo {author} {\bibfnamefont {P.}~\bibnamefont {Anzini}}, \bibinfo {author} {\bibfnamefont {Z.}~\bibnamefont {Filiberti}},\ and\ \bibinfo {author} {\bibfnamefont {A.}~\bibnamefont {Parola}},\ }\bibfield  {title} {\bibinfo {title} {Fluid flow at interfaces driven by thermal gradients},\ }\href {https://doi.org/10.1103/physreve.106.024116} {\bibfield  {journal} {\bibinfo  {journal} {Physical Review E}\ }\textbf {\bibinfo {volume} {106}},\ \bibinfo {pages} {024116} (\bibinfo {year} {2022})}\BibitemShut {NoStop}%
\bibitem [{\citenamefont {Anzini}\ \emph {et~al.}(2025)\citenamefont {Anzini}, \citenamefont {Filiberti},\ and\ \citenamefont {Parola}}]{Anzini2025}%
  \BibitemOpen
  \bibfield  {author} {\bibinfo {author} {\bibfnamefont {P.}~\bibnamefont {Anzini}}, \bibinfo {author} {\bibfnamefont {Z.}~\bibnamefont {Filiberti}},\ and\ \bibinfo {author} {\bibfnamefont {A.}~\bibnamefont {Parola}},\ }\bibfield  {title} {\bibinfo {title} {Temperature-driven flows in nanochannels: Theory and simulations},\ }\href {https://doi.org/10.1063/5.0251775} {\bibfield  {journal} {\bibinfo  {journal} {The Journal of Chemical Physics}\ }\textbf {\bibinfo {volume} {162}},\ \bibinfo {pages} {094501} (\bibinfo {year} {2025})}\BibitemShut {NoStop}%
\bibitem [{\citenamefont {Kinefuchi}\ \emph {et~al.}(2017)\citenamefont {Kinefuchi}, \citenamefont {Kotsubo}, \citenamefont {Osuka}, \citenamefont {Yoshimoto}, \citenamefont {Miyoshi}, \citenamefont {Takagi},\ and\ \citenamefont {Matsumoto}}]{Kinefuchi2017}%
  \BibitemOpen
  \bibfield  {author} {\bibinfo {author} {\bibfnamefont {I.}~\bibnamefont {Kinefuchi}}, \bibinfo {author} {\bibfnamefont {Y.}~\bibnamefont {Kotsubo}}, \bibinfo {author} {\bibfnamefont {K.}~\bibnamefont {Osuka}}, \bibinfo {author} {\bibfnamefont {Y.}~\bibnamefont {Yoshimoto}}, \bibinfo {author} {\bibfnamefont {N.}~\bibnamefont {Miyoshi}}, \bibinfo {author} {\bibfnamefont {S.}~\bibnamefont {Takagi}},\ and\ \bibinfo {author} {\bibfnamefont {Y.}~\bibnamefont {Matsumoto}},\ }\bibfield  {title} {\bibinfo {title} {Incident energy dependence of the scattering dynamics of water molecules on silicon and graphite surfaces: the effect on tangential momentum accommodation},\ }\href {https://doi.org/10.1007/s10404-017-1850-6} {\bibfield  {journal} {\bibinfo  {journal} {Microfluidics and Nanofluidics}\ }\textbf {\bibinfo {volume} {21}},\ \bibinfo {pages} {15} (\bibinfo {year} {2017})}\BibitemShut {NoStop}%
\bibitem [{\citenamefont {Fu}\ \emph {et~al.}(2017)\citenamefont {Fu}, \citenamefont {Merabia},\ and\ \citenamefont {Joly}}]{Fu2017}%
  \BibitemOpen
  \bibfield  {author} {\bibinfo {author} {\bibfnamefont {L.}~\bibnamefont {Fu}}, \bibinfo {author} {\bibfnamefont {S.}~\bibnamefont {Merabia}},\ and\ \bibinfo {author} {\bibfnamefont {L.}~\bibnamefont {Joly}},\ }\bibfield  {title} {\bibinfo {title} {What controls thermo-osmosis? {M}olecular simulations show the critical role of interfacial hydrodynamics},\ }\href {https://doi.org/10.1103/physrevlett.119.214501} {\bibfield  {journal} {\bibinfo  {journal} {Physical Review Letters}\ }\textbf {\bibinfo {volume} {119}},\ \bibinfo {pages} {214501} (\bibinfo {year} {2017})}\BibitemShut {NoStop}%
\bibitem [{\citenamefont {Wang}\ \emph {et~al.}(2020{\natexlab{b}})\citenamefont {Wang}, \citenamefont {Liu}, \citenamefont {Jing}, \citenamefont {Mohamad},\ and\ \citenamefont {Prezhdo}}]{Wang2020a}%
  \BibitemOpen
  \bibfield  {author} {\bibinfo {author} {\bibfnamefont {X.}~\bibnamefont {Wang}}, \bibinfo {author} {\bibfnamefont {M.}~\bibnamefont {Liu}}, \bibinfo {author} {\bibfnamefont {D.}~\bibnamefont {Jing}}, \bibinfo {author} {\bibfnamefont {A.}~\bibnamefont {Mohamad}},\ and\ \bibinfo {author} {\bibfnamefont {O.}~\bibnamefont {Prezhdo}},\ }\bibfield  {title} {\bibinfo {title} {Net unidirectional fluid transport in locally heated nanochannel by thermo-osmosis},\ }\href {https://doi.org/10.1021/acs.nanolett.0c04331} {\bibfield  {journal} {\bibinfo  {journal} {Nano Letters}\ }\textbf {\bibinfo {volume} {20}},\ \bibinfo {pages} {8965} (\bibinfo {year} {2020}{\natexlab{b}})}\BibitemShut {NoStop}%
\bibitem [{\citenamefont {Wang}\ \emph {et~al.}(2021)\citenamefont {Wang}, \citenamefont {Liu}, \citenamefont {Jing},\ and\ \citenamefont {Prezhdo}}]{Wang2021}%
  \BibitemOpen
  \bibfield  {author} {\bibinfo {author} {\bibfnamefont {X.}~\bibnamefont {Wang}}, \bibinfo {author} {\bibfnamefont {M.}~\bibnamefont {Liu}}, \bibinfo {author} {\bibfnamefont {D.}~\bibnamefont {Jing}},\ and\ \bibinfo {author} {\bibfnamefont {O.}~\bibnamefont {Prezhdo}},\ }\bibfield  {title} {\bibinfo {title} {Generating shear flows without moving parts by thermo-osmosis in heterogeneous nanochannels},\ }\href {https://doi.org/10.1021/acs.jpclett.1c02795} {\bibfield  {journal} {\bibinfo  {journal} {The Journal of Physical Chemistry Letters}\ }\textbf {\bibinfo {volume} {12}},\ \bibinfo {pages} {10099} (\bibinfo {year} {2021})}\BibitemShut {NoStop}%
\bibitem [{\citenamefont {Fan}\ \emph {et~al.}(2024)\citenamefont {Fan}, \citenamefont {Li}, \citenamefont {Tan}, \citenamefont {Zhang}, \citenamefont {Liu},\ and\ \citenamefont {Liu}}]{Fan2024}%
  \BibitemOpen
  \bibfield  {author} {\bibinfo {author} {\bibfnamefont {W.}~\bibnamefont {Fan}}, \bibinfo {author} {\bibfnamefont {J.}~\bibnamefont {Li}}, \bibinfo {author} {\bibfnamefont {Y.}~\bibnamefont {Tan}}, \bibinfo {author} {\bibfnamefont {Y.}~\bibnamefont {Zhang}}, \bibinfo {author} {\bibfnamefont {W.}~\bibnamefont {Liu}},\ and\ \bibinfo {author} {\bibfnamefont {Z.}~\bibnamefont {Liu}},\ }\bibfield  {title} {\bibinfo {title} {Refinement of the thermo-osmotic velocity calculation methodology and investigation of control strategies},\ }\href {https://doi.org/10.1016/j.ijheatmasstransfer.2024.126153} {\bibfield  {journal} {\bibinfo  {journal} {International Journal of Heat and Mass Transfer}\ }\textbf {\bibinfo {volume} {235}},\ \bibinfo {pages} {126153} (\bibinfo {year} {2024})}\BibitemShut {NoStop}%
\bibitem [{\citenamefont {Qi}\ \emph {et~al.}(2024)\citenamefont {Qi}, \citenamefont {Li}, \citenamefont {Wang},\ and\ \citenamefont {Xia}}]{Qi2024}%
  \BibitemOpen
  \bibfield  {author} {\bibinfo {author} {\bibfnamefont {K.}~\bibnamefont {Qi}}, \bibinfo {author} {\bibfnamefont {Z.}~\bibnamefont {Li}}, \bibinfo {author} {\bibfnamefont {J.}~\bibnamefont {Wang}},\ and\ \bibinfo {author} {\bibfnamefont {G.}~\bibnamefont {Xia}},\ }\bibfield  {title} {\bibinfo {title} {Direction reverse of the thermo-osmosis for a liquid in a nanochannel},\ }\href {https://doi.org/10.1063/5.0239503} {\bibfield  {journal} {\bibinfo  {journal} {Physics of Fluids}\ }\textbf {\bibinfo {volume} {36}},\ \bibinfo {pages} {112028} (\bibinfo {year} {2024})}\BibitemShut {NoStop}%
\bibitem [{\citenamefont {Chen}\ \emph {et~al.}(2021)\citenamefont {Chen}, \citenamefont {Sedighi},\ and\ \citenamefont {Jivkov}}]{Chen2021a}%
  \BibitemOpen
  \bibfield  {author} {\bibinfo {author} {\bibfnamefont {W.~Q.}\ \bibnamefont {Chen}}, \bibinfo {author} {\bibfnamefont {M.}~\bibnamefont {Sedighi}},\ and\ \bibinfo {author} {\bibfnamefont {A.~P.}\ \bibnamefont {Jivkov}},\ }\bibfield  {title} {\bibinfo {title} {Thermo-osmosis in hydrophilic nanochannels: mechanism and size effect},\ }\href {https://doi.org/10.1039/d0nr06687g} {\bibfield  {journal} {\bibinfo  {journal} {Nanoscale}\ }\textbf {\bibinfo {volume} {13}},\ \bibinfo {pages} {1696} (\bibinfo {year} {2021})}\BibitemShut {NoStop}%
\bibitem [{\citenamefont {Herrero}\ \emph {et~al.}(2022)\citenamefont {Herrero}, \citenamefont {F{\'{e}}liciano}, \citenamefont {Merabia},\ and\ \citenamefont {Joly}}]{Herrero2022}%
  \BibitemOpen
  \bibfield  {author} {\bibinfo {author} {\bibfnamefont {C.}~\bibnamefont {Herrero}}, \bibinfo {author} {\bibfnamefont {M.~D.~S.}\ \bibnamefont {F{\'{e}}liciano}}, \bibinfo {author} {\bibfnamefont {S.}~\bibnamefont {Merabia}},\ and\ \bibinfo {author} {\bibfnamefont {L.}~\bibnamefont {Joly}},\ }\bibfield  {title} {\bibinfo {title} {Fast and versatile thermo-osmotic flows with a pinch of salt},\ }\href {https://doi.org/10.1039/d1nr06998e} {\bibfield  {journal} {\bibinfo  {journal} {Nanoscale}\ }\textbf {\bibinfo {volume} {14}},\ \bibinfo {pages} {626} (\bibinfo {year} {2022})}\BibitemShut {NoStop}%
\bibitem [{\citenamefont {Chen}\ \emph {et~al.}(2023)\citenamefont {Chen}, \citenamefont {Jivkov},\ and\ \citenamefont {Sedighi}}]{Chen2023b}%
  \BibitemOpen
  \bibfield  {author} {\bibinfo {author} {\bibfnamefont {W.~Q.}\ \bibnamefont {Chen}}, \bibinfo {author} {\bibfnamefont {A.~P.}\ \bibnamefont {Jivkov}},\ and\ \bibinfo {author} {\bibfnamefont {M.}~\bibnamefont {Sedighi}},\ }\bibfield  {title} {\bibinfo {title} {Thermo-osmosis in charged nanochannels: Effects of surface charge and ionic strength},\ }\href {https://doi.org/10.1021/acsami.3c02559} {\bibfield  {journal} {\bibinfo  {journal} {ACS Applied Materials \& Interfaces}\ }\textbf {\bibinfo {volume} {15}},\ \bibinfo {pages} {34159} (\bibinfo {year} {2023})}\BibitemShut {NoStop}%
\bibitem [{\citenamefont {Ouadfel}\ \emph {et~al.}(2024)\citenamefont {Ouadfel}, \citenamefont {Merabia}, \citenamefont {Yamaguchi},\ and\ \citenamefont {Joly}}]{Ouadfel2024}%
  \BibitemOpen
  \bibfield  {author} {\bibinfo {author} {\bibfnamefont {M.}~\bibnamefont {Ouadfel}}, \bibinfo {author} {\bibfnamefont {S.}~\bibnamefont {Merabia}}, \bibinfo {author} {\bibfnamefont {Y.}~\bibnamefont {Yamaguchi}},\ and\ \bibinfo {author} {\bibfnamefont {L.}~\bibnamefont {Joly}},\ }\bibfield  {title} {\bibinfo {title} {Equilibrium and non-equilibrium molecular dynamics simulation of thermo-osmosis: Enhanced effects on polarized graphene surfaces},\ }\href {https://doi.org/10.1080/00268976.2024.2392016} {\bibfield  {journal} {\bibinfo  {journal} {Molecular Physics}\ }\textbf {\bibinfo {volume} {---}},\ \bibinfo {pages} {e2392016} (\bibinfo {year} {2024})}\BibitemShut {NoStop}%
\bibitem [{\citenamefont {Yabunaka}\ and\ \citenamefont {Fujitani}(2024)}]{Yabunaka2024}%
  \BibitemOpen
  \bibfield  {author} {\bibinfo {author} {\bibfnamefont {S.}~\bibnamefont {Yabunaka}}\ and\ \bibinfo {author} {\bibfnamefont {Y.}~\bibnamefont {Fujitani}},\ }\bibfield  {title} {\bibinfo {title} {Thermo-osmosis of a near-critical binary fluid mixture: A general formulation and universal flow direction},\ }\href {https://doi.org/10.1103/physreve.109.064610} {\bibfield  {journal} {\bibinfo  {journal} {Physical Review E}\ }\textbf {\bibinfo {volume} {109}},\ \bibinfo {pages} {064610} (\bibinfo {year} {2024})}\BibitemShut {NoStop}%
\bibitem [{\citenamefont {Thompson}\ and\ \citenamefont {Troian}(1997)}]{Thompson1997}%
  \BibitemOpen
  \bibfield  {author} {\bibinfo {author} {\bibfnamefont {P.~A.}\ \bibnamefont {Thompson}}\ and\ \bibinfo {author} {\bibfnamefont {S.~M.}\ \bibnamefont {Troian}},\ }\bibfield  {title} {\bibinfo {title} {A general boundary condition for liquid flow at solid surfaces},\ }\href {https://doi.org/10.1038/38686} {\bibfield  {journal} {\bibinfo  {journal} {Nature}\ }\textbf {\bibinfo {volume} {389}},\ \bibinfo {pages} {360} (\bibinfo {year} {1997})}\BibitemShut {NoStop}%
\bibitem [{\citenamefont {Craig}\ \emph {et~al.}(2001)\citenamefont {Craig}, \citenamefont {Neto},\ and\ \citenamefont {Williams}}]{Craig2001}%
  \BibitemOpen
  \bibfield  {author} {\bibinfo {author} {\bibfnamefont {V.~S.~J.}\ \bibnamefont {Craig}}, \bibinfo {author} {\bibfnamefont {C.}~\bibnamefont {Neto}},\ and\ \bibinfo {author} {\bibfnamefont {D.~R.~M.}\ \bibnamefont {Williams}},\ }\bibfield  {title} {\bibinfo {title} {Shear-dependent boundary slip in an aqueous {N}ewtonian liquid},\ }\href {https://doi.org/10.1103/physrevlett.87.054504} {\bibfield  {journal} {\bibinfo  {journal} {Physical Review Letters}\ }\textbf {\bibinfo {volume} {87}},\ \bibinfo {pages} {054504} (\bibinfo {year} {2001})}\BibitemShut {NoStop}%
\bibitem [{\citenamefont {Barrat}\ and\ \citenamefont {Bocquet}(1999{\natexlab{a}})}]{Barrat1999a}%
  \BibitemOpen
  \bibfield  {author} {\bibinfo {author} {\bibfnamefont {J.-L.}\ \bibnamefont {Barrat}}\ and\ \bibinfo {author} {\bibfnamefont {L.}~\bibnamefont {Bocquet}},\ }\bibfield  {title} {\bibinfo {title} {Large slip effect at a nonwetting fluid-solid interface},\ }\href {https://doi.org/10.1103/physrevlett.82.4671} {\bibfield  {journal} {\bibinfo  {journal} {Physical Review Letters}\ }\textbf {\bibinfo {volume} {82}},\ \bibinfo {pages} {4671} (\bibinfo {year} {1999}{\natexlab{a}})}\BibitemShut {NoStop}%
\bibitem [{\citenamefont {Zhu}\ and\ \citenamefont {Granick}(2001)}]{Zhu2001}%
  \BibitemOpen
  \bibfield  {author} {\bibinfo {author} {\bibfnamefont {Y.}~\bibnamefont {Zhu}}\ and\ \bibinfo {author} {\bibfnamefont {S.}~\bibnamefont {Granick}},\ }\bibfield  {title} {\bibinfo {title} {Rate-dependent slip of newtonian liquid at smooth surfaces},\ }\href {https://doi.org/10.1103/physrevlett.87.096105} {\bibfield  {journal} {\bibinfo  {journal} {Physical Review Letters}\ }\textbf {\bibinfo {volume} {87}},\ \bibinfo {pages} {096105} (\bibinfo {year} {2001})}\BibitemShut {NoStop}%
\bibitem [{\citenamefont {Baudry}\ \emph {et~al.}(2001)\citenamefont {Baudry}, \citenamefont {Charlaix}, \citenamefont {Tonck},\ and\ \citenamefont {Mazuyer}}]{Baudry2001}%
  \BibitemOpen
  \bibfield  {author} {\bibinfo {author} {\bibfnamefont {J.}~\bibnamefont {Baudry}}, \bibinfo {author} {\bibfnamefont {E.}~\bibnamefont {Charlaix}}, \bibinfo {author} {\bibfnamefont {A.}~\bibnamefont {Tonck}},\ and\ \bibinfo {author} {\bibfnamefont {D.}~\bibnamefont {Mazuyer}},\ }\bibfield  {title} {\bibinfo {title} {Experimental evidence for a large slip effect at a nonwetting fluid-solid interface},\ }\href {https://doi.org/10.1021/la0009994} {\bibfield  {journal} {\bibinfo  {journal} {Langmuir}\ }\textbf {\bibinfo {volume} {17}},\ \bibinfo {pages} {5232} (\bibinfo {year} {2001})}\BibitemShut {NoStop}%
\bibitem [{\citenamefont {L\'eger}(2002)}]{Leger2002}%
  \BibitemOpen
  \bibfield  {author} {\bibinfo {author} {\bibfnamefont {L.}~\bibnamefont {L\'eger}},\ }\bibfield  {title} {\bibinfo {title} {Friction mechanisms and interfacial slip at fluid solid interfaces},\ }\href {https://doi.org/10.1088/0953-8984/15/1/303} {\bibfield  {journal} {\bibinfo  {journal} {Journal of Physics: Condensed Matter}\ }\textbf {\bibinfo {volume} {15}},\ \bibinfo {pages} {S19} (\bibinfo {year} {2002})}\BibitemShut {NoStop}%
\bibitem [{\citenamefont {Cottin-Bizonne}\ \emph {et~al.}(2002)\citenamefont {Cottin-Bizonne}, \citenamefont {Jurine}, \citenamefont {Baudry}, \citenamefont {Crassous}, \citenamefont {Restagno},\ and\ \citenamefont {Charlaix}}]{CottinBizonne2002}%
  \BibitemOpen
  \bibfield  {author} {\bibinfo {author} {\bibfnamefont {C.}~\bibnamefont {Cottin-Bizonne}}, \bibinfo {author} {\bibfnamefont {S.}~\bibnamefont {Jurine}}, \bibinfo {author} {\bibfnamefont {J.}~\bibnamefont {Baudry}}, \bibinfo {author} {\bibfnamefont {J.}~\bibnamefont {Crassous}}, \bibinfo {author} {\bibfnamefont {F.}~\bibnamefont {Restagno}},\ and\ \bibinfo {author} {\bibfnamefont {E.}~\bibnamefont {Charlaix}},\ }\bibfield  {title} {\bibinfo {title} {Nanorheology: An investigation of the boundary condition at hydrophobic and hydrophilic interfaces},\ }\href {https://doi.org/10.1140/epje/i2001-10112-9} {\bibfield  {journal} {\bibinfo  {journal} {The European Physical Journal E}\ }\textbf {\bibinfo {volume} {9}},\ \bibinfo {pages} {47} (\bibinfo {year} {2002})}\BibitemShut {NoStop}%
\bibitem [{\citenamefont {Xie}\ \emph {et~al.}(2020)\citenamefont {Xie}, \citenamefont {Fu}, \citenamefont {Niehaus},\ and\ \citenamefont {Joly}}]{Xie2020a}%
  \BibitemOpen
  \bibfield  {author} {\bibinfo {author} {\bibfnamefont {Y.}~\bibnamefont {Xie}}, \bibinfo {author} {\bibfnamefont {L.}~\bibnamefont {Fu}}, \bibinfo {author} {\bibfnamefont {T.}~\bibnamefont {Niehaus}},\ and\ \bibinfo {author} {\bibfnamefont {L.}~\bibnamefont {Joly}},\ }\bibfield  {title} {\bibinfo {title} {Liquid-solid slip on charged walls: The dramatic impact of charge distribution},\ }\href {https://doi.org/10.1103/physrevlett.125.014501} {\bibfield  {journal} {\bibinfo  {journal} {Physical Review Letters}\ }\textbf {\bibinfo {volume} {125}},\ \bibinfo {pages} {014501} (\bibinfo {year} {2020})}\BibitemShut {NoStop}%
\bibitem [{\citenamefont {Chen}\ \emph {et~al.}(2022)\citenamefont {Chen}, \citenamefont {Li}, \citenamefont {Omori}, \citenamefont {Yamaguchi}, \citenamefont {Ikuta},\ and\ \citenamefont {Takahashi}}]{Chen2022d}%
  \BibitemOpen
  \bibfield  {author} {\bibinfo {author} {\bibfnamefont {K.-T.}\ \bibnamefont {Chen}}, \bibinfo {author} {\bibfnamefont {Q.-Y.}\ \bibnamefont {Li}}, \bibinfo {author} {\bibfnamefont {T.}~\bibnamefont {Omori}}, \bibinfo {author} {\bibfnamefont {Y.}~\bibnamefont {Yamaguchi}}, \bibinfo {author} {\bibfnamefont {T.}~\bibnamefont {Ikuta}},\ and\ \bibinfo {author} {\bibfnamefont {K.}~\bibnamefont {Takahashi}},\ }\bibfield  {title} {\bibinfo {title} {Slip length measurement in rectangular graphene nanochannels with a 3d flow analysis},\ }\href {https://doi.org/10.1016/j.carbon.2021.12.048} {\bibfield  {journal} {\bibinfo  {journal} {Carbon}\ }\textbf {\bibinfo {volume} {189}},\ \bibinfo {pages} {162} (\bibinfo {year} {2022})}\BibitemShut {NoStop}%
\bibitem [{\citenamefont {Barrat}\ and\ \citenamefont {Bocquet}(1999{\natexlab{b}})}]{Barrat1999}%
  \BibitemOpen
  \bibfield  {author} {\bibinfo {author} {\bibfnamefont {J.-L.}\ \bibnamefont {Barrat}}\ and\ \bibinfo {author} {\bibfnamefont {L.}~\bibnamefont {Bocquet}},\ }\bibfield  {title} {\bibinfo {title} {Influence of wetting properties on hydrodynamic boundary conditions at a fluid/solid interface},\ }\href {https://doi.org/10.1039/a809733j} {\bibfield  {journal} {\bibinfo  {journal} {Faraday Discussions}\ }\textbf {\bibinfo {volume} {112}},\ \bibinfo {pages} {119} (\bibinfo {year} {1999}{\natexlab{b}})}\BibitemShut {NoStop}%
\bibitem [{\citenamefont {Hansen}\ \emph {et~al.}(2011)\citenamefont {Hansen}, \citenamefont {Todd},\ and\ \citenamefont {Daivis}}]{Hansen2011}%
  \BibitemOpen
  \bibfield  {author} {\bibinfo {author} {\bibfnamefont {J.~S.}\ \bibnamefont {Hansen}}, \bibinfo {author} {\bibfnamefont {B.~D.}\ \bibnamefont {Todd}},\ and\ \bibinfo {author} {\bibfnamefont {P.~J.}\ \bibnamefont {Daivis}},\ }\bibfield  {title} {\bibinfo {title} {Prediction of fluid velocity slip at solid surfaces},\ }\href {https://doi.org/10.1103/physreve.84.016313} {\bibfield  {journal} {\bibinfo  {journal} {Physical Review E}\ }\textbf {\bibinfo {volume} {84}},\ \bibinfo {pages} {016313} (\bibinfo {year} {2011})}\BibitemShut {NoStop}%
\bibitem [{\citenamefont {Hadjiconstantinou}\ and\ \citenamefont {Swisher}(2022)}]{Hadjiconstantinou2022}%
  \BibitemOpen
  \bibfield  {author} {\bibinfo {author} {\bibfnamefont {N.~G.}\ \bibnamefont {Hadjiconstantinou}}\ and\ \bibinfo {author} {\bibfnamefont {M.~M.}\ \bibnamefont {Swisher}},\ }\bibfield  {title} {\bibinfo {title} {On the equivalence of nonequilibrium and equilibrium measurements of slip in molecular dynamics simulations},\ }\href {https://doi.org/10.1103/physrevfluids.7.114203} {\bibfield  {journal} {\bibinfo  {journal} {Physical Review Fluids}\ }\textbf {\bibinfo {volume} {7}},\ \bibinfo {pages} {114203} (\bibinfo {year} {2022})}\BibitemShut {NoStop}%
\bibitem [{\citenamefont {Corral-Casas}\ \emph {et~al.}(2024)\citenamefont {Corral-Casas}, \citenamefont {Chen}, \citenamefont {Borg},\ and\ \citenamefont {Gibelli}}]{CorralCasas2024}%
  \BibitemOpen
  \bibfield  {author} {\bibinfo {author} {\bibfnamefont {C.}~\bibnamefont {Corral-Casas}}, \bibinfo {author} {\bibfnamefont {Y.}~\bibnamefont {Chen}}, \bibinfo {author} {\bibfnamefont {M.~K.}\ \bibnamefont {Borg}},\ and\ \bibinfo {author} {\bibfnamefont {L.}~\bibnamefont {Gibelli}},\ }\bibfield  {title} {\bibinfo {title} {Density and confinement effects on fluid velocity slip},\ }\href {https://doi.org/10.1103/physrevfluids.9.034201} {\bibfield  {journal} {\bibinfo  {journal} {Physical Review Fluids}\ }\textbf {\bibinfo {volume} {9}},\ \bibinfo {pages} {034201} (\bibinfo {year} {2024})}\BibitemShut {NoStop}%
\bibitem [{\citenamefont {Lichter}\ \emph {et~al.}(2004)\citenamefont {Lichter}, \citenamefont {Roxin},\ and\ \citenamefont {Mandre}}]{Lichter2004}%
  \BibitemOpen
  \bibfield  {author} {\bibinfo {author} {\bibfnamefont {S.}~\bibnamefont {Lichter}}, \bibinfo {author} {\bibfnamefont {A.}~\bibnamefont {Roxin}},\ and\ \bibinfo {author} {\bibfnamefont {S.}~\bibnamefont {Mandre}},\ }\bibfield  {title} {\bibinfo {title} {Mechanisms for liquid slip at solid surfaces},\ }\href {https://doi.org/10.1103/physrevlett.93.086001} {\bibfield  {journal} {\bibinfo  {journal} {Physical Review Letters}\ }\textbf {\bibinfo {volume} {93}},\ \bibinfo {pages} {086001} (\bibinfo {year} {2004})}\BibitemShut {NoStop}%
\bibitem [{\citenamefont {Wang}\ and\ \citenamefont {Hadjiconstantinou}(2019)}]{Wang2019}%
  \BibitemOpen
  \bibfield  {author} {\bibinfo {author} {\bibfnamefont {G.~J.}\ \bibnamefont {Wang}}\ and\ \bibinfo {author} {\bibfnamefont {N.~G.}\ \bibnamefont {Hadjiconstantinou}},\ }\bibfield  {title} {\bibinfo {title} {Universal molecular-kinetic scaling relation for slip of a simple fluid at a solid boundary},\ }\href {https://doi.org/10.1103/physrevfluids.4.064201} {\bibfield  {journal} {\bibinfo  {journal} {Physical Review Fluids}\ }\textbf {\bibinfo {volume} {4}},\ \bibinfo {pages} {064201} (\bibinfo {year} {2019})}\BibitemShut {NoStop}%
\bibitem [{\citenamefont {Hadjiconstantinou}(2021)}]{Hadjiconstantinou2021}%
  \BibitemOpen
  \bibfield  {author} {\bibinfo {author} {\bibfnamefont {N.}~\bibnamefont {Hadjiconstantinou}},\ }\bibfield  {title} {\bibinfo {title} {An atomistic model for the {N}avier slip condition},\ }\href {https://doi.org/10.1017/jfm.2020.1103} {\bibfield  {journal} {\bibinfo  {journal} {Journal of Fluid Mechanics}\ }\textbf {\bibinfo {volume} {912}},\ \bibinfo {pages} {1} (\bibinfo {year} {2021})}\BibitemShut {NoStop}%
\bibitem [{\citenamefont {Shan}\ \emph {et~al.}(2022)\citenamefont {Shan}, \citenamefont {Wang}, \citenamefont {Wang}, \citenamefont {Zhang},\ and\ \citenamefont {Guo}}]{Shan2022}%
  \BibitemOpen
  \bibfield  {author} {\bibinfo {author} {\bibfnamefont {B.}~\bibnamefont {Shan}}, \bibinfo {author} {\bibfnamefont {P.}~\bibnamefont {Wang}}, \bibinfo {author} {\bibfnamefont {R.}~\bibnamefont {Wang}}, \bibinfo {author} {\bibfnamefont {Y.}~\bibnamefont {Zhang}},\ and\ \bibinfo {author} {\bibfnamefont {Z.}~\bibnamefont {Guo}},\ }\bibfield  {title} {\bibinfo {title} {Molecular kinetic modelling of nanoscale slip flow using a continuum approach},\ }\href {https://doi.org/10.1017/jfm.2022.186} {\bibfield  {journal} {\bibinfo  {journal} {Journal of Fluid Mechanics}\ }\textbf {\bibinfo {volume} {939}},\ \bibinfo {pages} {1} (\bibinfo {year} {2022})}\BibitemShut {NoStop}%
\bibitem [{\citenamefont {Hadjiconstantinou}(2024)}]{Hadjiconstantinou2024}%
  \BibitemOpen
  \bibfield  {author} {\bibinfo {author} {\bibfnamefont {N.~G.}\ \bibnamefont {Hadjiconstantinou}},\ }\bibfield  {title} {\bibinfo {title} {Molecular mechanics of liquid and gas slip flow},\ }\href {https://doi.org/10.1146/annurev-fluid-121021-014808} {\bibfield  {journal} {\bibinfo  {journal} {Annual Review of Fluid Mechanics}\ }\textbf {\bibinfo {volume} {56}},\ \bibinfo {pages} {435} (\bibinfo {year} {2024})}\BibitemShut {NoStop}%
\bibitem [{\citenamefont {Bhatnagar}\ \emph {et~al.}(1954)\citenamefont {Bhatnagar}, \citenamefont {Gross},\ and\ \citenamefont {Krook}}]{Bhatnagar1954}%
  \BibitemOpen
  \bibfield  {author} {\bibinfo {author} {\bibfnamefont {P.~L.}\ \bibnamefont {Bhatnagar}}, \bibinfo {author} {\bibfnamefont {E.~P.}\ \bibnamefont {Gross}},\ and\ \bibinfo {author} {\bibfnamefont {M.}~\bibnamefont {Krook}},\ }\bibfield  {title} {\bibinfo {title} {A model for collision processes in gases. {I}. small amplitude processes in charged and neutral one-component systems},\ }\href {https://doi.org/10.1103/PhysRev.94.511} {\bibfield  {journal} {\bibinfo  {journal} {Physical Review}\ }\textbf {\bibinfo {volume} {94}},\ \bibinfo {pages} {511} (\bibinfo {year} {1954})}\BibitemShut {NoStop}%
\bibitem [{\citenamefont {Welander}(1954)}]{Welander1954}%
  \BibitemOpen
  \bibfield  {author} {\bibinfo {author} {\bibfnamefont {P.}~\bibnamefont {Welander}},\ }\bibfield  {title} {\bibinfo {title} {On the temperature jump in a rarefied gas},\ }\href@noop {} {\bibfield  {journal} {\bibinfo  {journal} {Arkiv Fysik}\ }\textbf {\bibinfo {volume} {7}},\ \bibinfo {pages} {507} (\bibinfo {year} {1954})}\BibitemShut {NoStop}%
\bibitem [{\citenamefont {Aoki}\ \emph {et~al.}(2022)\citenamefont {Aoki}, \citenamefont {Giovangigli},\ and\ \citenamefont {Kosuge}}]{Aoki2022}%
  \BibitemOpen
  \bibfield  {author} {\bibinfo {author} {\bibfnamefont {K.}~\bibnamefont {Aoki}}, \bibinfo {author} {\bibfnamefont {V.}~\bibnamefont {Giovangigli}},\ and\ \bibinfo {author} {\bibfnamefont {S.}~\bibnamefont {Kosuge}},\ }\bibfield  {title} {\bibinfo {title} {Boundary conditions for the boltzmann equation~from gas-surface interaction kinetic models},\ }\href {https://doi.org/10.1103/physreve.106.035306} {\bibfield  {journal} {\bibinfo  {journal} {Physical Review E}\ }\textbf {\bibinfo {volume} {106}},\ \bibinfo {pages} {035306} (\bibinfo {year} {2022})}\BibitemShut {NoStop}%
\bibitem [{\citenamefont {Aoki}\ \emph {et~al.}(2024)\citenamefont {Aoki}, \citenamefont {Giovangigli}, \citenamefont {Golse},\ and\ \citenamefont {Kosuge}}]{Aoki2024}%
  \BibitemOpen
  \bibfield  {author} {\bibinfo {author} {\bibfnamefont {K.}~\bibnamefont {Aoki}}, \bibinfo {author} {\bibfnamefont {V.}~\bibnamefont {Giovangigli}}, \bibinfo {author} {\bibfnamefont {F.}~\bibnamefont {Golse}},\ and\ \bibinfo {author} {\bibfnamefont {S.}~\bibnamefont {Kosuge}},\ }\bibfield  {title} {\bibinfo {title} {The physisorbate-layer problem arising in kinetic theory of gas–surface interaction},\ }\href {https://doi.org/10.1007/s10955-024-03270-3} {\bibfield  {journal} {\bibinfo  {journal} {Journal of Statistical Physics}\ }\textbf {\bibinfo {volume} {191}},\ \bibinfo {pages} {53} (\bibinfo {year} {2024})}\BibitemShut {NoStop}%
\bibitem [{\citenamefont {Ohwada}\ \emph {et~al.}(1989{\natexlab{b}})\citenamefont {Ohwada}, \citenamefont {Sone},\ and\ \citenamefont {Aoki}}]{Ohwada1989}%
  \BibitemOpen
  \bibfield  {author} {\bibinfo {author} {\bibfnamefont {T.}~\bibnamefont {Ohwada}}, \bibinfo {author} {\bibfnamefont {Y.}~\bibnamefont {Sone}},\ and\ \bibinfo {author} {\bibfnamefont {K.}~\bibnamefont {Aoki}},\ }\bibfield  {title} {\bibinfo {title} {Numerical analysis of the shear and thermal creep flows of a rarefied gas over a plane wall on the basis of the linearized {B}oltzmann equation for hard-sphere molecules},\ }\href {https://doi.org/10.1063/1.857304} {\bibfield  {journal} {\bibinfo  {journal} {Physics of Fluids}\ }\textbf {\bibinfo {volume} {1}},\ \bibinfo {pages} {1588} (\bibinfo {year} {1989}{\natexlab{b}})}\BibitemShut {NoStop}%
\bibitem [{\citenamefont {Holway}(1966)}]{Holway1966}%
  \BibitemOpen
  \bibfield  {author} {\bibinfo {author} {\bibfnamefont {L.~H.}\ \bibnamefont {Holway}},\ }\bibfield  {title} {\bibinfo {title} {New statistical models for kinetic theory: Methods of construction},\ }\href {https://doi.org/10.1063/1.1761920} {\bibfield  {journal} {\bibinfo  {journal} {Physics of Fluids}\ }\textbf {\bibinfo {volume} {9}},\ \bibinfo {pages} {1658} (\bibinfo {year} {1966})}\BibitemShut {NoStop}%
\bibitem [{\citenamefont {Shakhov}(1972)}]{Shakhov1972}%
  \BibitemOpen
  \bibfield  {author} {\bibinfo {author} {\bibfnamefont {E.~M.}\ \bibnamefont {Shakhov}},\ }\bibfield  {title} {\bibinfo {title} {Generalization of the krook kinetic relaxation equation},\ }\href {https://doi.org/10.1007/bf01029546} {\bibfield  {journal} {\bibinfo  {journal} {Fluid Dynamics}\ }\textbf {\bibinfo {volume} {3}},\ \bibinfo {pages} {95} (\bibinfo {year} {1972})}\BibitemShut {NoStop}%
\bibitem [{\citenamefont {Israelachvili}(2011)}]{Israelachvili2011}%
  \BibitemOpen
  \bibfield  {author} {\bibinfo {author} {\bibfnamefont {J.~N.}\ \bibnamefont {Israelachvili}},\ }\href@noop {} {\emph {\bibinfo {title} {Intermolecular and {S}urface {F}orces}}},\ \bibinfo {edition} {3rd}\ ed.\ (\bibinfo  {publisher} {Elsevier},\ \bibinfo {year} {2011})\BibitemShut {NoStop}%
\bibitem [{\citenamefont {Sone}\ and\ \citenamefont {Takata}(1992)}]{Sone1992}%
  \BibitemOpen
  \bibfield  {author} {\bibinfo {author} {\bibfnamefont {Y.}~\bibnamefont {Sone}}\ and\ \bibinfo {author} {\bibfnamefont {S.}~\bibnamefont {Takata}},\ }\bibfield  {title} {\bibinfo {title} {Discontinuity of the velocity distribution function in a rarefied gas around a convex body and the s layer at the bottom of the knudsen layer},\ }\href {https://doi.org/10.1080/00411459208203796} {\bibfield  {journal} {\bibinfo  {journal} {Transport Theory and Statistical Physics}\ }\textbf {\bibinfo {volume} {21}},\ \bibinfo {pages} {501} (\bibinfo {year} {1992})}\BibitemShut {NoStop}%
\bibitem [{\citenamefont {Sugimoto}\ and\ \citenamefont {Sone}(1992)}]{Sugimoto1992}%
  \BibitemOpen
  \bibfield  {author} {\bibinfo {author} {\bibfnamefont {H.}~\bibnamefont {Sugimoto}}\ and\ \bibinfo {author} {\bibfnamefont {Y.}~\bibnamefont {Sone}},\ }\bibfield  {title} {\bibinfo {title} {Numerical analysis of steady flows of a gas evaporating from its cylindrical condensed phase on the basis of kinetic theory},\ }\href {https://doi.org/10.1063/1.858313} {\bibfield  {journal} {\bibinfo  {journal} {Physics of Fluids}\ }\textbf {\bibinfo {volume} {4}},\ \bibinfo {pages} {419} (\bibinfo {year} {1992})}\BibitemShut {NoStop}%
\bibitem [{\citenamefont {Tsuji}\ and\ \citenamefont {Aoki}(2013)}]{Tsuji2013}%
  \BibitemOpen
  \bibfield  {author} {\bibinfo {author} {\bibfnamefont {T.}~\bibnamefont {Tsuji}}\ and\ \bibinfo {author} {\bibfnamefont {K.}~\bibnamefont {Aoki}},\ }\bibfield  {title} {\bibinfo {title} {Moving boundary problems for a rarefied gas: Spatially one-dimensional case},\ }\href {https://doi.org/10.1016/j.jcp.2013.05.017} {\bibfield  {journal} {\bibinfo  {journal} {Journal of Computational Physics}\ }\textbf {\bibinfo {volume} {250}},\ \bibinfo {pages} {574} (\bibinfo {year} {2013})}\BibitemShut {NoStop}%
\bibitem [{\citenamefont {Niimi}(1971)}]{Niimi1971}%
  \BibitemOpen
  \bibfield  {author} {\bibinfo {author} {\bibfnamefont {H.}~\bibnamefont {Niimi}},\ }\bibfield  {title} {\bibinfo {title} {Thermal creep flow of rarefied gas between two parallel plates},\ }\href@noop {} {\bibfield  {journal} {\bibinfo  {journal} {Journal of the Physical Society of Japan}\ }\textbf {\bibinfo {volume} {30}},\ \bibinfo {pages} {572} (\bibinfo {year} {1971})}\BibitemShut {NoStop}%
\bibitem [{\citenamefont {Takata}\ and\ \citenamefont {Taguchi}(2017)}]{Takata2017}%
  \BibitemOpen
  \bibfield  {author} {\bibinfo {author} {\bibfnamefont {S.}~\bibnamefont {Takata}}\ and\ \bibinfo {author} {\bibfnamefont {S.}~\bibnamefont {Taguchi}},\ }\bibfield  {title} {\bibinfo {title} {Gradient divergence of fluid-dynamic quantities in rarefied gases on smooth boundaries},\ }\href {https://doi.org/10.1007/s10955-017-1850-7} {\bibfield  {journal} {\bibinfo  {journal} {Journal of Statistical Physics}\ }\textbf {\bibinfo {volume} {168}},\ \bibinfo {pages} {1319} (\bibinfo {year} {2017})}\BibitemShut {NoStop}%
\bibitem [{\citenamefont {Yamaguchi}\ and\ \citenamefont {Kikugawa}(2021)}]{Yamaguchi2021}%
  \BibitemOpen
  \bibfield  {author} {\bibinfo {author} {\bibfnamefont {H.}~\bibnamefont {Yamaguchi}}\ and\ \bibinfo {author} {\bibfnamefont {G.}~\bibnamefont {Kikugawa}},\ }\bibfield  {title} {\bibinfo {title} {Molecular dynamics study on flow structure inside a thermal transpiration flow field},\ }\href {https://doi.org/10.1063/5.0034146} {\bibfield  {journal} {\bibinfo  {journal} {Physics of Fluids}\ }\textbf {\bibinfo {volume} {33}},\ \bibinfo {pages} {012005} (\bibinfo {year} {2021})}\BibitemShut {NoStop}%
\bibitem [{\citenamefont {Allen}\ and\ \citenamefont {Tildesley}(1987)}]{Allen1987}%
  \BibitemOpen
  \bibfield  {author} {\bibinfo {author} {\bibfnamefont {M.~P.}\ \bibnamefont {Allen}}\ and\ \bibinfo {author} {\bibfnamefont {D.~J.}\ \bibnamefont {Tildesley}},\ }\href@noop {} {\emph {\bibinfo {title} {Computer Simulation of Liquids}}}\ (\bibinfo  {publisher} {Oxford University Press},\ \bibinfo {year} {1987})\BibitemShut {NoStop}%
\bibitem [{\citenamefont {Takata}\ \emph {et~al.}(2021)\citenamefont {Takata}, \citenamefont {Matsumoto},\ and\ \citenamefont {Hattori}}]{Takata2021}%
  \BibitemOpen
  \bibfield  {author} {\bibinfo {author} {\bibfnamefont {S.}~\bibnamefont {Takata}}, \bibinfo {author} {\bibfnamefont {T.}~\bibnamefont {Matsumoto}},\ and\ \bibinfo {author} {\bibfnamefont {M.}~\bibnamefont {Hattori}},\ }\bibfield  {title} {\bibinfo {title} {Kinetic model for the phase transition of the van der {W}aals fluid},\ }\href {https://doi.org/10.1103/physreve.103.062110} {\bibfield  {journal} {\bibinfo  {journal} {Physical Review E}\ }\textbf {\bibinfo {volume} {103}},\ \bibinfo {pages} {062110} (\bibinfo {year} {2021})}\BibitemShut {NoStop}%
\bibitem [{\citenamefont {Taguchi}\ and\ \citenamefont {Tsuji}(2020)}]{Taguchi2020}%
  \BibitemOpen
  \bibfield  {author} {\bibinfo {author} {\bibfnamefont {S.}~\bibnamefont {Taguchi}}\ and\ \bibinfo {author} {\bibfnamefont {T.}~\bibnamefont {Tsuji}},\ }\bibfield  {title} {\bibinfo {title} {On the motion of slightly rarefied gas induced by a discontinuous surface temperature},\ }\href {https://doi.org/10.1017/jfm.2020.332} {\bibfield  {journal} {\bibinfo  {journal} {Journal of Fluid Mechanics}\ }\textbf {\bibinfo {volume} {897}},\ \bibinfo {pages} {A16} (\bibinfo {year} {2020})}\BibitemShut {NoStop}%
\bibitem [{\citenamefont {Takata}\ and\ \citenamefont {Sone}(1995)}]{Takata1995a}%
  \BibitemOpen
  \bibfield  {author} {\bibinfo {author} {\bibfnamefont {S.}~\bibnamefont {Takata}}\ and\ \bibinfo {author} {\bibfnamefont {Y.}~\bibnamefont {Sone}},\ }\bibfield  {title} {\bibinfo {title} {Flow induced around a sphere with a non-uniform surface temperature in a rarefied gas, with application to the drag and thermal force problems of a spherical particle with an arbitrary thermal conductivity},\ }\href {https://www.scopus.com/inward/record.uri?eid=2-s2.0-0029178681&partnerID=40&md5=e9775eca23694ded0b5f774ebf0c67df} {\bibfield  {journal} {\bibinfo  {journal} {European Journal of Mechanics, B/Fluids}\ }\textbf {\bibinfo {volume} {14}},\ \bibinfo {pages} {487 – 518} (\bibinfo {year} {1995})},\ \bibinfo {note} {cited by: 30}\BibitemShut {NoStop}%
\bibitem [{\citenamefont {W{\"u}rger}(2010)}]{Wuerger2010}%
  \BibitemOpen
  \bibfield  {author} {\bibinfo {author} {\bibfnamefont {A.}~\bibnamefont {W{\"u}rger}},\ }\bibfield  {title} {\bibinfo {title} {Thermal non-equilibrium transport in colloids},\ }\href {https://doi.org/10.1088/0034-4885/73/12/126601} {\bibfield  {journal} {\bibinfo  {journal} {Reports on Progress in Physics}\ }\textbf {\bibinfo {volume} {73}},\ \bibinfo {pages} {126601} (\bibinfo {year} {2010})}\BibitemShut {NoStop}%
\bibitem [{\citenamefont {Bardos}\ \emph {et~al.}(1986)\citenamefont {Bardos}, \citenamefont {Caflisch},\ and\ \citenamefont {Nicolaenko}}]{Bardos1986}%
  \BibitemOpen
  \bibfield  {author} {\bibinfo {author} {\bibfnamefont {C.}~\bibnamefont {Bardos}}, \bibinfo {author} {\bibfnamefont {R.~E.}\ \bibnamefont {Caflisch}},\ and\ \bibinfo {author} {\bibfnamefont {B.}~\bibnamefont {Nicolaenko}},\ }\bibfield  {title} {\bibinfo {title} {The milne and {K}ramers problems for the {B}oltzmann equation of a hard sphere gas},\ }\href {https://doi.org/10.1002/cpa.3160390304} {\bibfield  {journal} {\bibinfo  {journal} {Communications on Pure and Applied Mathematics}\ }\textbf {\bibinfo {volume} {39}},\ \bibinfo {pages} {323} (\bibinfo {year} {1986})}\BibitemShut {NoStop}%
\bibitem [{\citenamefont {Coron}\ \emph {et~al.}(1988)\citenamefont {Coron}, \citenamefont {Golse},\ and\ \citenamefont {Sulem}}]{Coron1988}%
  \BibitemOpen
  \bibfield  {author} {\bibinfo {author} {\bibfnamefont {F.}~\bibnamefont {Coron}}, \bibinfo {author} {\bibfnamefont {F.}~\bibnamefont {Golse}},\ and\ \bibinfo {author} {\bibfnamefont {C.}~\bibnamefont {Sulem}},\ }\bibfield  {title} {\bibinfo {title} {A classification of well-posed kinetic layer problems},\ }\href {https://doi.org/10.1002/cpa.3160410403} {\bibfield  {journal} {\bibinfo  {journal} {Communications on Pure and Applied Mathematics}\ }\textbf {\bibinfo {volume} {41}},\ \bibinfo {pages} {409} (\bibinfo {year} {1988})}\BibitemShut {NoStop}%
\end{thebibliography}
\end{document}